\documentclass[12pt,a4paper,twoside]{article}
\usepackage{graphicx}
\usepackage{parskip}

\usepackage{yfonts}
\usepackage{amsmath}
\usepackage{amssymb}
\allowdisplaybreaks
\usepackage{xfrac}
\usepackage{setspace}
\usepackage{booktabs}
\usepackage{latexsym}
\usepackage{mathtools}
\usepackage{mathrsfs}
\usepackage{epstopdf}
\usepackage{cite}
\usepackage{url}

\usepackage{breakurl}
\usepackage[hyperfootnotes=false,breaklinks]{hyperref}
\hypersetup{
    colorlinks = false,
    linkcolor = {blue},
    anchorcolor = {black},
    citecolor = {RawSienna},
    filecolor = {cyan},
    menucolor = {red},
    runcolor = {cyan},
    urlcolor = {RawSienna},
		  }
\usepackage[nameinlink,capitalize]{cleveref}
\usepackage{multirow}
\usepackage{caption}
\usepackage[compact]{titlesec}
\usepackage{siunitx}
\usepackage{relsize}
\usepackage[mathscr]{euscript}
\usepackage{pbsi}
\usepackage{rotating}
\usepackage{enumitem}
\usepackage{tabularx}
\usepackage{longtable}
\usepackage{bold-extra}
\usepackage{fancyhdr}
\usepackage[normalem]{ulem}
\usepackage{color}
\usepackage[usenames,dvipsnames,svgnames,table]{xcolor}
\usepackage{bm}
\usepackage[top=2.54cm, bottom=2.54cm, left=2.54cm, right=2.54cm]{geometry}
\usepackage{pdflscape}
\usepackage[switch]{lineno}

\titlespacing\section{0pt}{10pt plus 4pt minus 2pt}{0pt plus 2pt minus 2pt}
\titlespacing\subsection{0pt}{10pt plus 4pt minus 2pt}{0pt plus 2pt minus 2pt}
\titlespacing\subsubsection{0pt}{10pt plus 4pt minus 2pt}{0pt plus 2pt minus 2pt}

\begin{document}
\titleformat*{\section}{\Large\bfseries}
\titleformat*{\subsection}{\large\em\bfseries}
\titleformat*{\subsubsection}{\normalsize\bfseries}
\definecolor{gway}{HTML}{5D5D5D}
\DeclarePairedDelimiter\abs{\lvert}{\rvert}
\makeatletter
\let\oldabs\abs
\def\abs{\@ifstar{\oldabs}{\oldabs*}}
\onehalfspacing
\renewcommand*{\thefootnote}{\fnsymbol{footnote}}
\vspace*{1cm}
\begin{center}
{\bf\em{An inflow-boundary-based Navier-Stokes wave tank: verification and validation for waves propagating over flat and inclined bottoms}}\\[20pt]
{\footnotesize {Shaswat Saincher$^\dag$ and Jyotirmay Banerjee$^\dag\footnote{Corresponding author at: Department of Mechanical Engineering, Sardar Vallabhbhai National Institute of Technology, Surat - 395007, Gujarat, India. Tel: +91 261 220 4145. Fax: +91 261 222 8394. Email: {\texttt{jbaner@med.svnit.ac.in}}}$}}\\
{\footnotesize\emph{$^{\dag}$Mechanical Engineering Department, S.V. National Institute of Technology, Surat, 395 007, India}}
\end{center}
\vspace*{1cm}
\parskip 12pt
\thispagestyle{empty}
\setcounter{page}{1}
\begin{spacing}{0.75}
\begin{center}
\begin{tabularx}{0.9\textwidth}{X}
{\scriptsize{Development of mass-source function based numerical wave tank (NWT) algorithms in the Navier-Stokes (NSE) framework is impeded by multiple design issues such as: (a) optimization of a number of variables characterizing the source region, (b) wave-vorticity interactions and (c) a mandatory requirement of modeling the domain on both sides of the wavemaker. In this paper, we circumvent these hurdles by proposing a volume-preserving inflow-boundary based Navier-Stokes wave tank. Wave generation and propagation is modeled in a two-phase PLIC-VOF set-up. Near-exact volume preservation is achieved (at arbitrarily large steepness) using kinematic stretching that is aimed towards balancing the streamwise momentum between points lying above and below the still water level. Numerical damping of steep waves is prevented by using blended third-order and limiter schemes for momentum advection. In addition, a mesh stair-stepping strategy has been adopted for modeling non-Cartesian immersed boundaries on a staggered variable arrangement. The proposed NWT model is rigorously benchmarked against various wave-propagation scenarios. These include the simulation of: (a) monochromatic waves of various steepnesses, (b) monochromatic waves superimposed with free harmonics, (c) irregular waves in deep water and (d) wave transformation occurring over a submerged trapezoidal bar. Excellent agreement with analytical, numerical and experimental data is reported with both validation as well as verification of the proposed NWT model being established.}}\\
{\scriptsize{\bf{Keywords:}} Navier-Stokes, numerical wave tank, kinematic stretching, blended advection schemes, immersed-boundary treatment}
\end{tabularx}
\end{center}
\end{spacing}
\setcounter{footnote}{0}
\renewcommand*{\thefootnote}{\number\value{footnote}}

\section{Introduction} \label{sec:intro}
\begin{figure}[!ht]
\begin {center}
\includegraphics[trim=0mm 0mm 0mm 0mm, clip, width = 15cm]{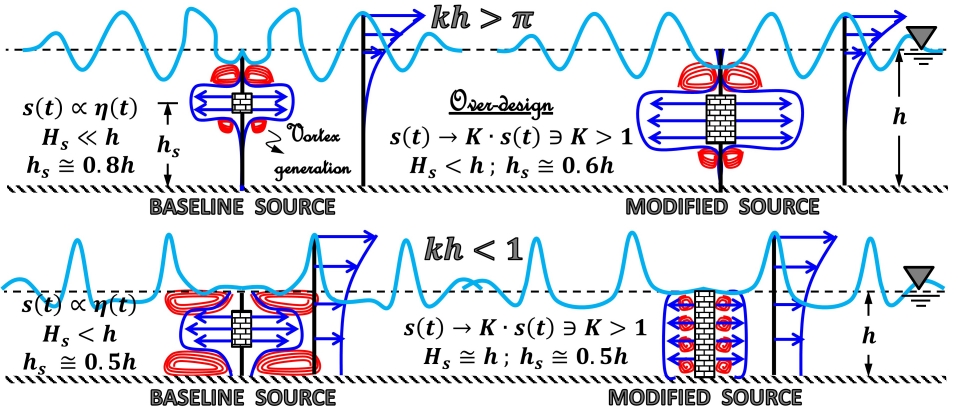}
\end {center}
\caption{\emph{Descriptive sketch illustrating a comparison between baseline\cite{lin99} and modified mass-source based {\emph{\textsf{WM}}} designs for wave generation in deep $(kh>\pi)$ \cite{peric15} and near-shallow water $(kh<1)$ \cite{sas17a}.}}
\label{fig:src_WM_designs}
\end{figure}

\begin{figure}[!ht]
\begin {center}
\includegraphics[trim=0mm 0mm 0mm 0mm, clip, width = 15cm]{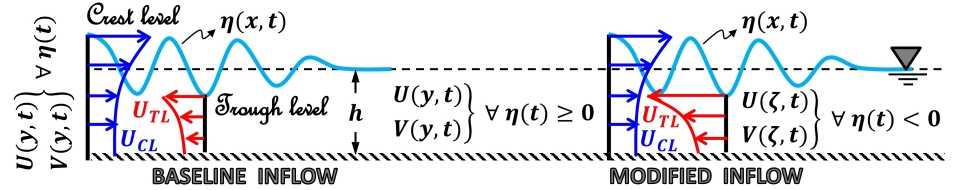}
\end {center}
\caption{\emph{Descriptive sketch highlighting the conceptual differences between baseline\cite{lin98} and (proposed kinematically stretched) modified inflow-boundary based {\emph{\textsf{WM}}} \cite{sas17b} designs; {\emph{CL}} and {\emph{TL}} denote crest and trough elevations respectively.}}
\label{fig:infkst}
\end{figure}

Following the advent of computational fluid dynamics (CFD), numerical wave tanks (NWTs) have emerged as a much needed secondary standard to wave flumes and basins. Since their inception, NWT algorithms have been applied towards addressing a variety of challenging problems such as wave-structure-interaction (WSI) \cite{sriram06b,queutey07}, wave-breaking-induced vortex dynamics and air-sea interaction \cite{lubin15,wei18}, fluid-ship-ice interactions \cite{kim19} and hydrodynamic appraisal of wave energy converters (WECs) \cite{kamath15}. These studies represent state-of-the-art in the field of NWT development and (a majority) have been carried out using {\emph{two-phase}} Navier-Stokes equations (NSE); especially for WSI and coastal problems. \\
Albeit the level of fidelity achieved (in comparison to Fully Non-linear Potential Theory (FNPT) models \cite{sriram06a,sriram06b}), wave generation in the NSE framework is challenging. As a matter of fact, the free surface elevation $(\eta)$ and velocity potential $(\phi)$ do not emerge as explicit variables in the NSE. Instead, $\eta$ and $\phi$ need to be implicitly linked to the continuity and momentum equations through free-surface modeling techniques and wave theories. The integration of analytic expressions for $\eta,\phi$ (from wave theory) into the NSE has to be realized by means of ``numerical wavemakers'' (abbreviated {\textsf{WM}} from this point onward). \\
Over the past two decades, several numerical wavemakers have been proposed such as inflow-boundary-based \cite{lin98,meskers02,sas17b}, mass-source-based \cite{lin99,peric15,sas17a}, relaxation-zone-based \cite{jacob12}, internal-inlet-based \cite{hafsia09}, moving-boundary-based \cite{anbarsooz13} and momentum-source-based \cite{choi09} techniques. Of these, the mass-source function and inflow-boundary techniques are (probably) the first and most extensively used NSE-based {\textsf{WM}}s. \\
In case of the mass-source {\textsf{WM}} \cite{lin99}, surface waves are generated through periodic ingestion/ejection of mass from a group of computational cells termed as the ``source region'' (cf. \autoref{fig:src_WM_designs}). 
Within the source region, zero-divergence condition of the velocity field $(\vec{V})$ is deliberately violated using: $\vec{\nabla}\bullet \vec{V}=s(t)$ where $s(t)\left(\propto \eta(t)\right)$ is strength of the source region. The mass-source {\textsf{WM}} is advantageous in that specification of $\eta(t)$ is sufficient for prescribing $s(t)$. Further, given that $\int\limits_T \eta(t)\cong 0$ for any (arbitrarily non-linear) wave theory, the source {\textsf{WM}} does not alter water-phase mass over a wave period $(T)$. Having said that, the mass-source {\textsf{WM}} is nonetheless faced with many design issues:
\begin{itemize}[noitemsep]
\item the mass-source {\textsf{WM}} has multiple design variables associated, namely, height $(H_s)$, width $(W_s)$ and placement $(h_s)$ (cf. \autoref{fig:src_WM_designs}); this complicates geometric optimization of the source region \cite{sas17a}. 
\item if a wide range of wave heights is to be simulated, the mass-source {\textsf{WM}} is only suitable in intermediate water $(1.0 \leq kh \leq 2.5)$ \cite{sas17a} where $U \propto y$. With increasing relative depth $(kh>2.5)$, $U \propto {\mathrm{e}}^{y}$ which forces one to continually increase $h_s$ (cf. \autoref{fig:src_WM_designs}) to prevent damping of source-injected momentum. Unfortunately, 
\begin{itemize}
\item there is a limit to increasing $h_s$ since source height $(H_s)$ must remain finite.
\item $h_s \uparrow\, ; H_s \downarrow$ induces wave distortion, especially when $H/\lambda > 0.03$ \cite{sas17a}. 
\end{itemize}
This limitation of the source-function {\textsf{WM}} in deep water was resolved by Peri\'{c} and Abdel-Maksoud \cite{peric15} (for $kh \approx 40$) by proposing a modified source design which involved decreasing $h_s$ and over-designing the source-strength: $s(t)\rightarrow K\cdot s(t) \ni K \gg 1$ (cf. \autoref{fig:src_WM_designs}). It was demonstrated that said design strategy is effective in eliminating wave distortion during steep wave generation in deep water \cite{peric15,sas17a}.  
\item owing to jet-like flows emanating from the source region \cite{peric15}, wave-vorticity interactions (WVI) become particularly intense in $kh<1$ which induce loss of source-injected momentum (to viscous effects), wave distortion and (eventual) height reduction \cite{lin99,sas17a}. The extent of WVI induced by the source {\textsf{WM}} in $kh<1$ appears to be governed by the order of streamwise momentum induced under wave crests $(U_{y=h})$ which is in turn a function of the shallow-water non-linearity $(H/h)$. 
\item aforementioned height damping and WVI in $kh<1$ cannot be eliminated through Peri\'{c} and Abdel-Maksoud's \cite{peric15} design strategy. Owing to negligible damping of wave momentum in near-shallow water $\left(\frac{\partial U}{\partial y} \sim 0 \leftarrow kh<1\right)$, strong WVI would occur regardless of where the source is placed along the water column. In such a scenario, even if $s(t)$ were over-designed to compensate for height reduction, it would in fact {\emph{increase}} WVI and aggravate height damping. In \cite{sas17a}, it has been quantitatively demonstrated by the authors that WVI in $kh<1$ can be arrested by stretching the source region to occupy the entire water depth ($H_s\cong h$; cf. \autoref{fig:src_WM_designs}). Then, wave damping (resulting from an increase in source area) could be arrested by mildly over-designing the source strength: $s(t)\rightarrow K\cdot s(t) \ni K>1$.
\item an obvious shortcoming of the mass-source {\textsf{WM}} is the requirement of modeling the domain on both sides of the wavemaker (cf. \autoref{fig:src_WM_designs}) which necessitates wave absorption at both ends of the NWT. It is a general practice that placement of the source region is offset from the center towards the western end of the NWT so as to facilitate a longer ``wave simulation region'' \cite{sas17a}. Hence, the portion of the NWT westward of the {\textsf{WM}} serves no research purpose and is a computational liability.                      
\end{itemize}
Given the above-mentioned shortcomings, various empirical design strategies have been proposed for the mass-source {\textsf{WM}} \cite{lin99,peric15,sas17a}. Despite these advances, the fact that {\textsf{WM}} design requires significant alterations with changing relative depth $(kh)$ and steepness $(H/\lambda)$ proves detrimental to the robustness of a proposed NWT algorithm. Aim of the present work is to demonstrate that said limitations of the mass-source technique could be overcome using an inflow-boundary-based {\textsf{WM}} formulation.\\
In case of the inflow-boundary technique, analytic expressions for velocities $U(t),V(t)$ and elevation $\eta(t)$ are directly specified at a vertical boundary of the NWT (cf. \autoref{fig:infkst}). Thus, it is sufficient to model the domain at one end of the {\textsf{WM}}. Further, unlike the mass-source {\textsf{WM}}, water volume is influxed (under crests)/effluxed (under troughs) rather than ejected/ingested through the boundary. Hence there is {\emph{no vortex formation}} in the {\textsf{WM}} near-field thereby precluding WVI. Most importantly, given that the specification of $U(t),V(t)$ and $\eta(t)$ is dictated by wave theory \cite{lin98,meskers02}, the baseline version of the inflow technique (cf. \autoref{fig:infkst}) has virtually {\emph{zero}} design variables requiring optimization. \\
However, the baseline inflow {\textsf{WM}} is susceptible to inducing {\emph{wave setup}} \cite{horko07,sas17b}. Said susceptibility is attributable to the fact that, with increasing non-linearity of the waves, there emerges a streamwise momentum discrepancy $|U_{CL}| \gg |U_{TL}|$ (subscripts $CL$ and $TL$ represent crest and trough elevations respectively) such that water volume influxed under crests is considerably greater than volume effluxed under troughs. This could be mathematically stated as: $\int\limits_{T}\int\limits_{h+\eta(t)} U(y,t)\,\mathrm{d}y\,\mathrm{d}t = \mathcal{V}_{+}$ where $\mathcal{V}_{+}$ is the water volume added through the inflow boundary over a wave period; $\mathcal{V}_{+} > 0$ holds for any (arbitrarily non-linear) wave theory \cite{dean91,svendsen06}. Over multiple wave generation cycles, the added volume becomes sufficiently large to induce wave setup in the NWT. 
\\
To the best of the authors' knowledge, there haven't been diligent efforts towards improving volume preservation characteristics of inflow-boundary-based {\textsf{WM}}s in the literature. In this paper, we propose to improve volume preservation characteristics of the baseline inflow {\textsf{WM}} using a novel Kinematic Stretching Technique (abbreviated KST from this point onward). Broadly speaking, a KST alters $U(y,t)$ (predicted by wave theory) by replacing $y$ with a ``stretched'' coordinate $\zeta(\propto \eta(t))$ which effectively stretches/compresses the velocity profile at points (considered along the wave) lying above/below the SWL. The central motivation underlying the development of KSTs is towards improving Airy theory-based estimations of hydrodynamic loads induced by irregular waves on offshore structures \cite{wheeler70,mohdzaki14}. In this regard, several attempts have been made to improve the estimation of $U(y,t)$ above the SWL using KSTs such as: (a) Wheeler stretching \cite{wheeler70}, (b) Vertical stretching \cite{horko07}, (c) Extrapolation stretching \cite{meskers02}, (d) Effective node elevation method \cite{mohdzaki14} and (e) Effective water depth method \cite{mohdzaki14}. It can be verified that each of the aforementioned KSTs introduces: $U_{CL}^{KST}<U_{CL}^{Airy}\,\,;\,\,\left|U_{TL}^{KST}\right| \geq \left|U_{TL}^{Airy}\right|$. These modifications would invariably result in a reduction in ${\mathcal{V}}_{+}$. This indicates that a KST could be employed to design inflow-boundary based {\textsf{WM}}s with superior volume conservation characteristics. However, an existing KST-based inflow-boundary {\textsf{WM}} would prove severely limited during generation of strongly non-linear waves which necessitate higher order wave theories (abbreviated HoT from this point onward) for accurate kinematic description. Given that $U_{CL}^{Airy} < U_{CL}^{HoT}$, it can be anticipated that $U_{CL}^{KST} \ll U_{CL}^{HoT}$. Hence, applying existing KSTs to strongly non-linear wave generation would under-predict $U_{CL}$ thereby defeating the very purpose of a HoT and (possibly) leading to height reduction. Another concern is regarding the {\emph{extent}} upto which $\mathcal{V}_+$ could be minimized for strongly non-linear waves through existing KSTs. \\      
From the above discussion, a need emerges to develop a KST-based inflow boundary {\textsf{WM}} that: (a) is applicable to non-linear waves governed by HoTs, (b) does not alter $U^{HoT}_{CL}$ and (c) preserves water volume for any given set of wave characteristics $h,H,T$. In the present work, we achieve this by proposing the ``modified inflow technique'' (cf. \autoref{fig:infkst}) as a robust wave generation methodology for NSE-based NWTs \cite{sas17b}. In case of modified inflow, $U(y,t)\,\forall\, y\geq h$ is predicted using a HoT (namely {\texttt{Stokes V}}) whilst $U(y,t)\,\forall\, y<h$ is predicted using a modified form of Wheeler stretching (KST). Hence, during wave generation, only the troughs are kinematically strengthened leaving the crest velocities (predicted by HoT) unaltered. It is demonstrated that the modified inflow {\textsf{WM}} can be used to achieve $\mathcal{V}_{+} \cong 0$ for any (arbitrarily non-linear) target wave design. When compared against the mass-source {\textsf{WM}}, the modified inflow technique is particularly advantageous in that the latter involves only one {\textsf{WM}} design variable necessitating parametric optimization compared to four in case of the former. Robustness of the proposed NWT model is further improved through the inclusion of blended schemes for momentum advection to prevent numerical damping of steep waves $(H/\lambda>0.03)$. The proposed NWT algorithm is benchmarked against various monochromatic and polychromatic wave generation as well as wave-transformation scenarios. It is demonstrated that the simulations show excellent agreement with analytic predictions as well as existing numerical findings and experimental measurements reported in the literature. \\
Rest of the paper is structured as follows: the mathematical model of the wave tank is described in \autoref{sec:mathmodl}, validation against monochromatic wave generation is presented in \autoref{sec:monoch_val}; uncertainty quantification of monochromatic wave simulations is reported in \autoref{sec:GCI}; benchmarking for polychromatic wave generation over flat bottoms is presented in \autoref{sec:polywaves} whilst benchmarking against wave-transformation over an inclined bottom is reported in \autoref{sec:wavetrans}. Salient aspects of the work are summarized in \autoref{sec:summary}.  

\section{Numerical wave tank} \label{sec:mathmodl}
The present work is aimed towards improving robustness of our existing NWT algorithm \cite{sas17a} through a reduction in number of {\textsf{WM}} design variables which is accomplished by replacing the mass-source generator with the modified-inflow technique \cite{sas17b}. Mathematical model of the NWT is detailed in the following subsections. 
\subsection{Governing equations and solution methodology} \label{ssec:gdes}
Wave hydrodynamics in the NWT has been modeled using the two-phase Navier-Stokes equations following a ``one-fluid'' approach \cite{pros07}. Given that air and water are individually incompressible (within the range of velocities induced by wave motion), the NSE are retained in a ``non-conservative'' form to avoid generation of unrealistically large velocities at the interface \cite{pros07}. The NSE would thus be comprised of the continuity and momentum equations which are represented as,     
\begin{flalign} \label{eq:NSE}
&\vec{\nabla}\bullet \vec{V}=0 \nonumber \\ 
&\dfrac{\partial\vec{V}}{\partial t}+ \left(\vec{V}\bullet \vec{\nabla}\right)\vec{V}=-\dfrac{1}{\rho^*}\vec{\nabla}p + \dfrac{1}{\rho^*} \vec{\nabla}\bullet\left(\mu^* \vec{\nabla} \vec{V}\right) + \vec{g}
\end{flalign} 
Here, $U$ and $V$ denote the streamwise $(x)$ and vertical $(y)$ components of velocity $\vec{V}$, $p$ is the pressure, $t$ is time, $\vec{\mathrm{d}A}$ is the area of surface surrounding the control volume $\mathrm{d}\forall$, $\rho^*$ and $\mu^*$ are the mixture density and viscosity respectively and $g$ is the acceleration due to gravity. The mixture properties $\rho^*$ and $\mu^*$ are in turn determined using the volume of fluid (VOF) method \cite{youngs82},
\begin{equation} \label{eq:mu_rho}
\rho^*=f\rho_w + (1-f)\rho_a\,\,\,{\mathsf{and}}\,\,\, \mu^*=f\mu_w + (1-f)\mu_a
\end{equation}
where subscripts $w$ and $a$ denote water and air phases respectively and $f$ is the volume fraction. Transport of $f$ is governed by the pure advection equation,
\begin{equation} \label{eq:VOF}
\dfrac{\partial f}{\partial t} + \vec{\nabla}\bullet\left(\vec{V} f\right) = 0
\end{equation}
The transport \cref{eq:NSE} and \cref{eq:VOF} have been discretized on a finite-volume staggered grid \cite{pros07} with the solution advanced explicitly in time. At the beginning of a new $(n+1)$ time level, $\rho^*$ and $\mu^*$ are determined from $f-$field of the previous $(n)$ time level which is followed by solution of \cref{eq:VOF} using the $n^{th}$ level $\vec{V}-$field. \cref{eq:VOF} is solved geometrically using Youngs PLIC-VOF technique \cite{youngs82} which is based on the recurrence of three steps; interface reconstruction, interface advection and material redistribution:
\begin{itemize}
\item in the first step, the air-water interface is reconstructed from the $n^{th}$ level $f-$field which is used to evaluate the interface normal $\vec{\mathfrak{n}}$ on a $3\times 3$ cell stencil $\left(\vec{\mathfrak{n}} = \vec{\nabla}f\right)$. Once $\vec{\mathfrak{n}}$ is obtained, volume conservation is invoked to estimate the placement of the interface within a cell. Here, the interfacial location is directly obtained by means of pre-computed analytical solutions of a set of sixteen possible orientation cases. 
\item in the second step, the reconstructed fluid region is advected across the velocity field $\vec{V}$. Advection is performed in the Eulerian framework through geometric calculation of fluid fluxes across the cell faces. Again, the fluxes are directly obtained from a set of thirty-two pre-calculated analytical solutions. \cref{eq:VOF} is advanced in an operator-split manner to eliminate ``double-fluxing errors'' and to allow a maximum Courant number of $C_{max} \leq 1$ for $f-$transport. Second-order accuracy in fluid advection is retained by alternating the splitting direction every time step.
\item as a final step, a conservative redistribution algorithm is run after each split to eliminate overshoots $(f>1)$ and undershoots $(f<0)$ in the volume fraction field.
\end{itemize} 
Thus, barring the (iterative) redistribution algorithm, the casewise selection structure ensures extremely fast interface reconstruction and advection computations in the NWT algorithm \cite{sas17a}. The PLIC-VOF solution yields the $f^{n+1}$ field which is followed by solution of the NSE using the velocity projection method through a predictor-corrector approach. The following discrete form is adopted for \cref{eq:NSE} during the predictor $(\star)$ step;
\begin{equation} \label{eq:predictor}
\mathbb{V}^{\star}_{\mathcal{I}}=\mathbb{V}^{n}_{\mathcal{I}}+\dfrac{\Delta t}{\forall_\mathcal{I}} \left\{\mathscr{A}\mathscr{F}^{n}_{\mathcal{I}}+\mathscr{B}\mathscr{F}^{n-1}_{\mathcal{I}}
+\mathscr{C}\mathscr{F}^{n-2}_{\mathcal{I}}\right\}-\dfrac{\Delta t}{\rho^{*}_{I-\sfrac{1}{2}}} \mathcal{G}^{\parallel}_{p^n_I}+\Delta t\left[\textsf{SRC}\right]
\end{equation}  
where $\mathbb{V}$ represents any velocity component, $I$ denotes a pressure cell, $\mathcal{I}$ denotes a momentum cell, $\forall_{\mathcal{I}}$ is volume of a momentum cell, $\Delta t$ is time-step size, $\mathcal{G}^{\parallel}_{p^n_I}$ is the gradient of pressure at $\mathcal{I}$ along $\mathbb{V}$ and $\left[\textsf{SRC}\right]=0\leftarrow \mathbb{V}\equiv U$ whilst $\left[\textsf{SRC}\right]=-g \leftarrow \mathbb{V}\equiv V$. The function $\mathscr{F}$ in \cref{eq:predictor} further expands as,
\begin{equation} \label{eq:ADV_DIF}
\mathscr{F}^{n}_{\mathcal{I}}=[\textsf{ADV}]^n +[\textsf{DIF}]^n
\end{equation} 
Here, $[\textsf{ADV}]^n \equiv -\sum\limits_{\textsf{face}}\mathbb{V}^n_{a}\mathbb{V}^{n}_{\mathfrak{N}}A_{\textsf{face}}^{\pm}$ is momentum advection in discrete form where, $A_{\textsf{face}}^{\pm}$ is signed face area, $\mathbb{V}_a$ is the ``advected quantity'' determined using a blend of first and third order advection schemes \cite{brad00} (see \autoref{ssec:schbln} for details) whilst $\mathbb{V}_{\mathfrak{N}}$ is the ``advecting agency'' determined using linear interpolation. Further, momentum diffusion in discrete form is given by: $[\textsf{DIF}]^n \equiv\dfrac{1}{\rho^*_{I-\sfrac{1}{2}}}\left\{\sum\limits_{\textsf{face}}\mu^* \mathcal{G}^{\parallel}_{\mathbb{V}^n_{\mathcal{I}}}A_{\textsf{face}}^{\pm} + \sum\limits_{\textsf{face}}\mu^* \mathcal{G}^{\perp}_{\mathbb{V}^n_{\mathcal{I}}}A_{\textsf{face}}^{\pm}\right\}$ where $\mathcal{G}^{\parallel}_{\mathbb{V}^n_{\mathcal{I}}}$ is the gradient of $\mathbb{V}^n_{\mathcal{I}}$ parallel to itself whilst $\mathcal{G}^{\perp}_{\mathbb{V}^n_{\mathcal{I}}}$ is the gradient of $\mathbb{V}^n_{\mathcal{I}}$ normal to itself. Because momentum cells are back-staggered by half a cell-size, $\mu^*$ is directly available at pressure cells whilst evaluating parallel gradients $(\mathcal{G}^\parallel)$, however, the same has to be obtained through bi-linear interpolation for perpendicular gradients $(\mathcal{G}^\perp)$. It should be further noted that $\mathcal{G}$ has been determined using the second-order central difference scheme in all cases. In the present work, $\mathbb{V}^{\star}_{\mathcal{I}}$ in \cref{eq:predictor} is estimated using the forward Euler method \cite{drikakis05}: $\left(\begin{smallmatrix} \mathscr{A} \\ \mathscr{B} \\ \mathscr{C} \end{smallmatrix} \right) \equiv \left(\begin{smallmatrix} 1 \\ 0 \\ 0 \end{smallmatrix} \right)$ for monochromatic wave-propagation scenarios (\autoref{sec:monoch_val}) whilst the same is estimated using the third-order Adams-Bashforth method \cite{drikakis05}: $\left(\begin{smallmatrix} \mathscr{A} \\ \mathscr{B} \\ \mathscr{C} \end{smallmatrix} \right) \equiv \left(\begin{smallmatrix} 23/12 \\ -16/12 \\ 5/12 \end{smallmatrix} \right)$ for polychromatic wave-propagation (\autoref{sec:polywaves} and \autoref{sec:wavetrans}). \\
The predictor $(\star)$ step is followed by a corrector/velocity projection step which involves solution of the following equation of pressure correction;
\begin{equation} \label{eq:EOPC}
\sum\limits_{\textsf{face}}\dfrac{\mathcal{G}_{p'_I}}{\rho_{\textsf{face}}}A_{\textsf{face}}^{\pm}=\dfrac{1}{\Delta t}\sum\limits_{\textsf{face}}\mathbb{V}^{\star}_{\mathcal{I}}A_{\textsf{face}}^{\pm}
\end{equation}
where $p'_I$ is the correction in pressure necessary to make $\mathbb{V}^{\star}_{\mathcal{I}}$ solenoidal subject to a condition that the root-mean-square value of the imbalance between both sides of \cref{eq:EOPC} is less than $1e-06$. Said imbalance is iteratively reduced using the Gauss-Seidel method. Once the $p'_I$ field is obtained, the $n+1^{th}$ level velocity field is computed using,
\begin{equation}
\mathbb{V}^{n+1}_{\mathcal{I}}=\mathbb{V}^{\star}_{\mathcal{I}}-\dfrac{\Delta t}{\rho^{*}_{I-\sfrac{1}{2}}} \mathcal{G}^{\parallel}_{p'_I}
\end{equation}
which, owing to a staggered grid, is directly obtained at $\mathcal{I}$ without any interpolation; this ensures a tight coupling between velocity and pressure \cite{pros07}. Domain and mesh design strategies adopted for the NWT are presented in the next subsection.
\subsection{Domain designing and meshing strategy} \label{ssec:dommesh}
\begin{figure}[!ht]
\begin {center}
\begin {tabular}{r l}
{
\centering
\includegraphics[trim=15mm 0mm 20mm 5mm, clip, height = 6cm]{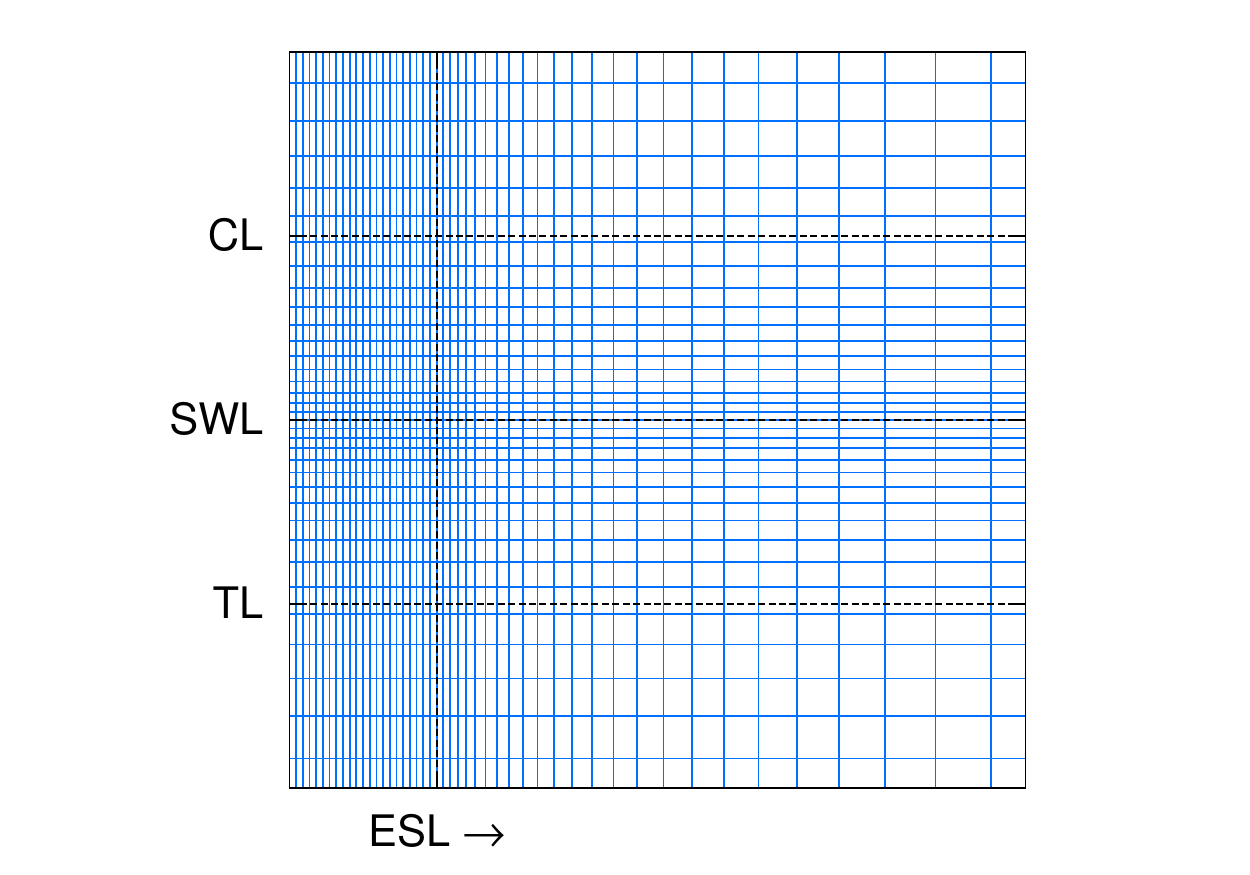}
} & {
\centering
\includegraphics[trim=25mm 0mm 20mm 5mm, clip, height = 6cm]{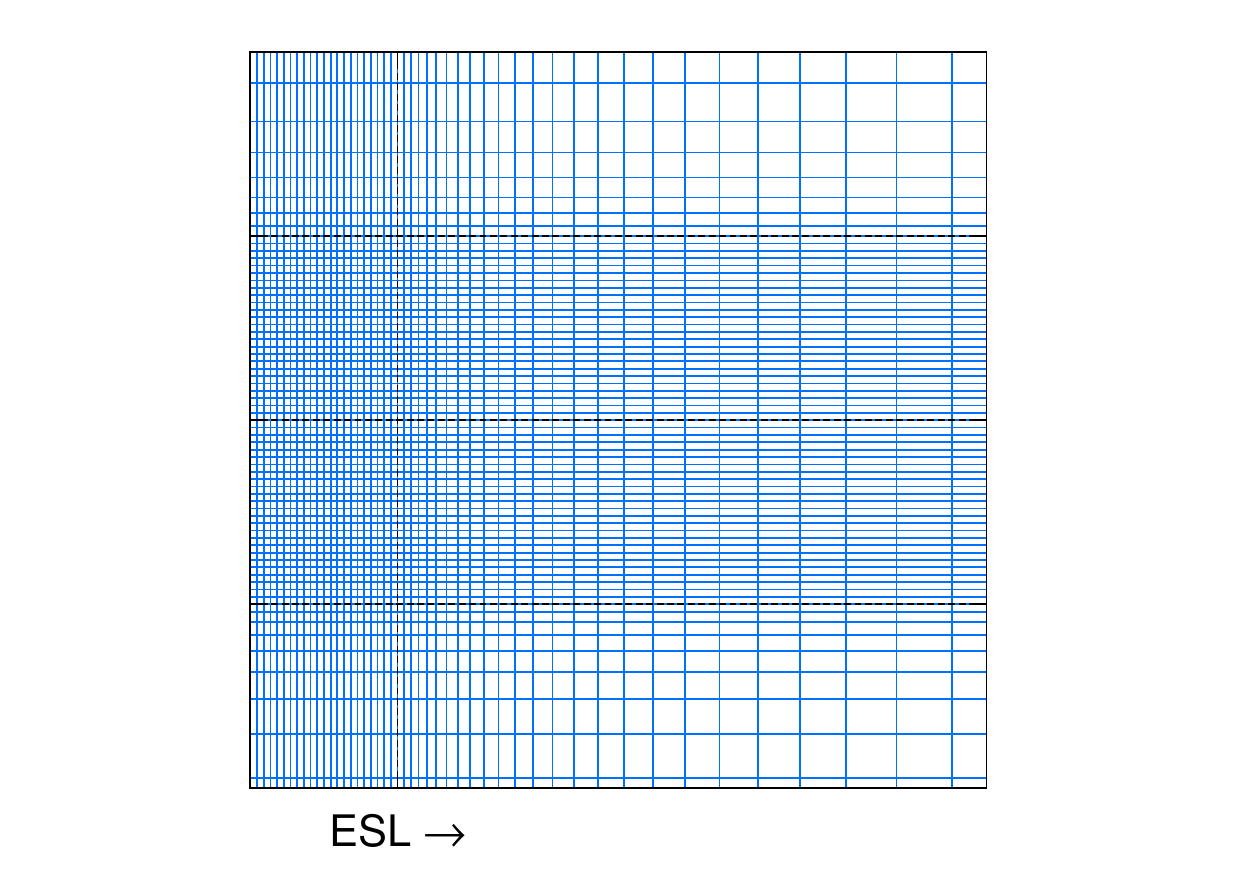}
}
\end {tabular}
\includegraphics[trim=0mm 0mm 0mm 0mm, clip, width = 15cm]{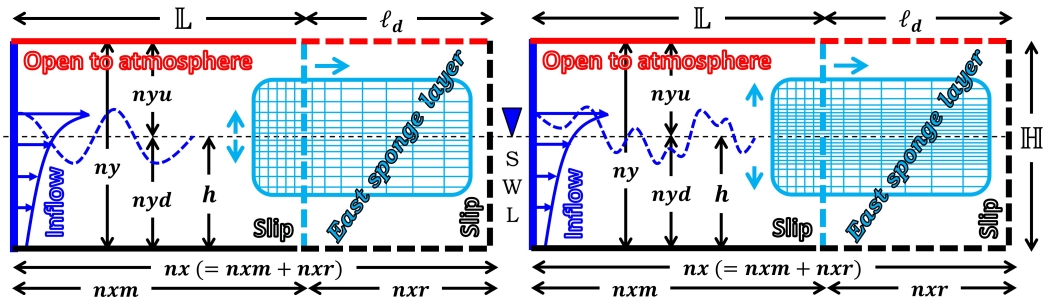}
\end {center}
\caption{\emph{Computational model (not to scale) of inflow-boundary based NWT illustrating {\emph{(top)}} vertical meshing strategies and {\emph{(bottom)}} domain and mesh design variables \cite{sas17b}.}}
\label{fig:dommesh}
\end{figure}

Computational model of the proposed inflow-boundary based NWT is shown in \autoref{fig:dommesh}. Given the fact that the NWT would be benchmarked against a variety of wave propagation scenarios, the domain and mesh configurations have been depicted here in a generalized manner with problem specific designs illustrated throughout \autoref{sec:monoch_val}-\autoref{sec:wavetrans}. Referring to \autoref{fig:dommesh}, $\mathbb{L}(\propto \lambda)$ is the length of the wave simulation region, $\ell_d(\propto \lambda)$ is length of the east sponge layer (ESL) and $\mathbb{H}(\sim h+2H)$ is height of the domain. For wave-transformation simulations, the ESL has been replaced by a $1:25$ dissipative beach whose numerical design is described in \autoref{ssec:wavetrans_setup}. In all cases, the wave simulation region is uniformly divided into $nxm$ cells. To achieve reflection-free wave absorption, the ESL is discretized using a successively stretched grid of $nxr(\sim 50- 100)$ cells. For the vertical mesh, the region $0 \leq y \leq h$ is divided into $nyd$ cells whilst $h\leq y \leq \mathbb{H}$ is divided into $nyu$ cells. As evident from \autoref{fig:dommesh}, two different vertical meshing strategies have been adopted.          
For monochromatic waves, the vertical mesh is successively stretched directly from $y=h$ \cite{sas17a}. For polychromatic waves, the vertical mesh is uniformly clustered within $\textsf{TL}\leq y \leq \textsf{CL}$ and the cells are successively stretched away from this region{\footnote{for polychromatic waves, we consider {\textsf{TL}} and {\textsf{CL}} as minimum and maximum displacements (respectively) of the free-surface about $y=h$ over $0\leq x \leq \mathbb{L}$.}} towards the boundaries (cf. \autoref{fig:dommesh}). Lastly, variables $nx_\lambda$ (cells per wavelength) and $ny_H$ (cells per wave-height) have been defined to analogize the mesh design process with steepness $(H/\lambda)$ of the incident waves as per guidelines proposed in \cite{sas17a}. The wave generator and absorber designs are described in the next subsection.
\subsection{Modified-inflow {\textsf{WM}} and wave absorption} \label{ssec:WM}
\begin {figure}[!ht]
\begin {center}
\begin {tabular}{c}
{
\centering
{\bf{(a)}}\includegraphics[trim=0mm 28mm 0mm 28mm, clip, width = 13cm]{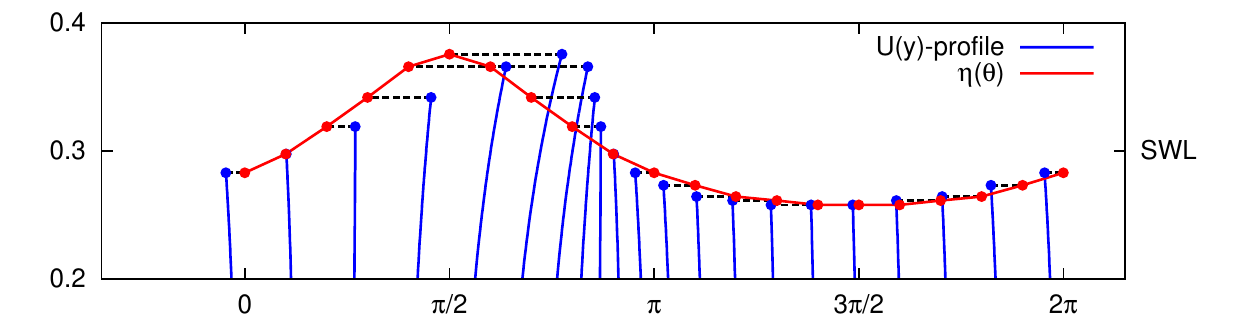}
}\\
{
\centering
{\bf{(b)}}\includegraphics[trim=0mm 28mm 0mm 28mm, clip, width = 13cm]{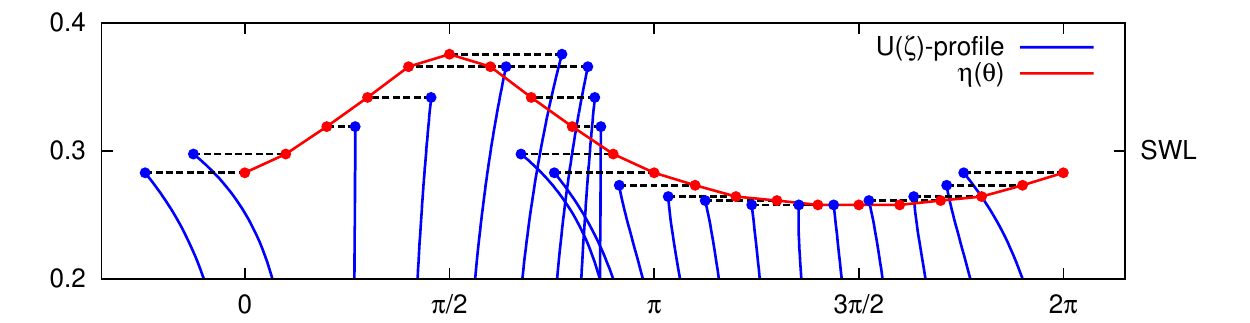}
}
\end {tabular}
\end {center}
\caption{\emph{Depthward profiles of $U-$velocity under a steep wave $(h=0.3\,m,T=1.5\,s,H=0.12\,m\,\vdash\,{H/\lambda}\approx 0.048)$ predicted using {\texttt{Stokes V}} based (a) baseline and (b) modified inflow $\left(\wp=0.335\,\,\forall\,\,y<h\right)$ methods; ordinate:depthward-position; abscissa:wave-phase.}}
\label{fig:modinf}
\end{figure}

As detailed in \autoref{sec:intro}, the baseline-inflow {\textsf{WM}} adds a net volume of water $\mathcal{V}_+$ into the NWT domain; for strongly non-linear waves, the added volume induces setup which hampers the fidelity of the simulation{\footnote{it is obvious that the magnitude of the induced wave setup increases with decreasing domain length.}}. The issue of volume addition is addressed here by proposing a {\textsf{WM}} based on a KST similar to Wheeler stretching \cite{wheeler70} but within the framework of a higher-order wave theory (namely {\texttt{Stokes V}}). At fifth order, the following velocity potential $\phi(\zeta,\theta)$ is prescribed at the inflow boundary,
\begin{flalign} \label{eq:modinf}
\dfrac{2\pi T \phi (\zeta,\theta)}{\lambda^2}=&\left({\mathcal{A}} A_{11}+{\mathcal{A}}^3 A_{13}+{\mathcal{A}}^5 A_{15}\right)\cosh(k\zeta)\cos(\theta)\nonumber\\
&+\left({\mathcal{A}}^2 A_{22} +{\mathcal{A}}^4 A_{24}\right)\cosh(2k\zeta)\sin(2\theta)-\left({\mathcal{A}}^3 A_{33}+{\mathcal{A}}^5 A_{35}\right)\cosh(3k\zeta)\cos(3\theta)\nonumber\\
&-{\mathcal{A}}^4 A_{44}\cosh(4k\zeta)\sin(4\theta)+{\mathcal{A}}^5 A_{55} \cosh(5k\zeta)\cos(5\theta)\,\leftarrow \theta\equiv kx-\omega t
\end{flalign}
where, $k$ is the circular wavenumber, ${\mathcal{A}}$ is a topological parameter, $\omega$ is the circular frequency and the coefficients $A_{11}$, $A_{13}$ etc. are weights assigned to component harmonics of the {\texttt{Stokes V}} wave. Here, the potential $\phi(y,\theta)$ in \cref{eq:modinf} has been modified by replacing the vertical coordinate $y$ with a stretched coordinate $\zeta$ such that $\zeta\equiv y\cdot \frac{h+\wp}{h+\eta(t)}$ where $\wp>0 \,\,\forall\,\, y<h\,;\,\wp=\eta(t)\,\,\forall\,\, y\geq h$. We term the proposed method as the {\emph{modified inflow technique}}. It is noteworthy that $\wp=\eta(t)$ yields baseline inflow \cite{lin98} (cf. \autoref{fig:modinf} (a)) whilst $\wp=0$ corresponds to Wheeler's method \cite{wheeler70} which kinematically over-designs the troughs but also under-designs the crests. In the case of modified inflow $(\wp>0)$, kinematic predictions of {\texttt{Stokes V}} theory are preserved above the SWL with kinematic over-design only applied below the SWL (cf. \autoref{fig:modinf}(b)). Further, prescription of $\wp$ is flexible and depends on the wave design in question; the added volume $\mathcal{V}_+$ can hence be directly controlled through $\wp$ \cite{sas17b}. Most importantly, the proposed modified inflow {\textsf{WM}} is characterized by only one design variable: $\wp$. In \autoref{sssec:lrgstp} it is demonstrated that the modified-inflow technique outperforms both baseline-inflow as well as Wheeler stretching in terms of the fidelity of waves being generated. 
\\
In addition to the wave generator, accuracy of NWT simulations is also contingent on reflection-free absorption of incident wave energy at the ``far end'' opposite to the {\textsf{WM}}. In the present work, sponge layers are used for efficiently absorbing waves propagating over horizontal bottoms. For waves propagating over inclined bottoms, physical beaches have been modeled as partially submerged/emergent boundaries for wave dissipation \cite{sas17b}. This is done so that wave-transformation simulations (reported in \autoref{sec:wavetrans}) bear a close resemblance to the experimental conditions of Beji and Battjes \cite{beji94} against whose measurements the former are validated. To avoid confusion, the numerical model of the dissipative beach is contextually described in \autoref{sec:wavetrans}. \\
Incident waves are absorbed in the ESL through introduction of spatially-varying sink terms \cite{sas17a} in $\mathbb{V}^{\star}_{\mathcal{I}}$ over a purposefully coarsened streamwise mesh (cf. \autoref{fig:dommesh}),
\begin{equation} \label{eq:ESL}
\mathbb{V}^{\star}_{\mathcal{I}} \rightarrow \mathbb{V}^{\star}_{\mathcal{I}}-\exp
\left(-\alpha\left(1-\left(\frac{x-x_a}{\ell_d}\right)\right)^R\right)
\mathbb{V}^{n}_{\mathcal{I}}
\end{equation}  
where $x_a(\equiv\mathbb{L})$ marks the beginning of the ESL, $x(>x_a)$ is measured along the direction of wave propagation, $\ell_d$ is the length of the ESL whilst $\alpha$ and $R$ are sponge layer strength and rate of absorption respectively. It should be noted that numerical optimization of ESL design parameters is not attempted here. Instead, $\ell_d \geq 4.0\lambda$; $\alpha=10$; $R=0.4- 1.0$ are directly adapted from our previous works \cite{sas17a,sas17b} for which excellent wave absorption performance was demonstrated over a wide range of steepness $(0.004 \leq H/\lambda \leq 0.052)$ and relative depths $(0.76\leq kh\leq 6.2)$. The need for and implementation of blended advection schemes in the present NWT framework is elucidated in the next subsection.     
\subsection{Addressing numerical damping of waves} \label{ssec:schbln}
In our previous work \cite{sas17a}, we have extensively discussed the mechanics of steep wave generation $(H/\lambda>0.03)$ in context to the mass-source {\textsf{WM}} being applied to both deep $(kh\approx 6.2)$ and near-shallow $(kh\approx 0.8)$ water conditions. This was motivated by an observation that in the NSE framework, keeping $h$ and $T$ fixed, the generated waves become increasingly susceptible to damping as $H\uparrow$ and also by an observation that, barring few studies (besides our own) \cite{peric15,windt17}, this issue (even though encountered in multiple studies) has received limited attention in the literature. Whilst said susceptibility to damping (especially the ``rate of height damping'' along the length of the NWT) is closely linked to type of numerical {\textsf{WM}} selected for simulation \cite{sas17a,windt17}, it is now realized that the former is also controlled by the strategy adopted for numerically modeling wave propagation (which is, for the most part, inviscid) in a viscous framework such as the NSE. The act of numerically approximating the NSE itself introduces several errors which cause the numerical model to exhibit certain traits that are actually absent from the physical system \cite{ferziger02}. In case of ocean waves, one such (numerical) artifact that might be mistaken for a physical trait is spurious/numerical diffusion $(\Gamma^{num})$. We justify this assertion through a pilot assessment of the magnitude of momentum advection $[\mathsf{ADV}]$ and diffusion $[\mathsf{DIF}]$ terms in the discrete NSE (cf. \cref{eq:predictor}) in comparison to gravity $[\mathsf{SRC}]$ which is the dominant restoring force \cite{dean91}. 
\begin {figure}[!ht]
\begin {center}
\begin {tabular}{c c}
{\textsc{Case} \textbf{A}} & {\textsc{Case} $\mathbf{C_a}$}\\
{
\centering
\includegraphics[trim=15mm 0mm 15mm 0mm, clip, height = 7cm]{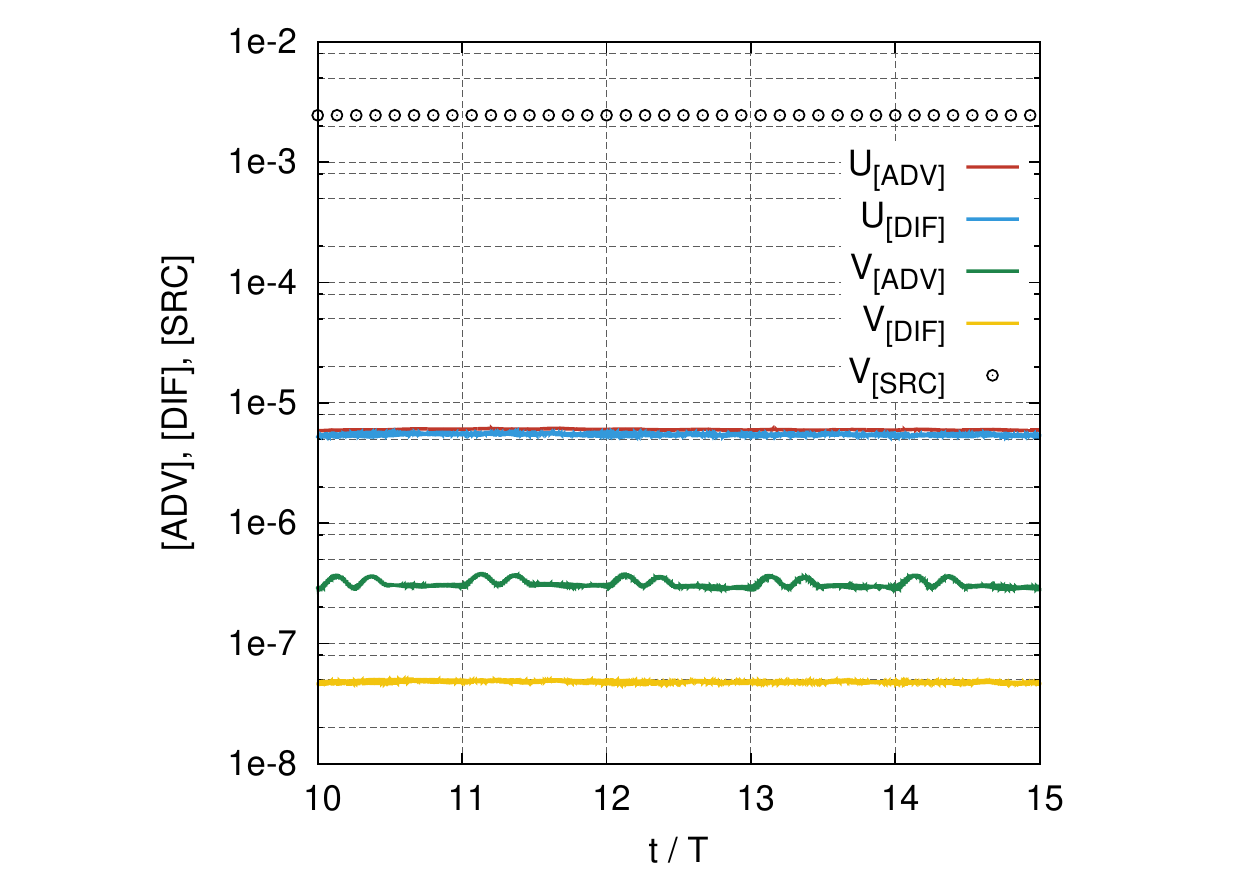}
}&
{
\centering
\includegraphics[trim=15mm 0mm 15mm 0mm, clip, height = 7cm]{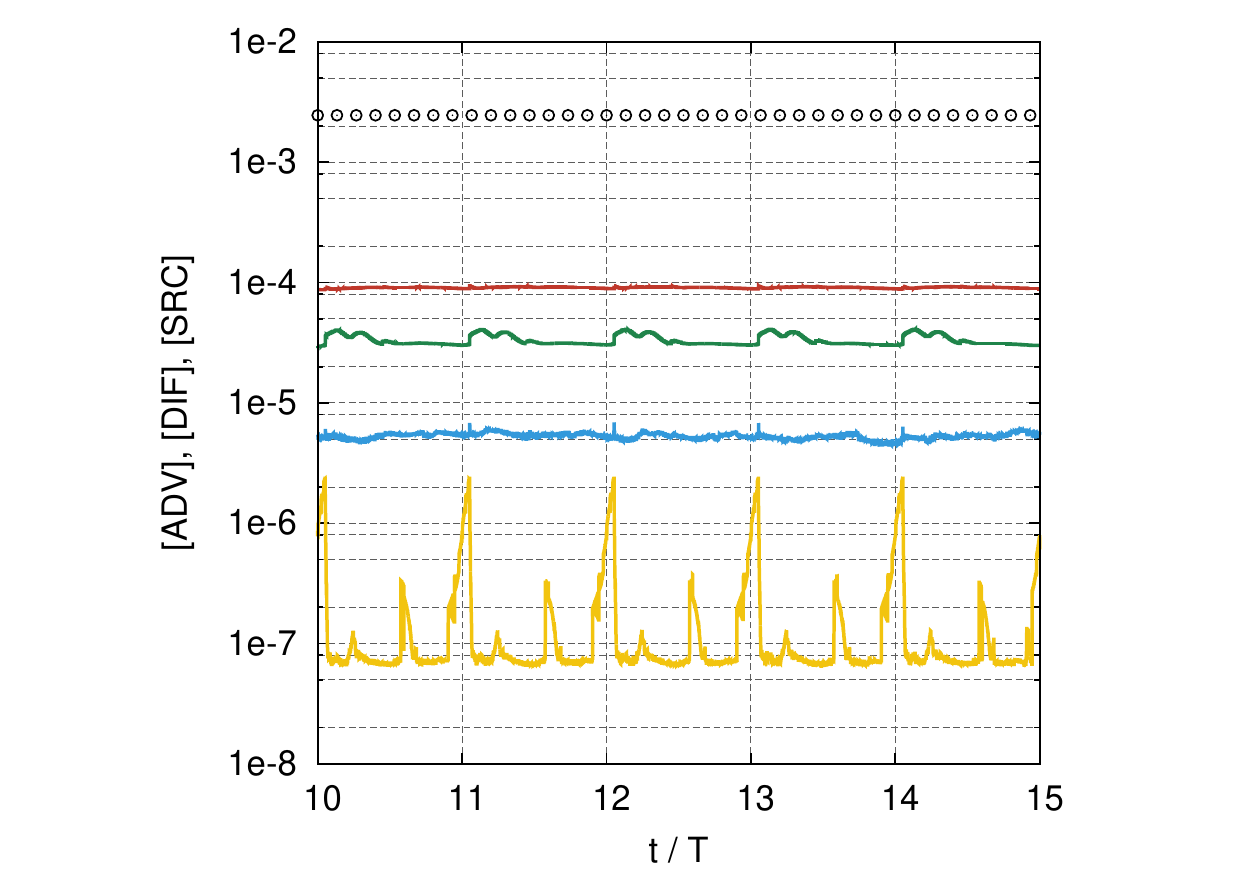}
}
\end {tabular}
\end {center}
\caption{\emph{Normalized time variation of $\mathsf{RMS}\left\{[\mathsf{ADV}]^n\right\}$, $\mathsf{RMS}\left\{[\mathsf{DIF}]^n\right\}$ and $\mathsf{RMS}\left\{[\mathsf{SRC}]^n\right\}$ (cf. \cref{eq:predictor}) evaluated over $0\leq x\leq \mathbb{L}$ for cases {\bf{A}} $(h=0.3\,m,T=1.5\,s,H=0.01\,m\,\vdash\,\mathsf{Ur}=2.033)$ and $\mathbf{C_a}$ $(h=0.3\,m,T=1.5\,s,H=0.12\,m\,\vdash\,\mathsf{Ur}=27.24)$.}}
\label{fig:numdamp}
\end{figure}
The evaluation, shown in \autoref{fig:numdamp}, is done for two different waves: case {\bf{A}} \cite{sas17a} and case $\mathbf{C_a}$ \cite{sas17b}; the latter is twelve times higher in comparison to the former. The results indicate a shift in balance between $[\mathsf{ADV}]$ and $[\mathsf{DIF}]$ terms such that the problem transforms from a (conventional) {\emph{advection-diffusion}} scenario at low steepness to an {\emph{advection-dominant}} scenario at large steepness. Hence, the selection of the numerical scheme used for approximating ``advected momentum'' $\mathbb{V}_a^n$ (cf. \cref{eq:ADV_DIF}), being inconsequential for case {\bf{A}}, is crucial for case $\mathbf{C_a}$. This is intimately linked to the fact that $\Gamma^{num}\propto \rho U \Delta x$ considering a FOU-based evaluation of $\mathbb{V}_a^n$ \cite{ferziger02}. Then, for ocean waves it can be proved that $\Gamma^{num}\propto \rho HT^{-1}\mathrm{e}^{H\lambda^{-1}}\Delta x$ using Airy theory{\footnote{in fact, streamwise velocity at the crest level can be approximated as: $U_{CL}^{Airy}\cong K_1 HT^{-1}\mathrm{e}^{K_2 H\lambda^{-1}}$ where $K_1\neq K_2 > \pi \leftarrow kh<\pi$ and $K_1=K_2=\pi\leftarrow kh\geq \pi$.}}; numerical diffusion increases rapidly with wave height. In fact, for a FOU-based evaluation of $\mathbb{V}_a^n$, case $\mathbf{C_a}$ would require a twelve times finer mesh compared to case {\bf{A}} to restrict $\Gamma^{num}$ within the same order of magnitude; this is computationally unjustifiable. \\
The trouble with nullifying $\Gamma^{num}$ in two-phase NSE-based NWT simulations is that FOU (which is inherently bounded and non-oscillatory \cite{ferziger02}) cannot simply be replaced by ``pure'' higher order schemes (such as CD, SOU or QUICK); the latter class of algorithms induce dispersive oscillations in $\mathbb{V}_a^n$, especially at the interface \cite{pros07}. In fact, in the authors' experience, even TVD schemes in pure form tend to produce spurious momentum near the interface (leading to wave distortion). In the present work, we circumvent the issue of spurious momentum generation by implementing a blend of higher-order (QUICK) and ``limiter'' (FOU) schemes to estimate advected momentum \cite{brad00},
\begin{equation} \label{eq:schbln}
\mathbb{V}_a^n=\mathscr{S}\left.\mathbb{V}_a^n\right|_{\textsf{FOU}}+(1-\mathscr{S})\left.\mathbb{V}_a^n\right|_{\textsf{QUICK}}
\end{equation}  
where $0.5< \mathscr{S}\leq 1$. Said blending strategy has been observed to be extremely effective at restricting numerical damping of steep waves without inducing topological distortion. This is demonstrated in the next section on monochromatic wave generation.

\section{Monochromatic waves: validation} \label{sec:monoch_val}
For regular wave generation, both sinusoidal $(\mathsf{Ur}<10)$ as well as trochoidal $(\mathsf{Ur} \gtrapprox 20)$ wave designs have been considered from literature. Three important capabilities of the proposed NWT are tested: (a) (topological) quality of generated waves, (b) minimization of setup \cite{sas17b} and (c) minimization of height error $\mathcal{HE}$ \cite{sas17a} (especially for $H/\lambda>0.03$). The characteristics considered for monochromatic wave generation simulations are listed in \autoref{tab:tab1}. These include sinusoidal high frequency {\textbf{(SH)}} waves, sinusoidal low frequency ``high'' {\textbf{(SLH)}} waves, steep waves {\bf{C}} and $\mathbf{C_a}$ and an extreme-steepness design $\mathbf{I_b}$. In terms of the Ursell number, case $\mathbf{C_a}$ represents the limit of application of {\texttt{Stokes}} theory \cite{svendsen06}. NWT simulations of said wave designs are presented next.
\begin{table}[h]
\caption{\emph{Characteristics selected for monochromatic wave generation simulations: $kh$ is relative depth; $H/\lambda$ is the steepness; $\mathsf{Ur}(=H\lambda^2/h^3)$ is the Ursell number.}}
\begin{tabularx}{\textwidth}{X c c c c c c c}
\toprule
{\textsc{Case}}&{$h\,(m)$}&{$T\,(s)$}&{$H\,(cm)$}&{$\lambda\,(m)$}&{$kh$}&{$H/\lambda$}&{$\mathsf{Ur}$}\\
\midrule
{\textbf{SH}\cite{beji94}}&{$0.40$}&{$1.25$}&{$2.50$}&{$2.055$}&{$1.22$}&{$\color{Cerulean}\bm{0.012}$}&{$\mathsf{1.65}$}\\
{\textbf{SLH}\cite{huang99}}&{$0.40$}&{$2.00$}&{$4.00$}&{$3.711$}&{$0.68$}&{$\color{Cerulean}\bm{0.011}$}&{$\mathsf{8.61}$}\\
{\textbf{C}\cite{sas17a}}&{$0.30$}&{$1.50$}&{$9.00$}&{$2.421$}&{$0.78$}&{$\color{red}\bm{0.037}$}&{$\mathsf{19.5}$}\\
{$\mathbf{C_a}$\cite{sas17b}}&{$0.30$}&{$1.50$}&{$12.0$}&{$2.476$}&{$0.76$}&{$\color{red}\bm{0.048}$}&{$\mathsf{27.2}$}\\
{$\mathbf{I_b}$}&{$0.75$}&{$0.70$}&{$6.00$}&{$0.807$}&{$5.84$}&{$\color{BlueViolet}\bm{0.074}$}&{$\mathsf{0.09}$}\\
\bottomrule
\end{tabularx}
\label{tab:tab1}
\end{table}
\subsection{Topology of generated waves: qualitative assessment} \label{ssec:topval_qual}
In the present section, the monochromatic wave-generation performance of the proposed NWT model is qualitatively assessed against {\texttt{Stokes V}} theory in terms of the topology of the generated waves by means of spatial and temporal free-surface elevation profiles.
\subsubsection{Small steepness waves $(H/\lambda\approx 0.01)$} \label{sssec:smlstp}
It can be appreciated that {\bf{SH}} and {\bf{SLH}} are ``borderline'' low steepness waves ($H/\lambda \approx 0.01$; cf. \autoref{tab:tab1}). In this situation, NSE-based wave generation can be treated as a combined advection-diffusion problem (cf. \autoref{ssec:schbln}).
\begin {figure}[!ht]
\begin {center}
\begin {tabular}{c c}
{
\centering
\includegraphics[trim=15mm 0mm 19mm 0mm, clip, height = 6.5cm]{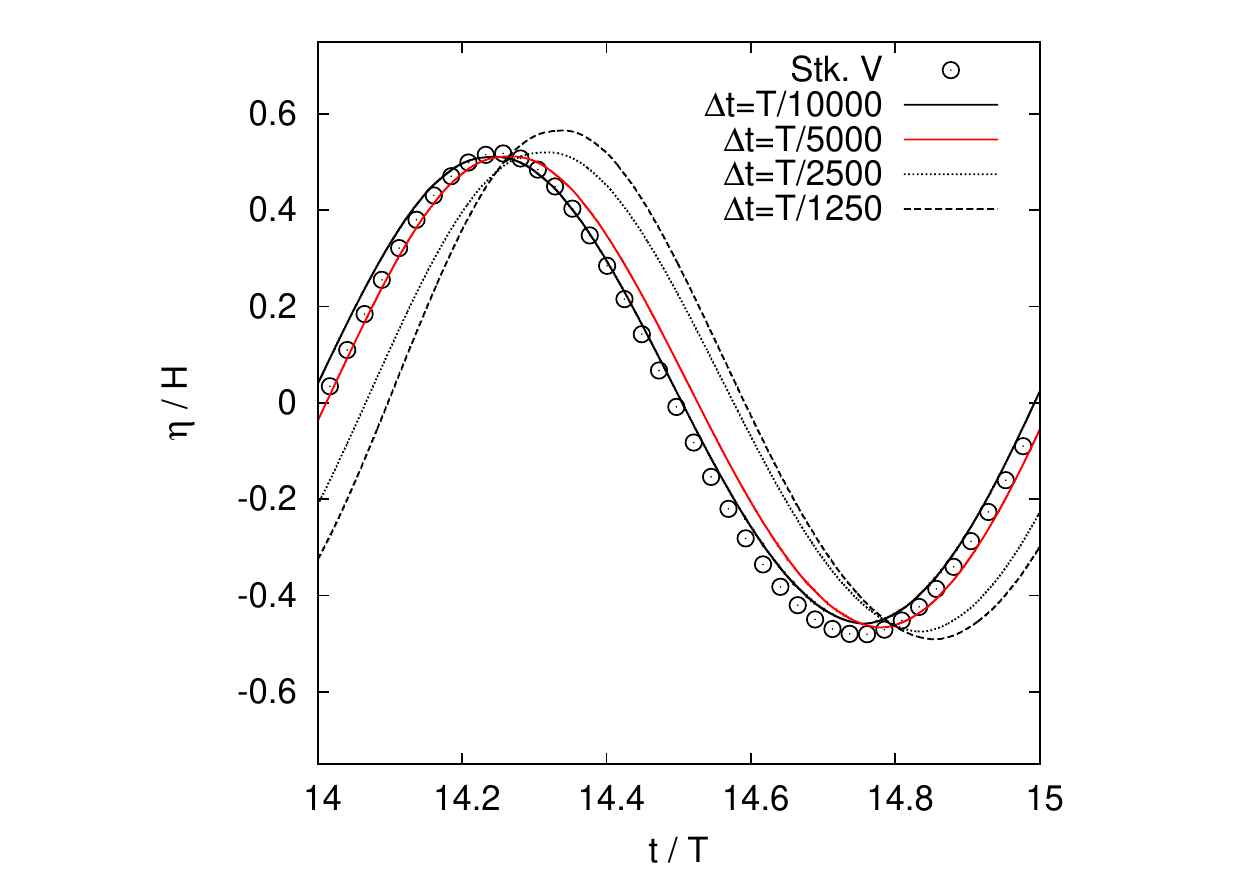}
} & {
\centering
\includegraphics[trim=15mm 0mm 19mm 0mm, clip, height = 6.5cm]{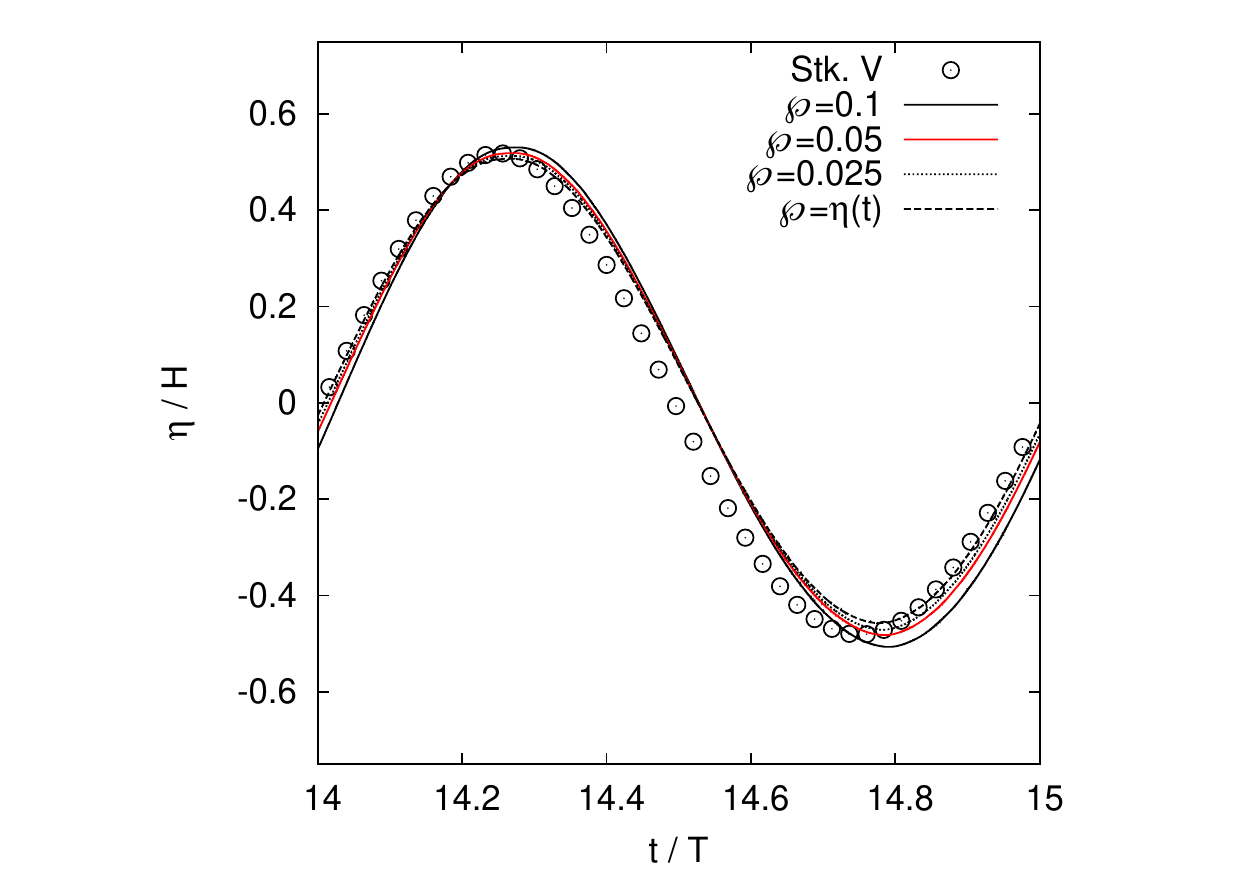}
}\\
\multicolumn{2}{c}{
\centering
\includegraphics[trim=2mm 36mm 5mm 37mm, clip, width = 15.5cm]{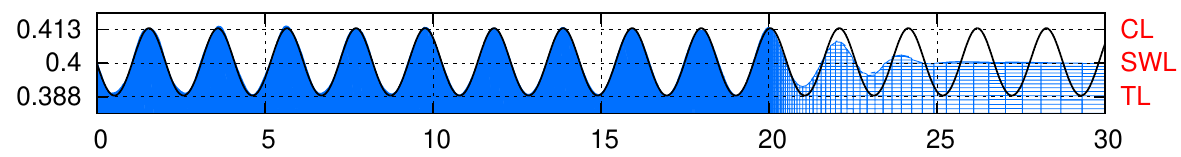}
}\\
\midrule
{
\centering
\includegraphics[trim=15mm 0mm 19mm 0mm, clip, height = 6.5cm]{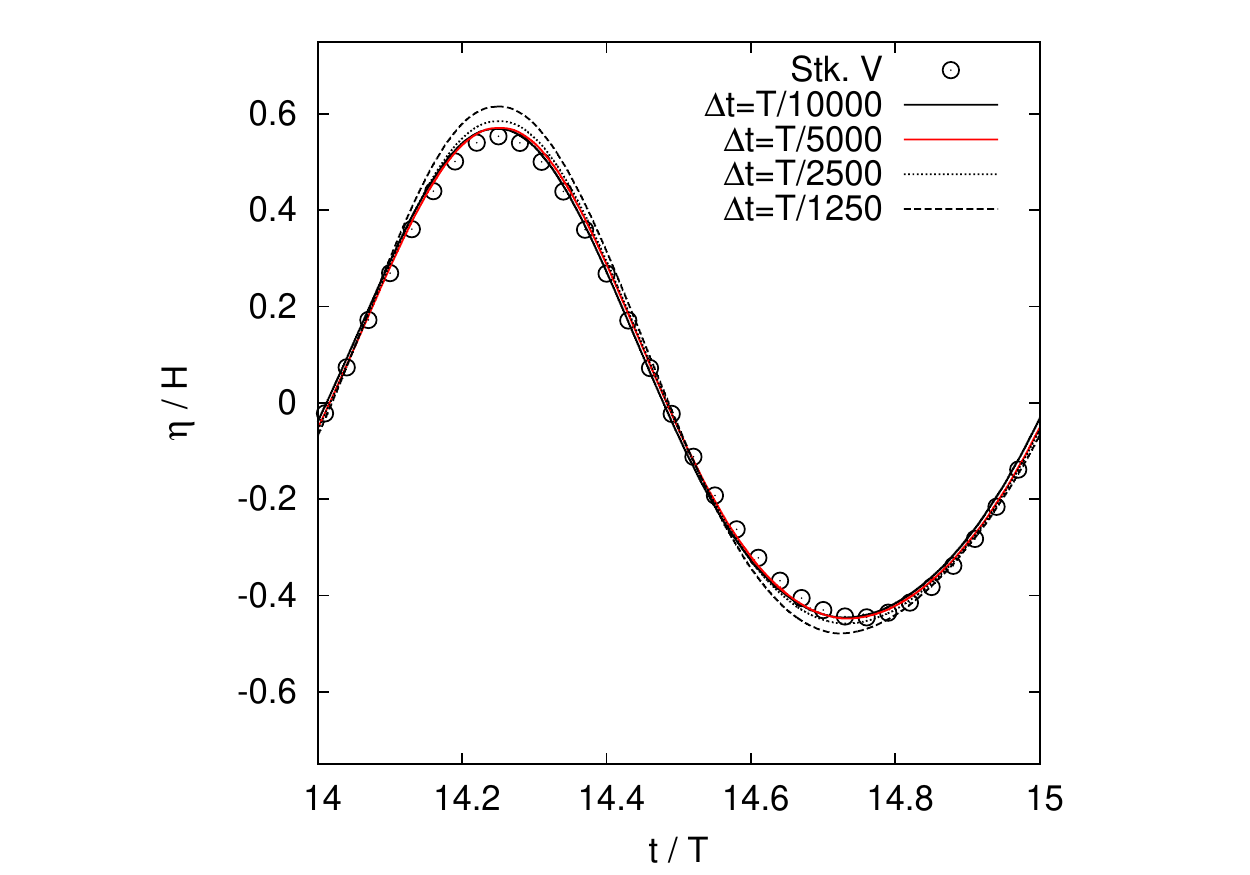}
} & {
\centering
\includegraphics[trim=15mm 0mm 19mm 0mm, clip, height = 6.5cm]{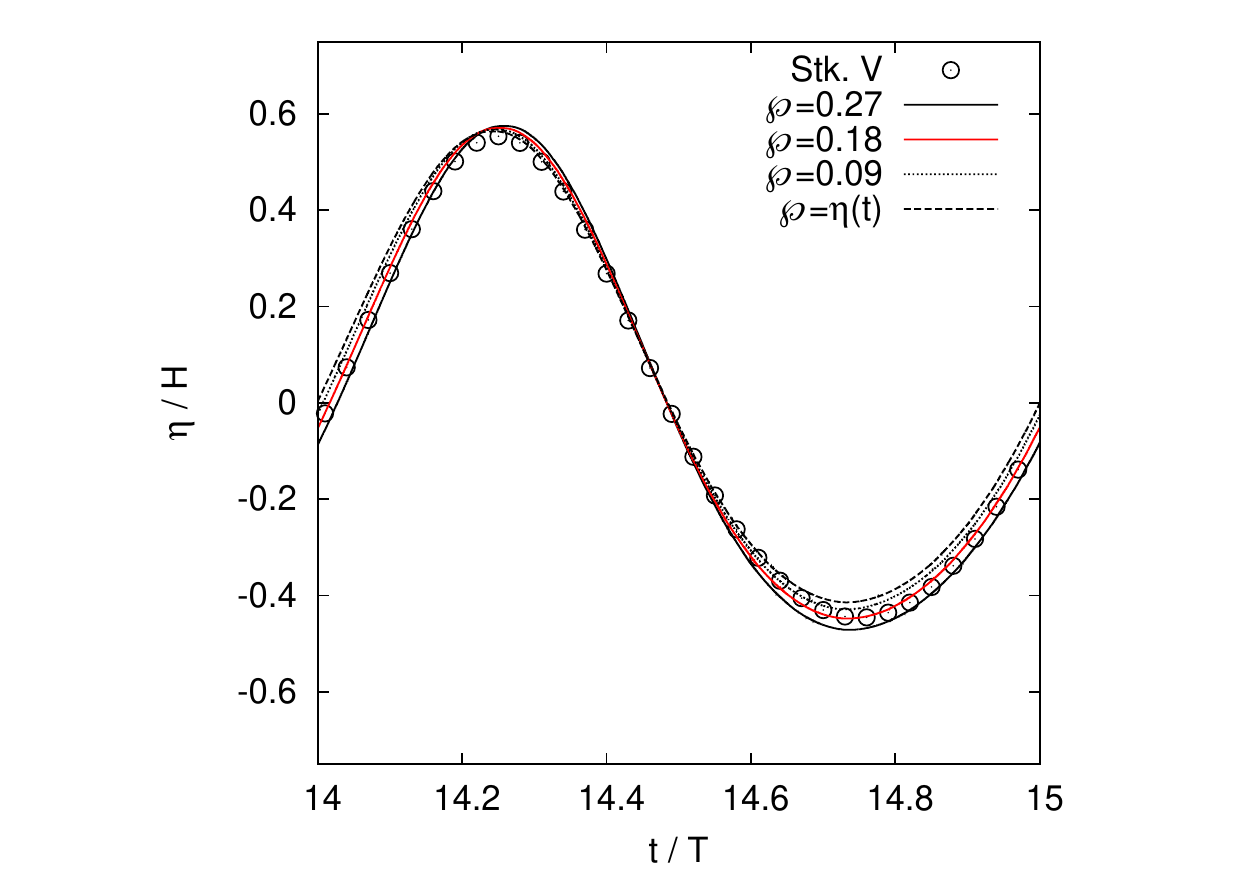}
}\\
\multicolumn{2}{c}{
\centering
\includegraphics[trim=5mm 36mm 5mm 37mm, clip, width = 15.5cm]{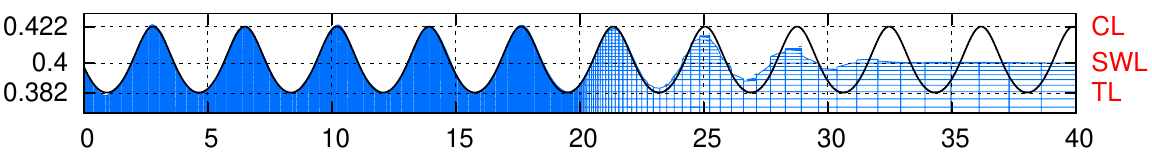}
}
\end {tabular}
\end {center}
\caption{\emph{Normalized $\eta(t)$ profiles (measured at $x=\mathsf{WM}+5\lambda$) illustrating parametric selection of $\Delta t$ and $\wp$ for sinusoidal waves and corresponding $\eta(x)$ profiles of inflow-generated {\emph{(top)}} {\bf{SH}} and {\emph{(bottom)}} {\bf{SLH}} waves at $t=20T$.}}
\label{fig:smlstptopl}
\end{figure}
Thus, pure FOU is retained ($\mathscr{S}=1$; cf. \cref{eq:schbln}) for momentum advection in both cases. Further, some simulation parameters (common to both wave designs) have been directly selected based on guidelines established in our previous work \cite{sas17a}: $\mathbb{L}=20.0\,m$, $\mathbb{H}=0.44\,m$, $\ell_d\sim 5\lambda$; $nx_\lambda\approx 84$, $ny_H\approx 18$; $R=0.4$. The pressure field $(p)$ in the water phase has been initialized following the hydrostatic law $p=\rho_w g (h-y) \,\,\forall\,\, y \leq h$. At this juncture, it is worth recalling that a refined temporal resolution is necessary for NSE-based wave generation at $H/\lambda\lessapprox 0.01$ \cite{sas17a}. It is further anticipated that $\wp_{\textbf{SH}} \neq \wp_{\textbf{SLH}} \leftarrow \mathsf{Ur}_{\textbf{SH}} \ll \mathsf{Ur}_{\textbf{SLH}}$ (cf. \autoref{tab:tab1}). Hence, optimum values of $\Delta t$ and $\wp$ have been determined through parametric selection which is depicted in \autoref{fig:smlstptopl}. It is evident that the topology of (comparatively shorter) {\bf{SH}} waves is strongly governed by $\Delta t$ whilst that of (comparatively non-linear) {\bf{SLH}} waves is largely governed by $\wp$. Nonetheless, $\Delta t=T/5000$ is observed to yield sufficiently accurate wave topology within $+5\lambda$ from the wavemaker whilst a near-exact agreement with {\texttt{Stokes V}} theory is achieved for $\Delta t=T/10000$ (cf. $\eta(x)$ profiles in \autoref{fig:smlstptopl}). Considering the placement of the toe of the structure from the {\textsf{WM}}, $\Delta t=T/5000$ is deemed sufficient for the wave-transformation simulations reported later in \autoref{sec:wavetrans}. Moreover, $\wp_{\textbf{SH}}=0.05$ and $\wp_{\textbf{SLH}}=0.18$ are observed to effectively nullify wave-setup such that $\mathcal{VE}< 0.01\%$ is achieved in both cases; interestingly, increasing $\wp$ beyond these values is observed to induce wave {\emph{set-down}} through volume subtraction (cf. \autoref{fig:smlstptopl}). The small-steepness wave simulations reported here demonstrate that criteria chosen for selecting spatio-temporal resolution based on {\emph{source-function}} {\textsf{WM}} simulations \cite{sas17a} are equally applicable to the inflow-based NWT. Thus, the selection of $nx_\lambda$, $ny_H$ and $\Delta t$ in a NWT is contingent on the mathematical framework (FNPT, NSE etc.) used for modeling the wave propagation and is rather independent of the wavemaker design. 
\subsubsection{Large steepness waves $(0.03 \leq H/\lambda \leq 0.05)$} \label{sssec:lrgstp}
The ability of the proposed modified inflow {\textsf{WM}} to accurately generate steep, trochoidal waves is assessed next by means of two designs {\bf{C}} and $\mathbf{C_a}$ which are described in \autoref{tab:tab1}. Nearly identical NWT setups have been considered in both cases \cite{sas17b}: $\mathbb{L}=19.0\,m$, $\mathbb{H}=0.6\,m$, $\ell_d\approx 4\lambda$; $nx_\lambda \approx 170$, $ny_H \approx 6$; $R=1$. Given the steep nature of the target wave designs, time is non-uniformly advanced using the forward Euler method: $T/\Delta t \geq 750 \ni C_{max} \leq 0.25$. Further, $\left(\begin{matrix} \wp \\ \mathscr{S} \end{matrix} \right) \equiv \left(\begin{matrix} 0.265 \\ 1 \end{matrix} \right)$ and $\left(\begin{matrix} \wp \\ \mathscr{S} \end{matrix} \right) \equiv \left(\begin{matrix} 0.335 \\ 0.5 \end{matrix} \right)$ are parametrically chosen for cases {\bf{C}} and $\mathbf{C_a}$ respectively with the pressure initialized using the hydrostatic law.  
\begin {figure}[!ht]
\begin {center}
\begin {tabular}{c}
{
\centering
{\bf{(a)}}\includegraphics[trim=0mm 25mm 0mm 22mm, clip, width = 14cm]{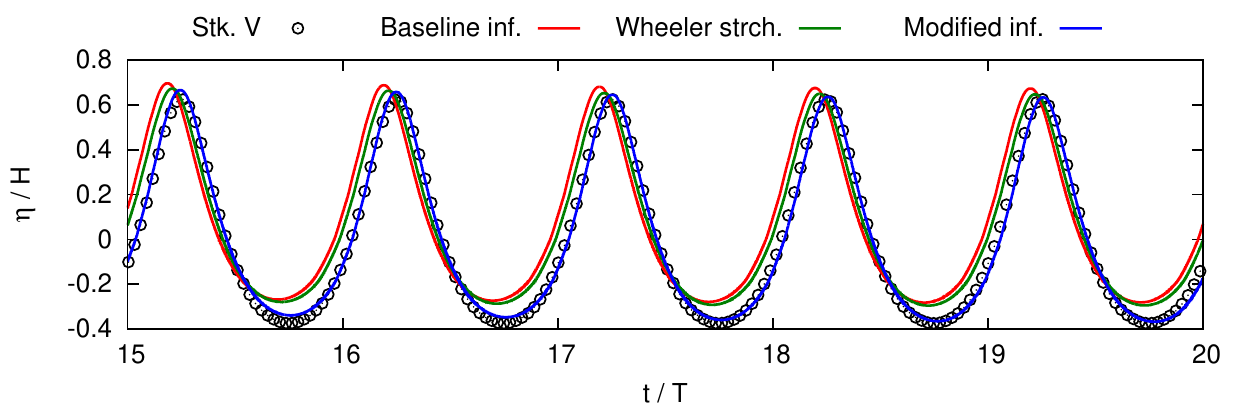}
}\\
{
\centering
{\bf{(b)}}\includegraphics[trim=0mm 25mm 0mm 22mm, clip, width = 14cm]{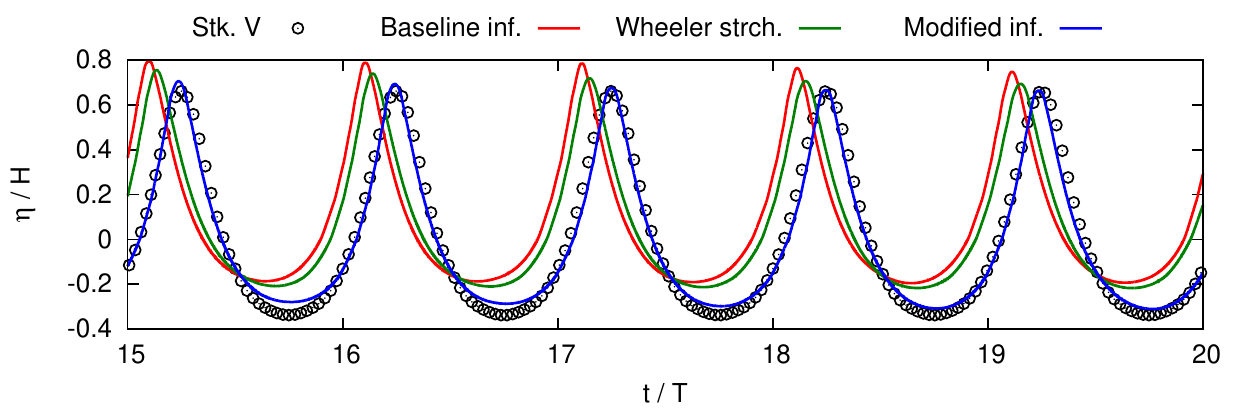}
}\\
{
\centering
{\bf{(c)}}\includegraphics[trim=-4mm 25mm 0mm 22mm, clip, width = 14cm]{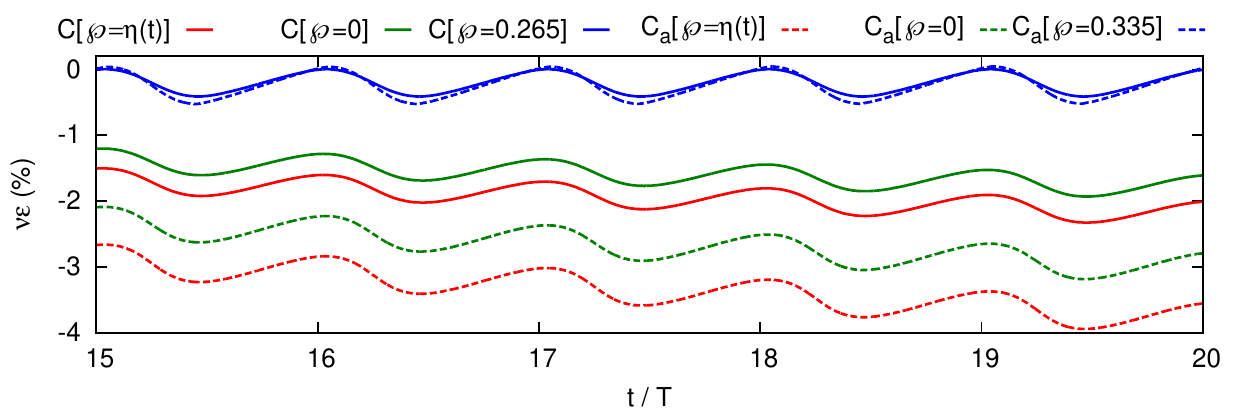}
}\\
{
\centering
{\bf{(d)}}\includegraphics[trim=-1mm 25mm 0mm 23mm, clip, width = 14.1cm]{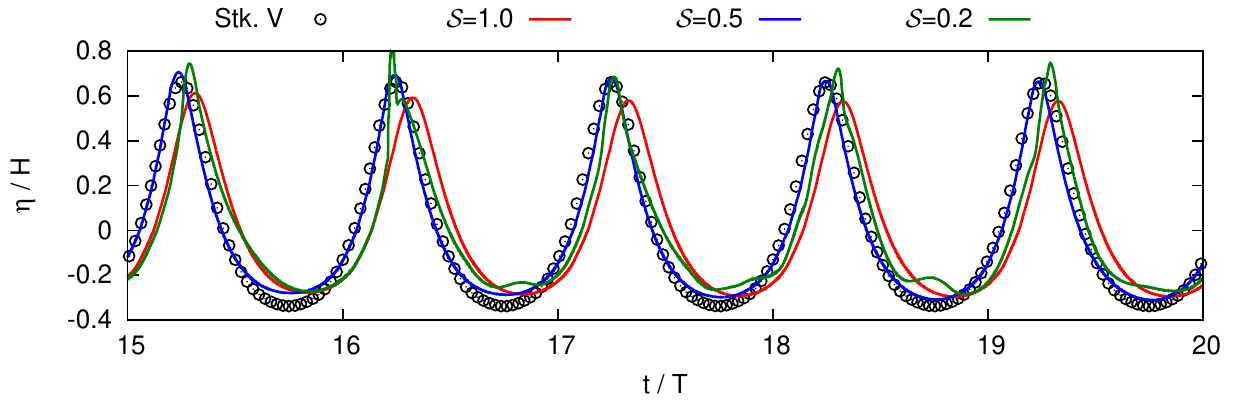}
}
\end {tabular}
\end {center}
\caption{\emph{Comparative assessment of various inflow-boundary {\textsf{WM}}s and momentum advection schemes for large-steepness wave generation: (a,b,d) $\eta(t)$ signals (measured at $x=\mathsf{WM}+6\lambda$) for cases (a) $\mathbf{C}$ and (b,d) $\mathbf{C_a}$ and (c) time variation of volume error $(\mathcal{VE})$ during the last five wave periods of the simulation.}}
\label{fig:larstptopl}
\end{figure}
Results of the NWT simulations are depicted in \autoref{fig:larstptopl}. In \autoref{fig:larstptopl}(a,b), local variation of free surface elevation $(\eta(t))$ is compared for baseline $(\wp=\eta(t);\zeta \equiv y)$, Wheeler stretched $(\wp=0;\zeta \equiv y)$ and modified $(\wp>0\,\, \forall\,\,y<h; \wp=\eta(t)\,\,\forall\,\, y\geq h)$ inflow-based {\textsf{WM}}s. Although Wheeler stretching over-designs $\left|U_{TL}\right|$, the technique only manages to correct wave-setup by a small amount. In contrast, the proposed ``modified inflow'' {\textsf{WM}} clearly outperforms the conventional inflow-boundary based {\textsf{WM}}s as setup induced in $\eta(t)$ due to volume addition is convincingly nullified in both cases. In addition, volume preservation characteristics of the three inflow-boundary {\textsf{WM}}s have been quantified in \autoref{fig:larstptopl}(c). Quantification is based on monitoring the percentage change in primary phase (water) volume $(\mathcal{VE})$ occurring within the NWT \cite{sas17a} during the last five wave periods of the simulation. A net volume addition over a wave period $(\mathcal{V}_+)$ is clearly observable in the baseline and Wheeler formulations whilst modified inflow {\textsf{WM}} is seen to exactly preserve volume over a wave period. \autoref{fig:larstptopl}(c) also establishes that $\mathcal{V}_+ \propto \mathsf{Ur}$ (cf. \autoref{tab:tab1}) which explains the stronger setup induced for case $\mathbf{C_a}$ (cf. \autoref{fig:larstptopl}(b)). It naturally follows that ``optimum'' $\wp$ required for (exactly) balancing $\mathcal{V}_+$ would also increase with Ursell number $(\wp \propto \mathsf{Ur})$ which is established from \autoref{fig:larstptopl}. \\
The ability of the proposed scheme-blending strategy (cf. \cref{eq:schbln}) to prevent numerical damping of steep waves is exhibited in \autoref{fig:larstptopl}(d) for case $\mathbf{C_a}\,(\mathsf{Ur}>27)$. Given that $H/\lambda \approx 0.05$ for case $\mathbf{C_a}$, pure FOU based momentum advection $(\mathscr{S}=1)$ leads to considerable height damping $(\mathcal{HE}\approx 15\%)$ as the waves propagate towards the ESL. In such a situation, the results substantiate the computational reasonableness of using a blended scheme $(\mathscr{S}=0.5)$ in lieu of mesh refinement $(nx_\lambda\approx 170)$ to control numerical damping. However, it is worth observing from \autoref{fig:larstptopl}(d) that large amounts of QUICK in the blend $(\mathscr{S}=0.2)$ introduces dispersion errors which induce wave distortion; $\mathscr{S}\geq 0.5$ is thus retained for all validation cases considered in \autoref{sec:polywaves}-\autoref{sec:wavetrans}. Here, FOU plays the important role of limiting spurious momentum generated due to a ``density discontinuity'' existing at the air-water interface \cite{pros07}.  
\subsubsection{Extreme steepness waves $(H/\lambda > 0.05)$} \label{sssec:extstp}
\begin {figure}[!ht]
\begin {center}
\begin {tabular}{c}
{
\centering
{\bf{(a)}}\includegraphics[trim=0mm 25mm 0mm 22mm, clip, width = 14cm]{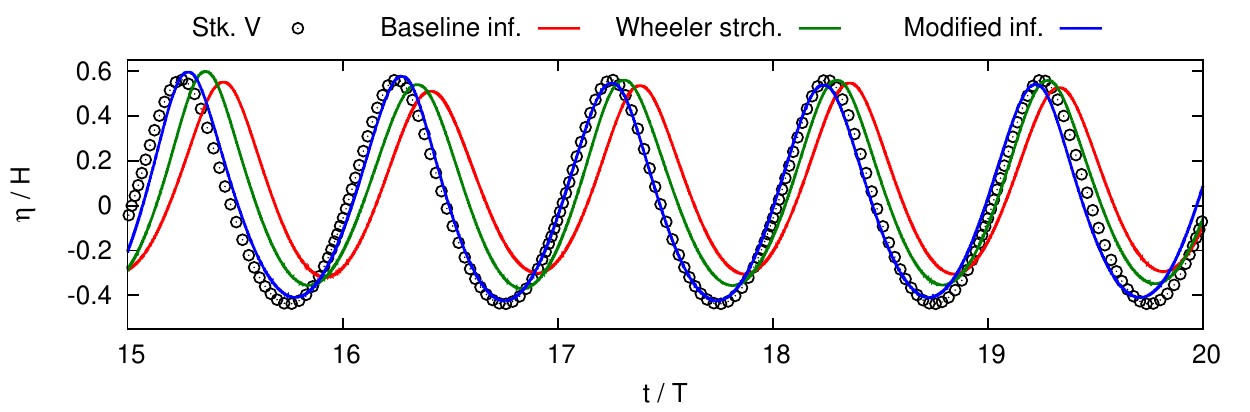}
}\\
{
\centering
{\bf{(b)}}\includegraphics[trim=-4mm 25mm 0mm 23mm, clip, width = 14cm]{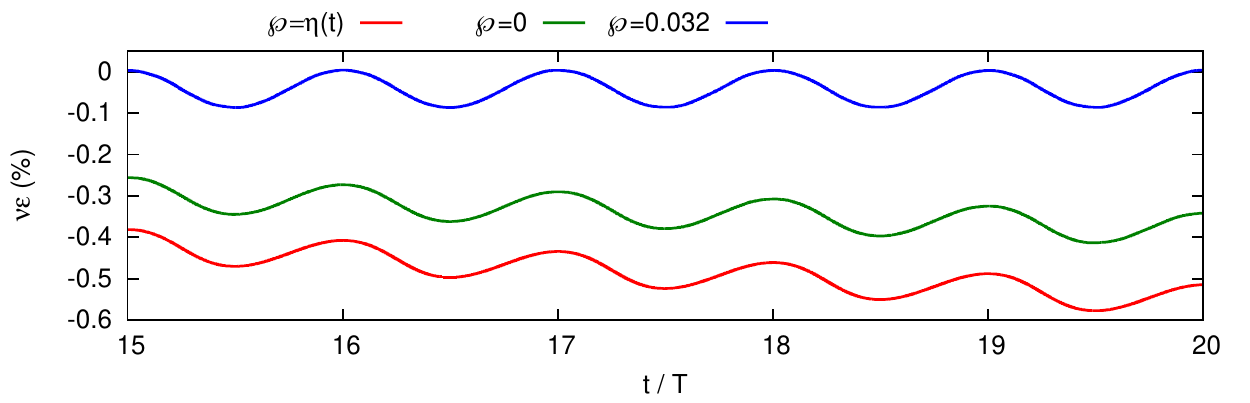}
}\\
{
\centering
{\bf{(c)}}\includegraphics[trim=0mm 25mm 0mm 23mm, clip, width = 14cm]{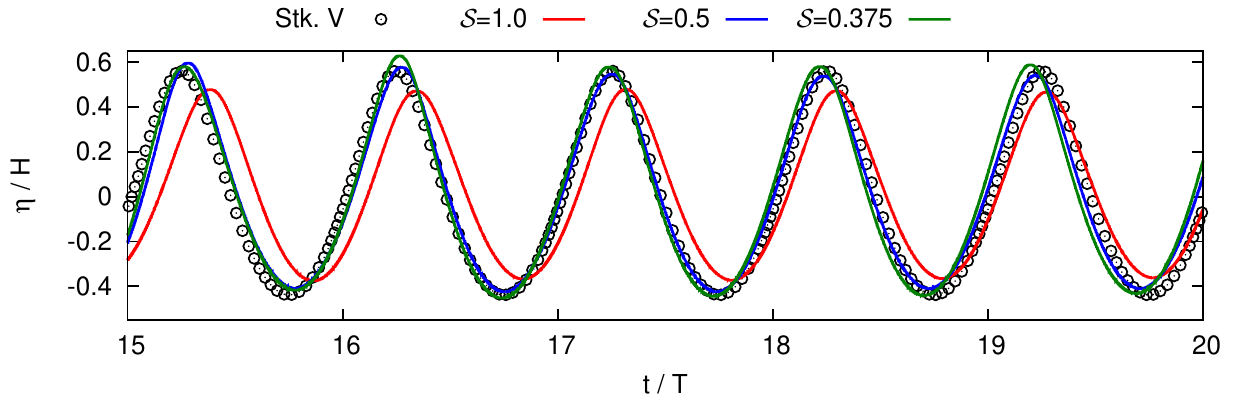}
}
\end {tabular}
\end {center}
\caption{\emph{Comparative assessment of various inflow-boundary based {\textsf{WM}}s and momentum advection schemes for extreme steepness (case $\mathbf{I_b}$) wave generation through (a,c) $\eta(t)$ signals (measured at $x=\mathsf{WM}+6\lambda$) and (b) time variation of volume error $(\mathcal{VE})$ during the last five wave periods of the simulation.}}
\label{fig:extstptopl}
\end{figure}

The final wave design considered for benchmarking the proposed NWT model against regular wave generation is the extreme-steepness case $\mathbf{I_b}$ in deep water (cf. \autoref{tab:tab1}). The following NWT setup has been considered for case $\mathbf{I_b}$ simulations: $\mathbb{L}=8.0\,m$, $\mathbb{H}=1.0\,m$, $\ell_d\approx 6\lambda$; $nx_\lambda \approx 177$, $ny_H \approx 66$; $R=0.4$. Time is non-uniformly advanced using the forward Euler method: $T/\Delta t \geq 700 \ni C_{max} \leq 0.25$. Further, $\left(\begin{matrix} \wp \\ \mathscr{S} \end{matrix} \right) \equiv \left(\begin{matrix} 0.032 \\ 0.5 \end{matrix} \right)$ are parametrically chosen with the pressure initialized using the hydrostatic law. Extreme-steepness wave-generation capabilities of various inflow-boundary-based {\textsf{WM}}s are depicted in \autoref{fig:extstptopl}. A thorough appreciation of the versatility of the modified-inflow {\textsf{WM}} can be gained by assessing the $\eta(t)$ profiles reported in \autoref{fig:extstptopl}(a).
Of the three inflow-boundary {\textsf{WM}}s considered, modified-inflow is observed to attain the closest agreement with {\texttt{Stokes V}} predictions. The baseline-inflow and Wheeler-stretching-based simulations are observed to suffer from both height damping as well as wave-setup. Interestingly, the height-damping observed in the case of Wheeler stretching $(\wp=0)$ is chiefly attributable to the fact that the technique under-designs streamwise momentum at points lying above the SWL. Hence, a preservation of {\texttt{Stokes V}} momentum prediction above the SWL combined with momentum over-design below the SWL helps arrest numerical damping of waves (even for $H/\lambda \approx 0.075$) in the case of the modified-inflow-based NWT. \\
Owing to $\mathsf{Ur}\sim 0.1$, the topology of case $\mathbf{I_b}$ is predominantly sinusoidal; this results in a comparatively lower net volume addition at $t=20T$ even in the case of the baseline inflow {\textsf{WM}} (cf. \autoref{fig:extstptopl}(b)). This substantiates the hypothesis previously put forth in \autoref{sssec:lrgstp} that $\mathcal{V}_+ \propto \mathsf{Ur}$. Quite interestingly, one may observe from \autoref{fig:extstptopl}(b) that the topological symmetry of the wave design gets reflected in the corresponding $\mathcal{VE}(t)$ record over a wave-generation cycle.  Owing to extreme steepness, the generated waves necessitate an ``optimally blended'' advection scheme which is evident from \autoref{fig:extstptopl}(c).  
\subsection{Topology of generated waves: quantitative assessment} \label{ssec:topval_quant}
Regular wave generation performance of the modified-inflow {\textsf{WM}} is quantified in the present section.  
\begin{table}[h]
\caption{Quantification of regular wave generation in the NWT at $t=20T$. Height $(\mathcal{HE})$, wavelength $(\mathcal{WE})$ and volume $(\mathcal{VE})$ errors are shown in terms of percentages (values $>5\%$ are {\uline{underlined}}; -ve values indicate under-prediction of $H,\lambda$ but also volume addition).}
\begin{tabularx}{\textwidth}{X c c c c c}
\toprule
{} & {\bf{SH}} & {\bf{SLH}} & {\bf{C}} & {$\mathbf{C_a}$} & {$\mathbf{I_b}$}\\
\midrule
{$\mathcal{HE}_1$} & {$-3.64e+00$} & {$2.18e+00$} & {\uline{$6.06e+00$}} & {$2.56e+00$} & {\uline{$-9.72e+00$}}\\
{$\mathcal{HE}_2$} & {$2.56e+00$} & {$-6.08e-01$} & {$-2.72e-01$} & {$-3.33e-02$} & {\uline{$-9.30e+00$}}\\
{$\mathcal{HE}_3$} & {$7.20e-01$} & {$2.55e+00$} & {\uline{$6.54e+00$}} & {\uline{$6.35e+00$}} & {\uline{$-7.96e+00$}}\\
{$\mathcal{HE}_4$} & {$-2.56e+00$} & {$-6.73e-01$} & {$2.24e+00$} & {$1.50e-01$} & {$-2.45e+00$}\\
{$\mathcal{HE}_5$} & {$1.52e+00$} & {$1.30e+00$} & {$-1.85e+00$} & {\uline{$-5.33e+00$}} & {$1.67e+00$}\\
{$\mathcal{HE}_6$} & {$-2.80e-01$} & {--} & {$-1.08e+00$} & {\uline{$-6.24e+00$}} & {$-3.82e+00$}\\
{$\mathcal{HE}_7$} & {$6.40e-01$} & {--} & {$-6.66e-01$} & {\uline{$-5.38e+00$}} & {$-4.81e+00$}\\
\midrule
{$\mathcal{WE}_1$} & {$-1.28e+00$} & {$8.03e-02$} & {$2.95e+00$} & {$4.66e+00$} & {$3.60e-01$}\\
{$\mathcal{WE}_2$} & {$4.81e-01$} & {$4.51e-01$} & {$1.55e+00$} & {$3.63e+00$} & {$3.49e+00$}\\
{$\mathcal{WE}_3$} & {$-2.11e-01$} & {$2.26e-01$} & {$-1.77e+00$} & {$-1.95e+00$} & {$1.99e+00$}\\
{$\mathcal{WE}_4$} & {$-1.14e+00$} & {$2.90e-01$} & {$-1.30e+00$} & {$-7.17e-01$} & {$1.59e+00$}\\
{$\mathcal{WE}_5$} & {$2.55e-01$} & {$-6.12e-01$} & {$7.97e-02$} & {$5.67e-01$} & {$6.83e-03$}\\
{$\mathcal{WE}_6$} & {$-6.06e-01$} & {--} & {$5.22e-01$} & {$5.75e-01$} & {$-3.13e+00$}\\
{$\mathcal{WE}_7$} & {$1.00e+00$} & {--} & {$9.21e-02$} & {$-3.50e-01$} & {$-3.11e+00$}\\
\midrule
{$\mathcal{VE}$} & {$1.07e-04$} & {$-5.97e-03$} & {$-5.41e-03$} & {$1.81e-02$} & {$3.91e-03$}\\
\bottomrule
\end{tabularx}
\label{tab:tab2}
\end{table}
The quantification is done in terms of percentage errors in height and wavelength of individual wave packets \cite{sas17a} as well as the net percentage change in water volume (over twenty wave-generation cycles) for all five wave designs considered in \autoref{sssec:smlstp}-\autoref{sssec:extstp}. It is noticeable that the proposed {\textsf{WM}} is capable of simulating a target wave-topology with sufficient accuracy. This statement is especially true for the wavelength $(\lambda)$ as $\mathcal{WE}<5\%$ in all cases. This in turn demonstrates that the kinematic over-design introduced below the SWL (cf. \autoref{fig:modinf}) does not hamper the ability of the {\textsf{WM}} to correctly capture amplitude dispersion. Achieving the target wave-height $(H)$ proved to be more challenging, especially for $H/\lambda \gtrapprox 0.05$ (cases $\mathbf{C_a}$ and $\mathbf{I_b}$). Nonetheless, $\mathcal{HE}\approx 5\%$ for $x>{\mathsf{WM}}+4\lambda$ is certainly acceptable considering the large steepness of the designs considered. It is also evident from \autoref{tab:tab2} that the modified-inflow {\textsf{WM}} exhibits excellent volume preservation characteristics: $\mathcal{VE}<0.02\%$ is sustained for all five designs even after twenty cycles of wave-generation. \\
In a novel attempt, the propagation characteristics of the generated waves are rigorously benchmarked against {\texttt{Stokes V}} predictions in the next section.  
\subsection{Graphical verification of the phase and group velocities} \label{ssec:C_Cg}
\begin {figure}[!ht]
\begin {center}
\begin {tabular}{c}
{\centering
{\textbf{(a)}}\includegraphics[trim=15mm 0mm 15mm 0mm, clip, height = 11cm]{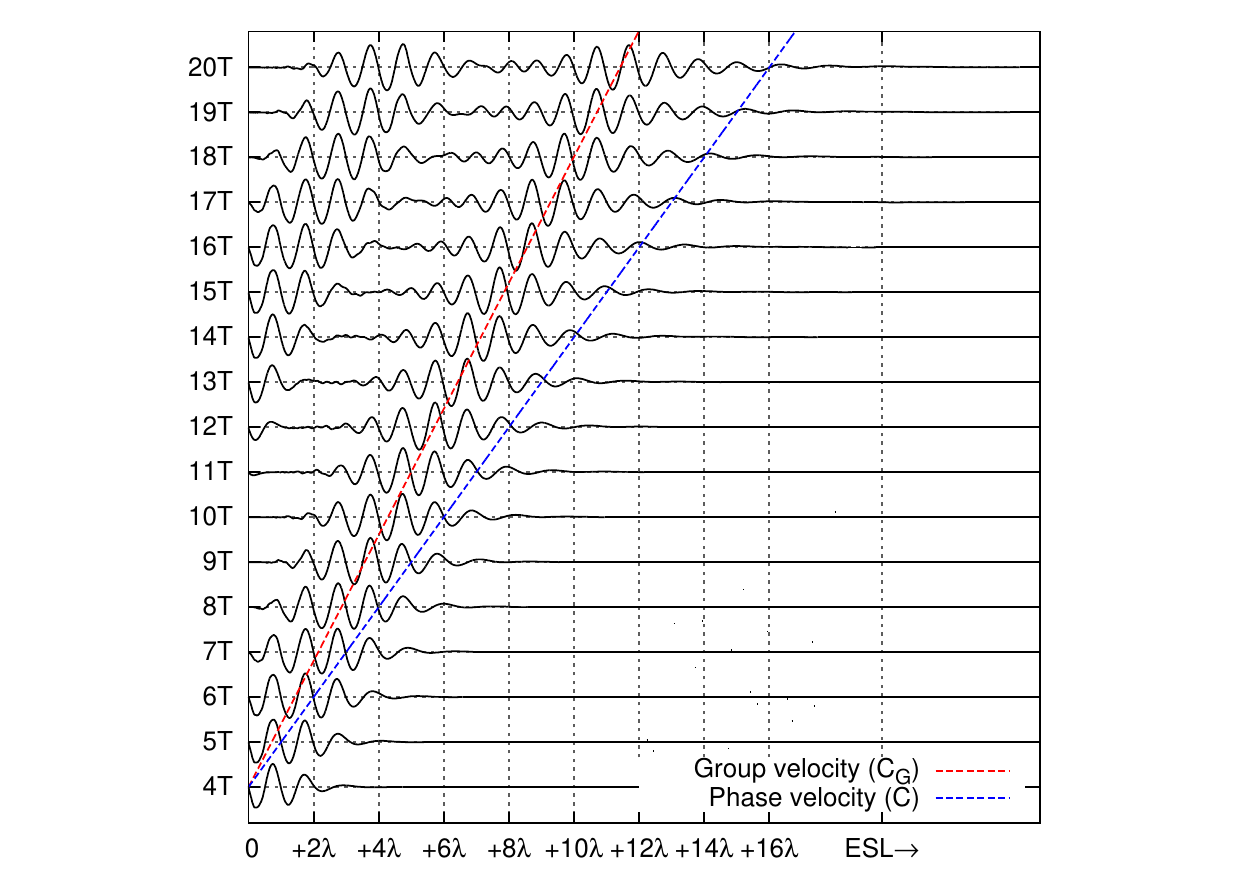}
}\\
{\centering
{\textbf{(b)}}\includegraphics[trim=15mm 0mm 15mm 0mm, clip, height = 11cm]{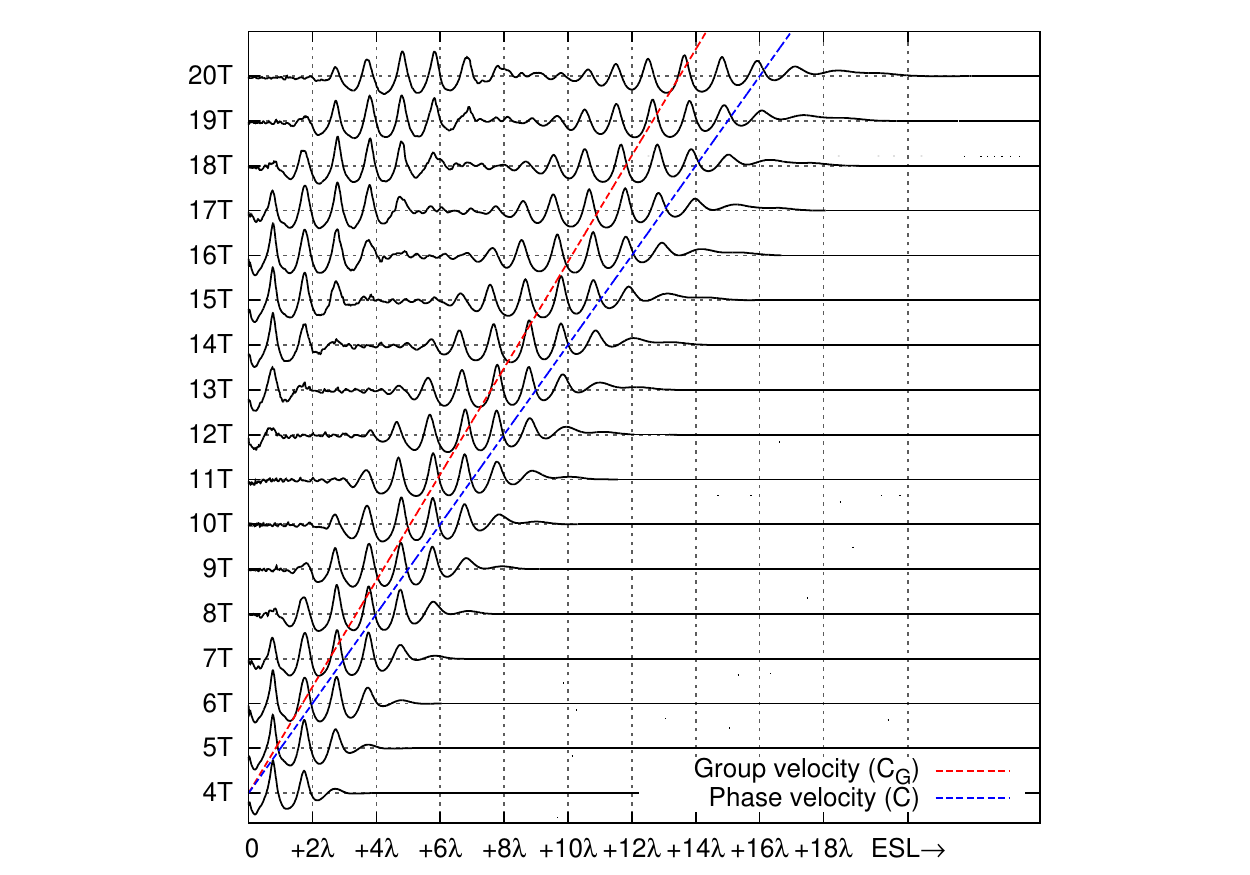}
}
\end {tabular}
\end {center}
\caption{\emph{Graphical representation and verification of the group velocity $\left(C_G\right)$ of (a) regular {\textbf{SH}} and (b) case {\textbf{C}} wave-groups by means of wave-waterfall diagrams.}} 
\label{fig:grpvel_SH_C}
\end{figure}

In the present section, the phase $(C)$ and group velocities $(C_G)$ of the waves generated by the modified-inflow {\textsf{WM}} are validated. Such an assessment is considerable more rigorous in comparison to topological validation (reported in \autoref{ssec:topval_qual} and \autoref{ssec:topval_quant}) because the $C,C_G$ validation mandates a simultaneous agreement between the wave profile, it's propagation speed $(C)$ and amplitude dispersion. In the present work, $C$ and $C_G$ of regular {\bf{SH}} and case {\bf{C}} waves have been graphically evaluated by means of ``wave-waterfall diagrams'' \cite{maguire11} and validated against {\texttt{Stokes V}} theory. The waterfall diagrams are essentially $x-t$ plots of $\eta(x)$ profiles with shoreward distance (NWT length) represented along the abscissa and discrete time instances (being integer multiples of the wave period $T$) of said $\eta(x)$ profiles represented along the ordinate. Then, {\textbf{the inverse slope of a line passing through the same wave phase across adjacent wave periods represents the phase speed of the waves.}} 
\begin{table}[h]
\caption{Variables governing individual and group propagation characteristics of waves.}
\begin{tabularx}{\textwidth}{X c c c c c c}
\toprule
{} & {$\lambda_V\,(m)$} & {$T\,(s)$} & {$kh$} & {$C\,(m/s)$} & {$C_G\,(m/s)$} & {$C/C_G$} \\
\midrule
{{\bf{SH}}\cite{beji94}} & {$2.055$} & {$1.25$} & {$1.22$} & {$1.644$} & {$1.174$} & {$1.400$}\\
{{\bf{C}}\cite{sas17a}} & {$2.421$} & {$1.50$} & {$0.78$} & {$1.614$} & {$1.361$} & {$1.186$}\\ 
\bottomrule
\end{tabularx}\\[5pt]
\label{tab:tab3}
\end{table}
\begin {figure}[!ht]
\begin {center}
\begin {tabular}{c}
{\centering
{\textbf{(a)}}\includegraphics[trim=15mm 0mm 15mm 0mm, clip, height = 11cm]{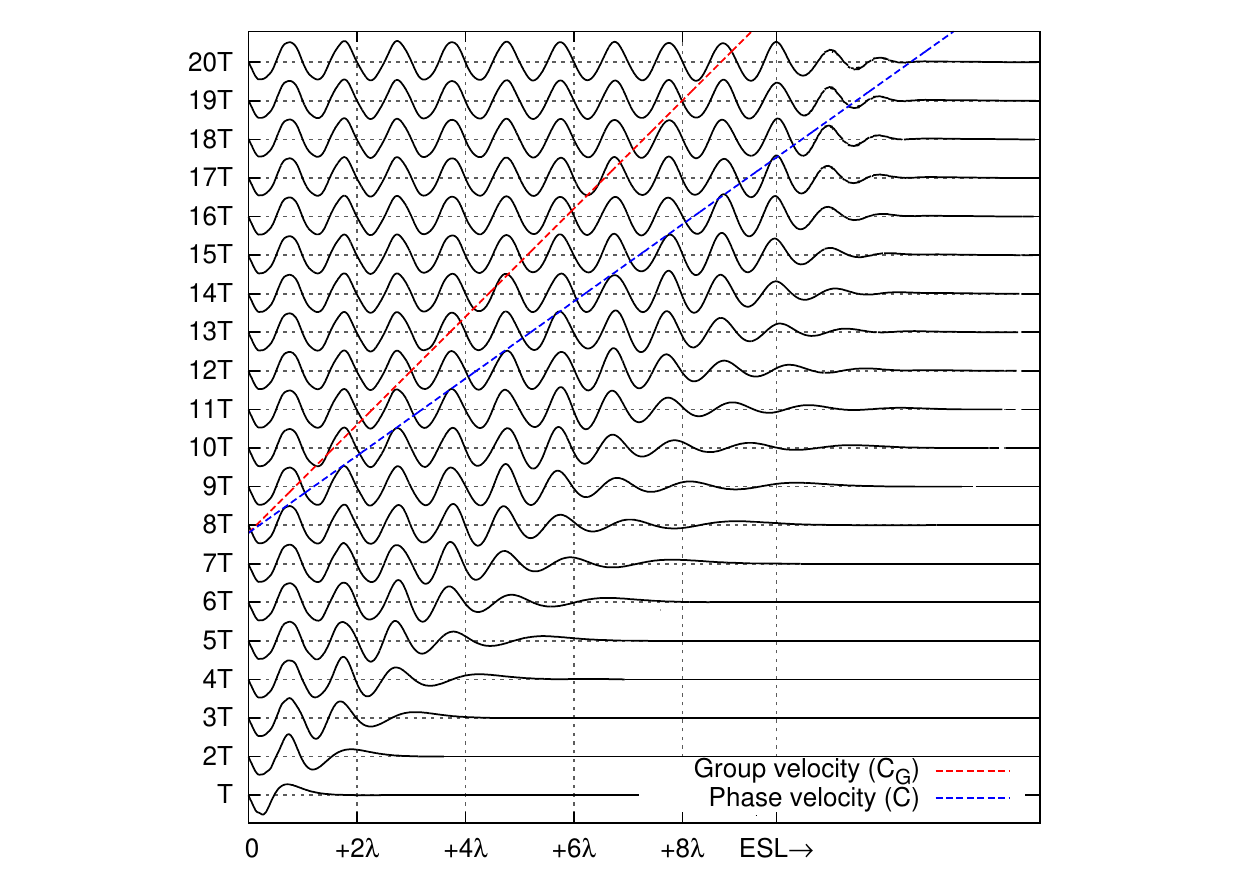}
}\\
{\centering
{\textbf{(b)}}\includegraphics[trim=15mm 0mm 15mm 0mm, clip, height = 11cm]{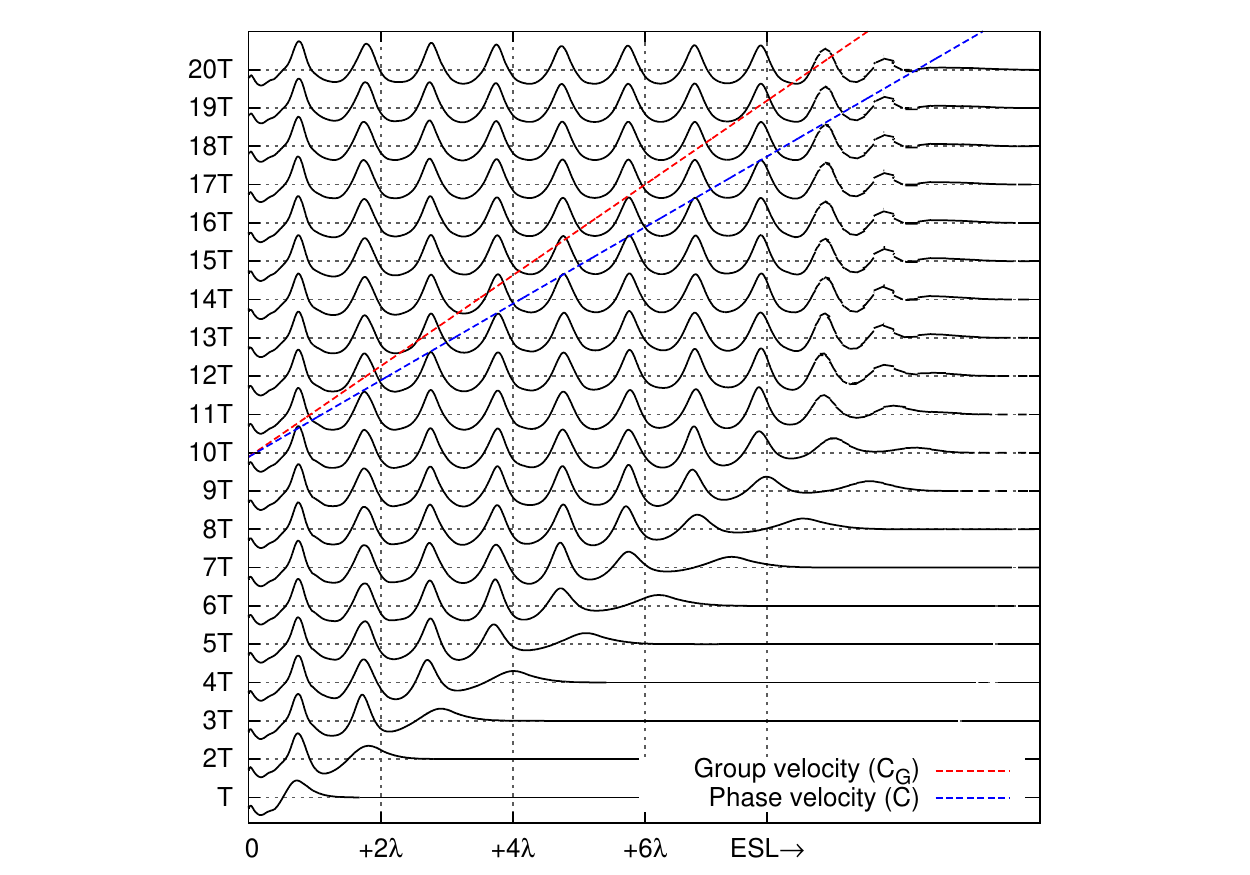}
}
\end {tabular}
\end {center}
\caption{\emph{Graphical representation and verification of the phase $\left(C\right)$ velocities of (a) regular {\textbf{SH}} and (b) case {\textbf{C}} waves by means of wave-waterfall diagrams.}} 
\label{fig:phvel_SH_C}
\end{figure}
Further, if the waves propagate in deep $(kh\geq \pi \vdash C=2C_G)$ or shallow water $(kh\leq \pi/10 \vdash C=C_G)$, the slope of the $C_G$ line would be respectively {\emph{twice}} or {\emph{equal}} to the slope of the $C$ line. Hence, the deep and shallow-water limits of wave theory present certain simplifications \cite{maguire11} to the analysis in that the same waterfall plot could be used for graphically representing both $C$ as well as $C_G$. In the present evaluation, however, the peculiar choice of wave design(s) precludes citation to such simplifying assumptions of relative depth (cf. \autoref{tab:tab3}). It is evident from \autoref{tab:tab3} that $C$ isn't an integer multiple of $C_G$ for either wave design. Thus, it is not possible to graphically verify $C_G$ on a waterfall diagram of a {\emph{regular wave train}} in either case. This limitation is evident from \autoref{fig:phvel_SH_C}. Whilst an excellent graphical validation is achieved for $C$ in both cases, the line representing $C_G^{-1}$ proves rather immaterial in either case. Given that $C_G$ is the propagation speed of the {\emph{envelope}} comprising a group of waves \cite{dean91}, it becomes necessary to generate a wave envelope that evolves in a spatio-temporal sense along the shoreward direction.  
\begin {figure}[!ht]
\begin {center}
\begin {tabular}{c}
{\centering
\includegraphics[trim=0mm 30mm 0mm 32mm, clip, width = 14cm]{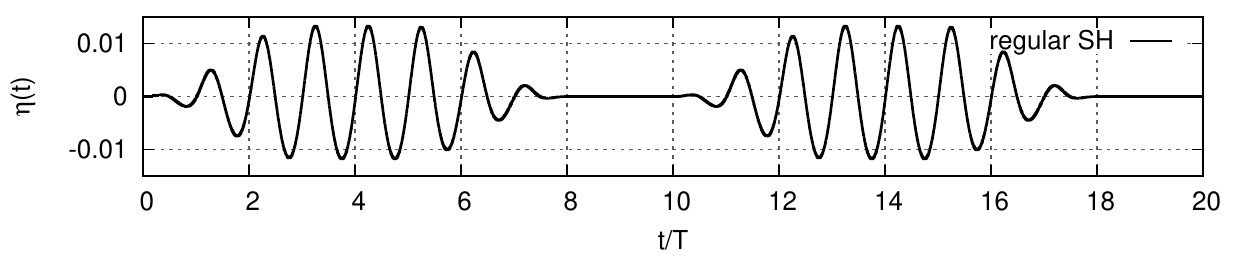}
}\\
{\centering
\includegraphics[trim=0mm 30mm 0mm 32mm, clip, width = 14cm]{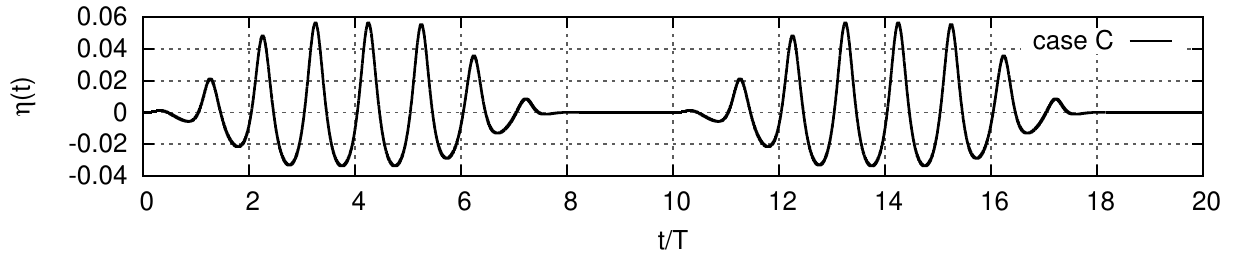}
}
\end {tabular}
\end {center}
\caption{\emph{Free-surface elevation signals (in meters) locally imposed at the inflow boundary for generating stable groups of regular {\bf{SH}} and case {\bf{C}} waves.}} 
\label{fig:grpvel_WM_signals}
\end{figure}
Once formed, shoreward evolution of such an envelope could then be traced by means of waterfall diagrams similar to those of \autoref{fig:phvel_SH_C}. In order to generate ``groups'' of regular {\bf{SH}} and case {\bf{C}} waves, the inflow {\textsf{WM}} was run on an {\texttt{zero$\rightarrow$up-ramping$\rightarrow$steady$\rightarrow$down-ramping$\rightarrow$zero}} cycle with a periodicity of $10T$ for twenty wave periods (cf. \autoref{fig:grpvel_WM_signals}). The up-ramping and down-ramping of the {\textsf{WM}} strength was achieved using appropriate cosine functions of the form: $\frac{1}{2}\left\{1\pm\cos\left\{\frac{\pi(t-\mathscr{X}T)}{3T}\right\}\right\}$, where $\mathscr{X}$ is an integer constant that was incremented by $10$ between consecutive cycles. Further, given that the beginning as well as end of a ``group generation cycle'' would invariably involve small steepness wave generation, $\Delta t$ was conservatively set to $T/10000$ in both cases to preserve numerical stability of the small waves \cite{sas17a}. In order to prevent height damping of case {\bf{C}} waves, a $50-50$ FOU-QUICK blend $(\mathscr{S}=0.5)$ was adopted for momentum advection. The regular {\bf{SH}} and case {\bf{C}} wave groups generated in the NWT (corresponding to the {\textsf{WM}} signals reported in \autoref{fig:grpvel_WM_signals}) are depicted by means of wave-waterfalls in \autoref{fig:grpvel_SH_C}. It can be appreciated that the wave groups are numerically stable and correctly propagate with the group speed $C_G$. 
Propagation with $C_G$ is most strikingly manifested in that the $C_G^{-1}$ line consistently follows the highest wave in the center of the group at all times. Thus, the $C_G^{-1}$ line intersects the envelope of the wave group at constant phase across consecutive wave periods. However, it should be noted that waves at the leading edge of the group are in a state of perpetual replacement owing to amplitude dispersion. Considering that this replacement occurs with the phase speed, the $C^{-1}$ line correctly intersects the leading edge of the group at constant phase across consecutive wave periods (cf. \autoref{fig:grpvel_SH_C}). Thus, the waves generated using the modified-inflow technique corroborate {\texttt{Stokes V}} predictions in both topological as well as a kinematic sense.
\subsection{Formation of evanescent modes} \label{ssec:evawaves} 
The proposed inflow-boundary-based NWT model would be eventually applied to wave transformation occurring over a submerged shoal. Thus, it is important to estimate the minimum separation necessary between the toe of the structure and the {\textsf{WM}} such that the waves incident on the structure are free from topological distortions induced during generation. 
\begin {figure}[!ht]
\begin {center}
\begin {tabular}{c}
{\centering
{\textbf{(a)}}\includegraphics[trim=0mm 17mm 5mm 17mm, clip, width = 15cm]{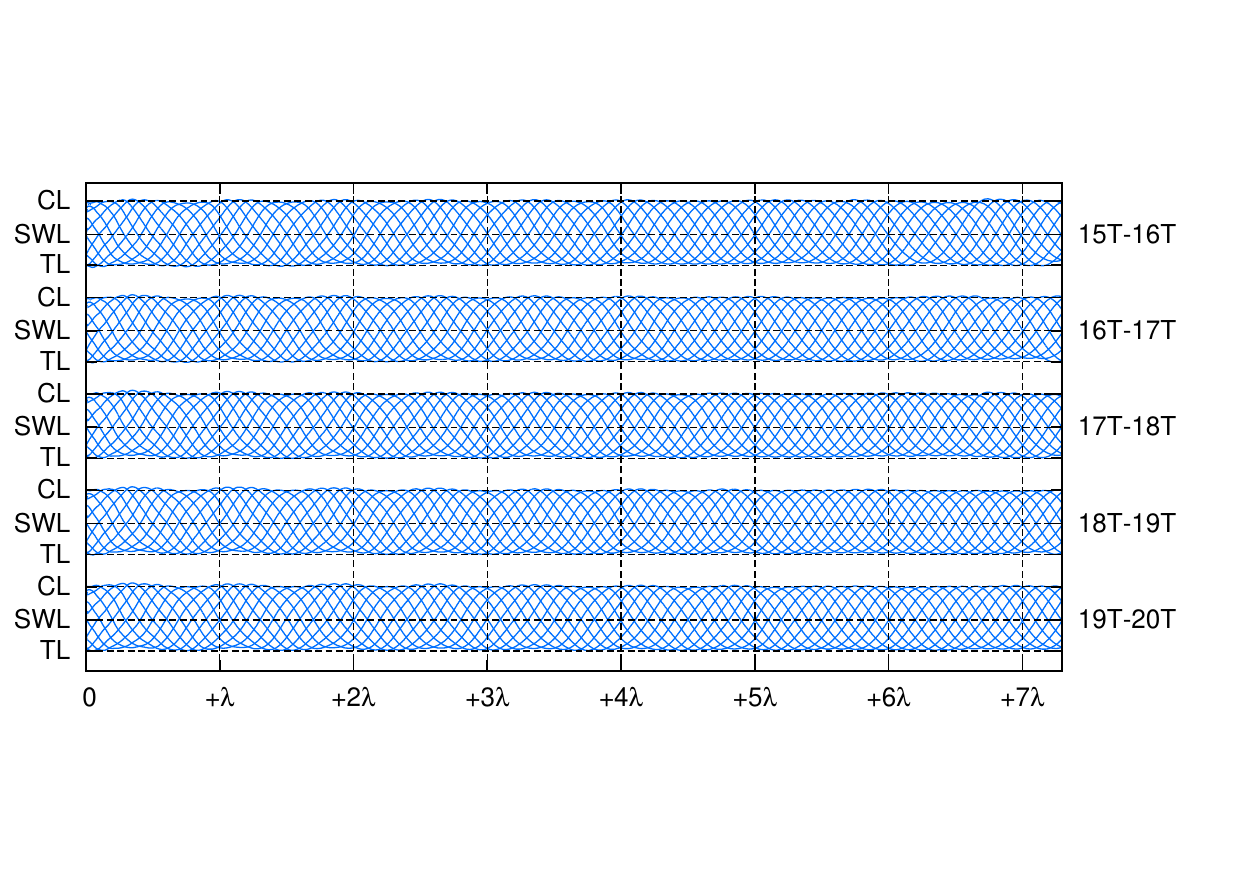}
}\\
{\centering
{\textbf{(b)}}\includegraphics[trim=0mm 17mm 5mm 17mm, clip, width = 15cm]{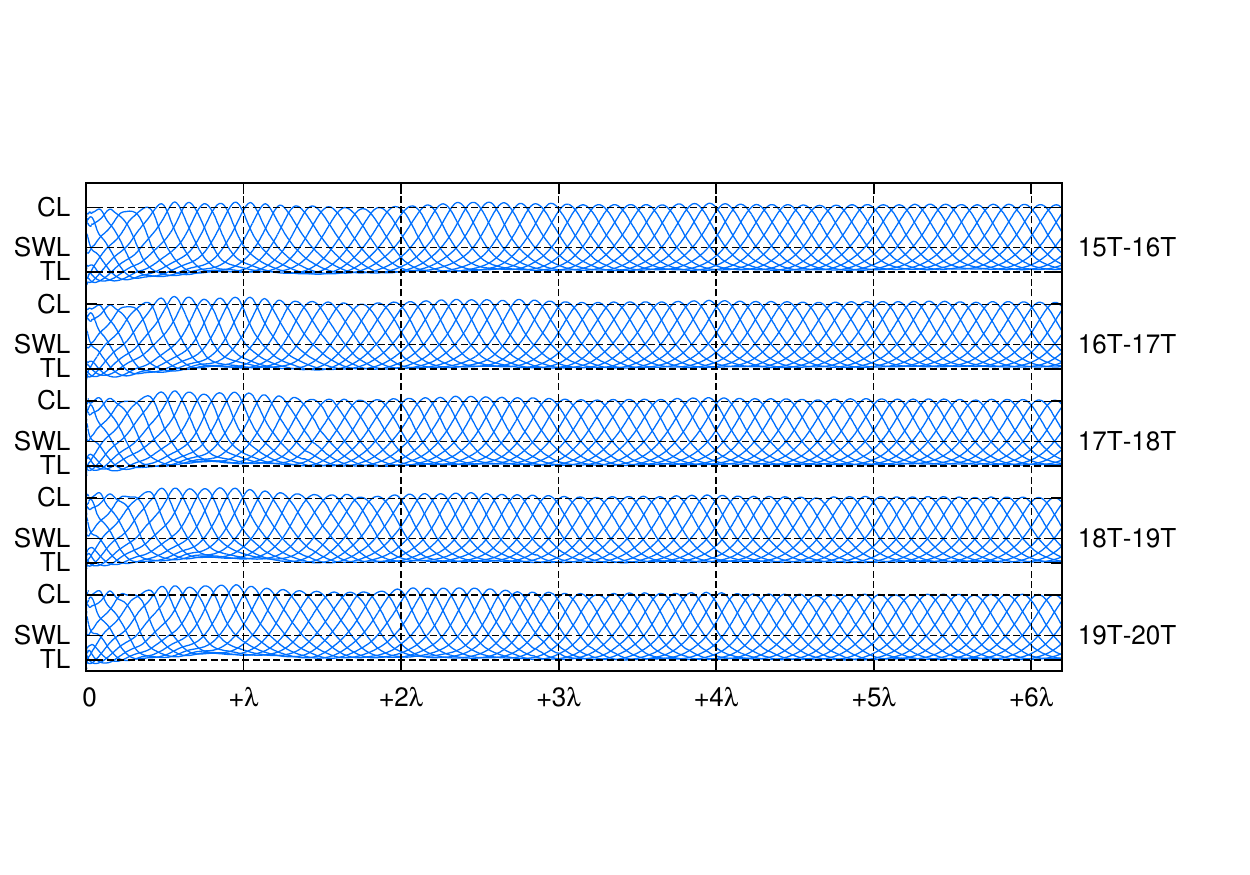}
}
\end {tabular}
\end {center}
\caption{\emph{Envelopes (reconstructed by successively plotting the PLIC-VOF interface ten times over the course of a wave-period \cite{sas17a}) representing (a) {\bf{SH}} and (b) case {\bf{C}} waves.}} 
\label{fig:eva_waves_SH_C}
\end{figure}
At the {\textsf{WM}}, the wave train gets distorted due to the emergence of evanescent modes which are essentially ``newborn standing waves''. The evanescent waves mark the region across which the velocity field transitions from ``wavemaker-induced'' to ``wave-induced''. A formal mathematical treatment aimed towards analyzing the spatial evolution of evanescent components induced by the modified-inflow technique is well beyond the scope of the present paper. Instead, the formation of evanescent modes at the inflow boundary is assessed graphically through generation of envelopes representing wave propagation during the final five periods $(15T-20T)$ in the NWT. The wave envelopes thus obtained have been plotted, for successive wave periods, in \autoref{fig:eva_waves_SH_C} for both regular {\bf{SH}} as well as case {\bf{C}} waves. In context to \autoref{fig:eva_waves_SH_C}, the effect of standing evanescent waves is manifested in the wave topology (thus envelope) getting displaced from the SWL \cite{maguire11,keaney15}. It is evident that the proposed modified-inflow {\textsf{WM}} {\emph{does}} exhibit evanescent waves (which are highly prominent in the case {\bf{C}} envelopes $(kh\approx 0.8)$). The existence of evanescent modes in case of the modified-inflow technique is attributed to the fact that, unlike the works of Maguire \cite{maguire11} and Keaney \cite{keaney15}, the explicit goal of designing the {\textsf{WM}} was volume preservation and {\emph{not}} elimination of evanescent waves. Nonetheless, given that the evanescent waves dampen exponentially as one moves away from the {\textsf{WM}} \cite{maguire11}, their effect is observed to be significant only within a distance of one {\emph{wavelength}} from the wavemaker. This is nonetheless significant when one considers the fact that $\lambda \approx 8h$ for case {\bf{C}} waves. Thus, it is advisable to place the structure atleast $+2\lambda$ from the {\textsf{WM}}, especially if $kh\leq 1$ for the incident waves. Steeper incident waves in $kh<1$ (such as cases {\bf{C}} or $\mathbf{C_a}$) might necessitate a minimum distance of $+4\lambda$ between the inflow-boundary and the (toe of the) structure.

\section{Monochromatic waves: uncertainty quantification} \label{sec:GCI}
\begin {figure}[!ht]
\begin {center}
\begin {tabular}{c c}
{$\Delta_4 \equiv 200\times 22$} & {$\Delta_3 \equiv 400\times 44$}\\
{\centering
\includegraphics[trim=5mm 30mm 0mm 35mm, clip, width = 7.5cm]{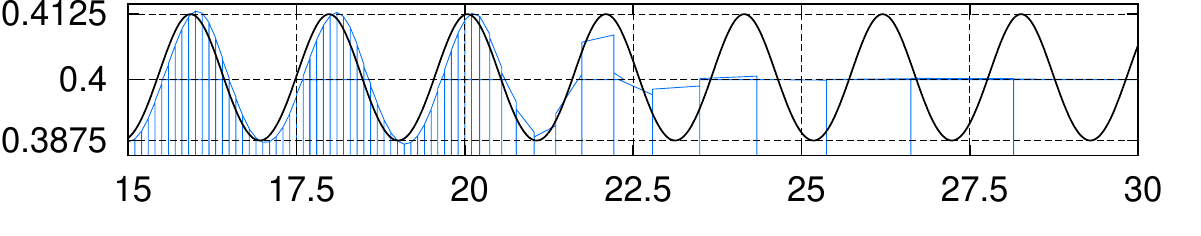}} & {
\centering
\includegraphics[trim=5mm 30mm 0mm 35mm, clip, width = 7.5cm]{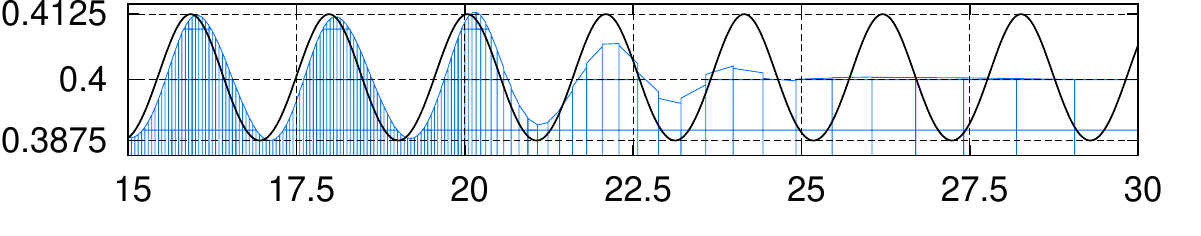}}\\
{$\Delta_2 \equiv 800\times 88$} & {$\Delta_1 \equiv 1600\times 176$}\\
{\centering
\includegraphics[trim=5mm 30mm 0mm 35mm, clip, width = 7.5cm]{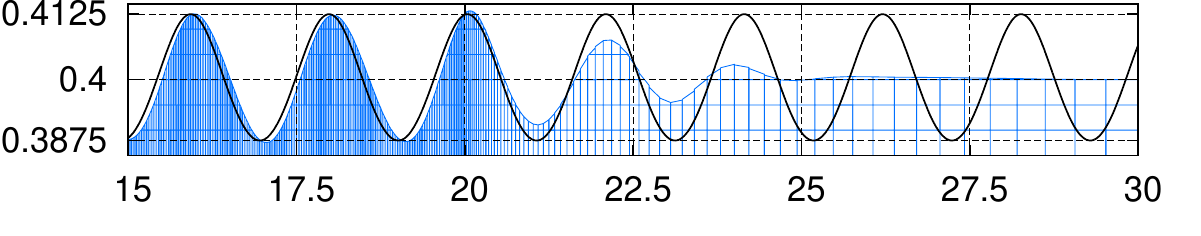}} & {
\centering
\includegraphics[trim=5mm 30mm 0mm 35mm, clip, width = 7.5cm]{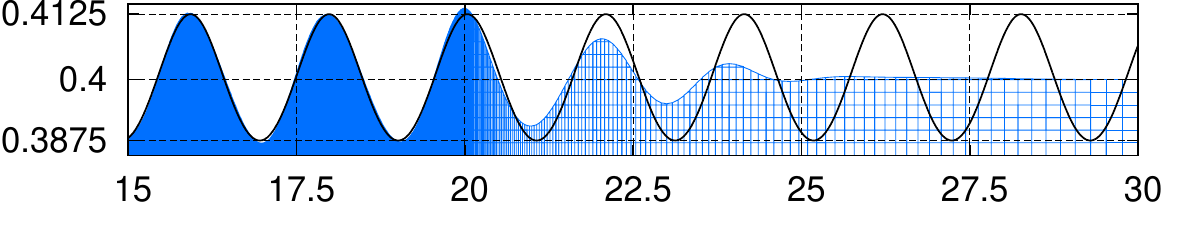}}\\
\midrule
{\centering
{\textbf{(a)}}\includegraphics[trim=10mm 0mm 15mm 0mm, clip, height = 6cm]{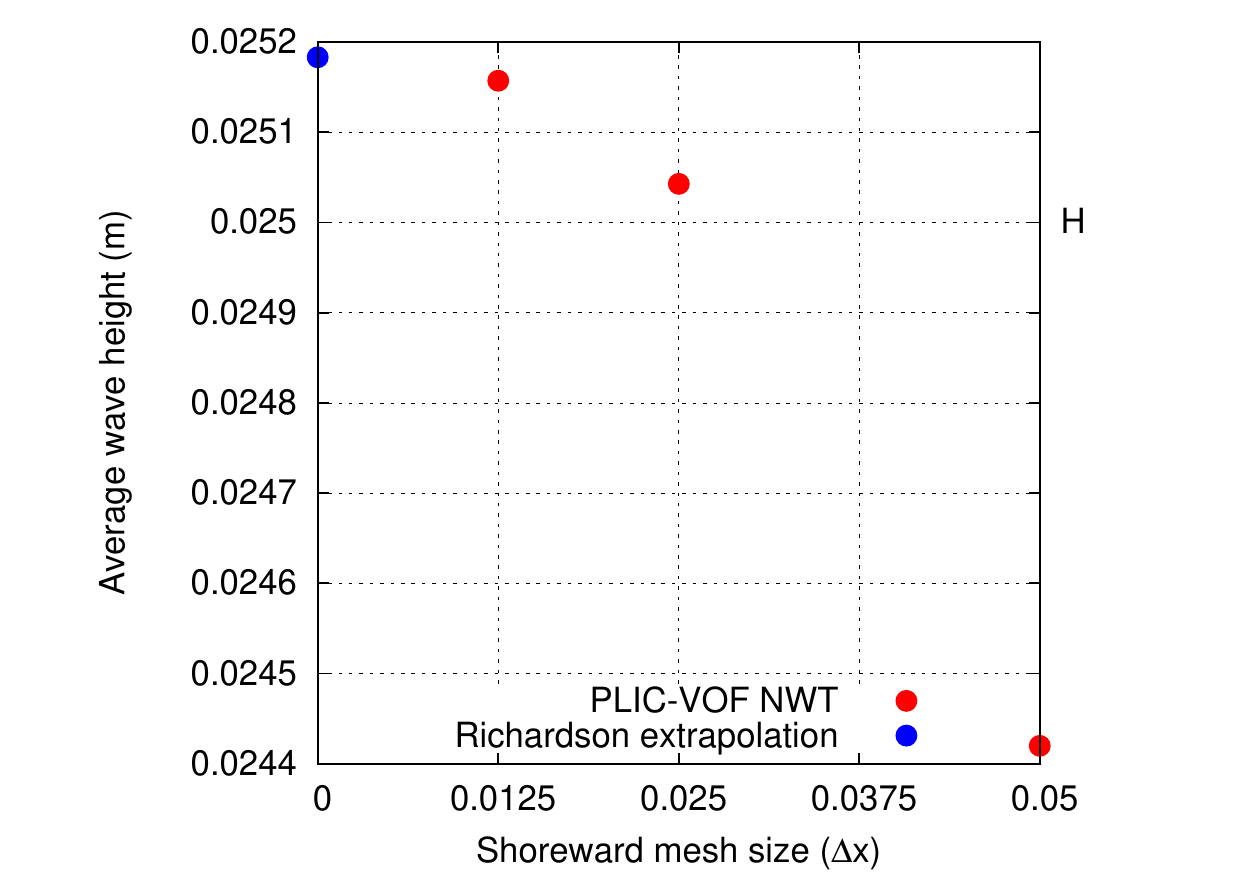}
} & {\centering
{\textbf{(b)}}\includegraphics[trim=10mm 0mm 15mm 0mm, clip, height = 6cm]{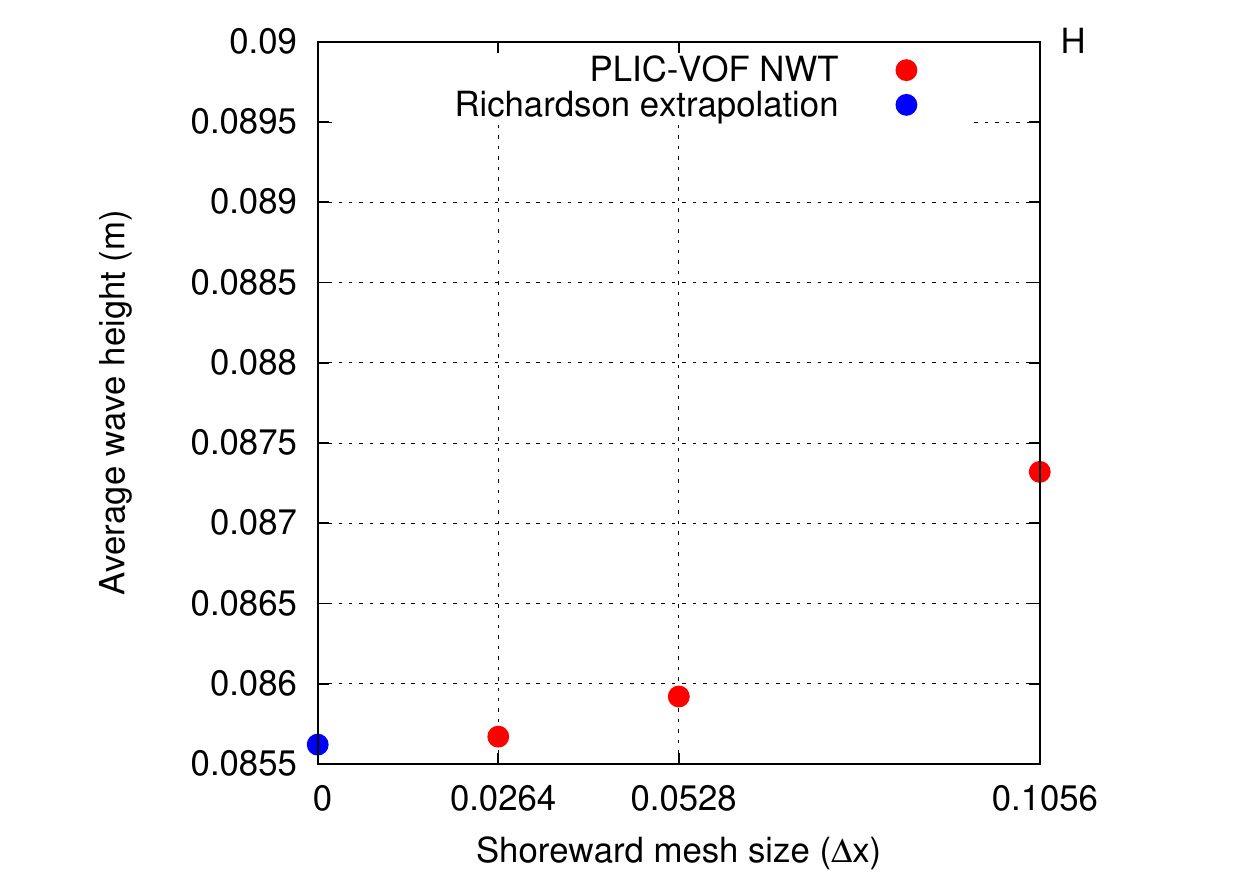}
}\\
\midrule
{$\Delta_4 \equiv 180\times 10$} & {$\Delta_3 \equiv 360\times 20$}\\
{\centering
\includegraphics[trim=5mm 30mm 0mm 35mm, clip, width = 7.5cm]{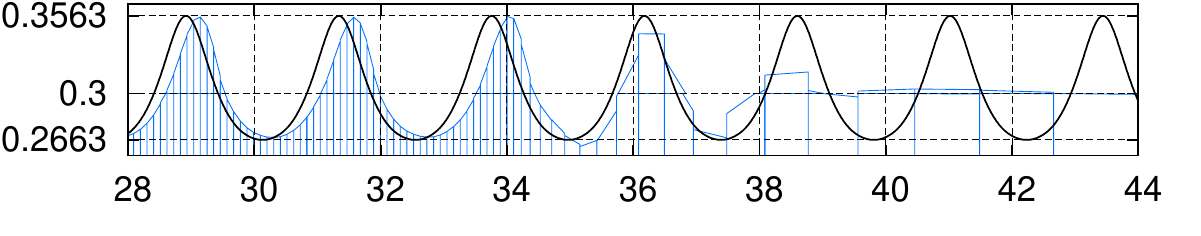}} & {
\centering
\includegraphics[trim=5mm 30mm 0mm 35mm, clip, width = 7.5cm]{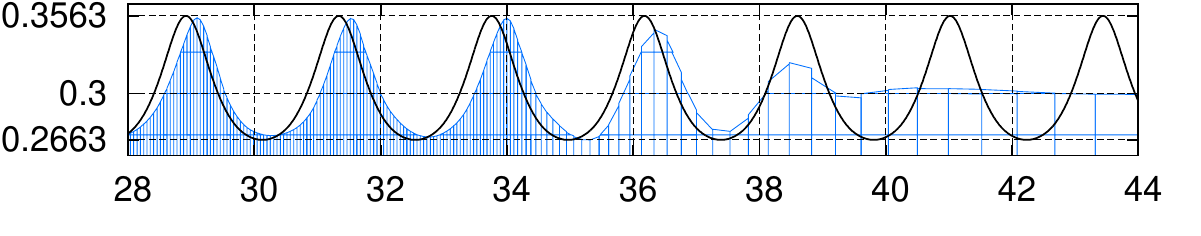}}\\
{$\Delta_2 \equiv 720\times 40$} & {$\Delta_1 \equiv 1440\times 80$}\\
{\centering
\includegraphics[trim=5mm 30mm 0mm 35mm, clip, width = 7.5cm]{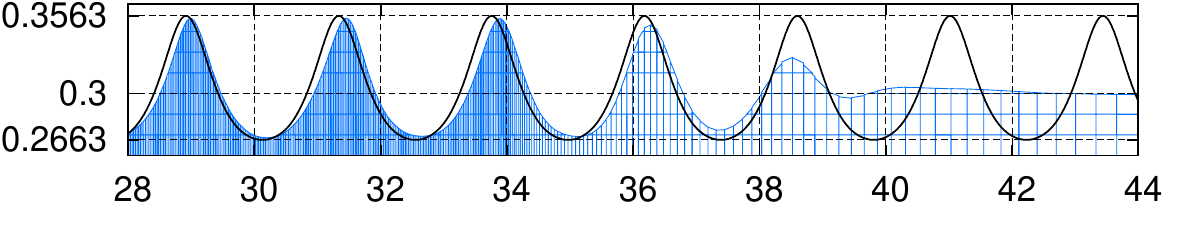}} & {
\centering
\includegraphics[trim=5mm 30mm 0mm 35mm, clip, width = 7.5cm]{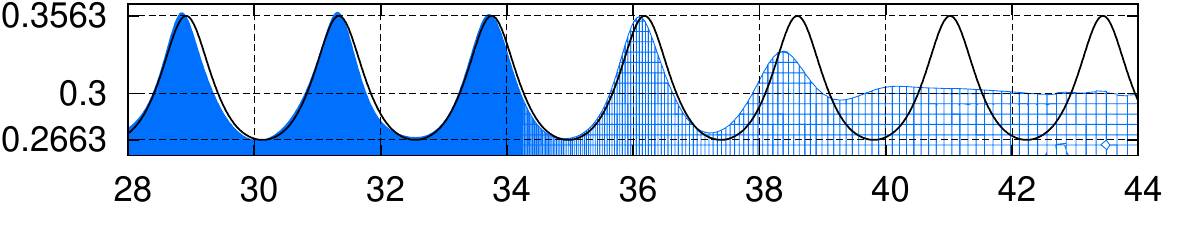}}
\end {tabular}
\end {center}
\caption{\emph{{\emph{(center)}}Demonstration of asymptotic solution convergence for (a) {\textbf{SH}} and (b) case {\textbf{C}} waves using the GCI methodology alongwith a reporting of the meshes considered and corresponding $\eta(x)$ profiles obtained for {\emph{(top)}} {\textbf{SH}} and {\emph{(bottom)}} case {\textbf{C}} waves.}} 
\label{fig:GCI_SH_C}
\end{figure}

The proposed PLIC-VOF NWT algorithm has been successfully validated against {\texttt{Stokes V}} theory for the topological quality as well as kinematics of propagation of the generated waves. Whilst a ``validation exercise'' ensures that the NWT algorithm correctly models the hydrodynamics of wave propagation, the same doesn't guarantee whether the simulations achieve a sufficiently high order of error convergence. Evaluation of the order of convergence of an algorithm falls within the purview of ``model verification'' which can be undertaken by means of a Grid-Convergence-Index (GCI) assessment. More specifically, a validation exercise ensures that the set of PDEs adopted correctly models the physics of the problem whilst a verification exercise ensures that the numerical solution (of the set of PDEs) approaches the exact solution as the spatio-temporal resolution is ``infinitely refined''. In the present work, the GCI methodology \cite{roache97} has been implemented as a code {\emph{verification}} exercise for the generation of regular {\textbf{SH}} and case {\textbf{C}} waves (cf. \autoref{fig:GCI_SH_C}). The mean wave height $\overline{H}$ of the waves, four wavelengths away from the wavemaker, has been selected as the parameter for GCI-based verification. The following procedure has been adopted for carrying out the GCI-based verification of the NWT algorithm:
\begin{enumerate}[topsep=0ex,itemsep=2pt,partopsep=0pt,parsep=0pt]
\item four sets of meshes $(\Delta_{1\rightarrow 4})$ have been selected for both regular {\bf{SH}} and case {\bf{C}} waves such that the cell size is uniform along both shoreward and depthward directions. The number of cells considered within the wave propagation region for the GCI-assessment are reported in \autoref{fig:GCI_SH_C} where, $\Delta_4$ and $\Delta_1$ represent the coarsest and finest meshes respectively. The grid has been successively refined by a fixed, integer rate $\mathscr{R}=2$. Further, the GCI methodology is only applicable to uniform meshes and hence the mesh structure presented here is vastly different from that adopted in the validation studies reported in \autoref{sssec:smlstp} and \autoref{sssec:lrgstp}. Regular {\bf{SH}} and case {\bf{C}} waves have been simulated on each mesh configuration for twenty wave periods; the corresponding $\eta(x)$ profiles are reported in \autoref{fig:GCI_SH_C}.
\item at $t=20T$, the average numerical wave height $\overline{H}$ has been evaluated within the region: $\mathsf{WM}+4\lambda \leq x \leq \mathbb{L}$. The $\overline{H}$ values thus obtained are reported in \autoref{tab:tab4}.
\item of the meshes $\Delta_{1\rightarrow 4}$, only three are necessary for the GCI analysis \cite{roache97}. Considering the fact that the order of convergence of the model has to be a real number, the values of $\overline{H}$ selected should be such that they: (a) pertain to three consecutive meshes and (b) do not contain inflexions. Based on this reasoning, $\overline{H}$ values pertaining to the mesh sets $\Delta_{1\rightarrow 3}$ and $\Delta_{2\rightarrow 4}$ have been selected for the GCI analysis of {\bf{SH}} and case {\bf{C}} waves respectively.
\item the formal order of convergence of the method is evaluated using the expression \cite{roache97}: $\mathfrak{p}={\ln\left(\frac{\overline{H}_{k+2}-\overline{H}_{k+1}}{\overline{H}_{k+1}-\overline{H}_k}\right)}\Big/{\ln(\mathscr{R})}$ where $k=1$ for {\bf{SH}} waves and $k=2$ for case {\bf{C}} waves. With reference to the $\overline{H}$ values reported in \autoref{tab:tab4}, one obtains $\mathfrak{p}=2.444$ and $\mathfrak{p}=2.485$ for {\textbf{SH}} and case {\textbf{C}} waves respectively. It is thus established that regular wave generation within the NWT is atleast second-order accurate.
\item the average wave height obtainable on an infinitely refined mesh $\overline{H}_0$ is predicted using Richardson extrapolation \cite{roache97}: $\overline{H}_0=\overline{H}_k+\frac{\overline{H}_k-\overline{H}_{k+1}}{\mathscr{R}^\mathfrak{p}-1}$. The extrapolation (depicted graphically in \autoref{fig:GCI_SH_C}(a,b) by means of $\overline{H}-\Delta x$ plots) reveals $\overline{H}_0=0.0252\,m$ for {\bf{SH}} waves and $\overline{H}_0=0.0856\,m$ for case {\bf{C}} waves.
\item the GCI has been evaluated for the medium and fine grids \cite{roache97}: $\mathrm{GCI}_{k+1,k+2}=F_s\cdot \left|\frac{\overline{H}_{k+1}-\overline{H}_{k+2}}{\overline{H}_{k+1}} \right|\Big/{\left(\mathscr{R}^\mathfrak{p}-1\right)} \times 100\%$ ; $\mathrm{GCI}_{k,k+1}=F_s\cdot \left|\frac{\overline{H}_k-\overline{H}_{k+1}}{\overline{H}_{k}} \right|\Big/{\left(\mathscr{R}^\mathfrak{p}-1\right)} \times 100\%$ using a ``moderately conservative'' safety factor \cite{roache97}: $F_s=1.25$. The medium and fine grid GCI values are reported in \autoref{tab:tab4} for both wave designs considered.
\item the medium and fine grid GCI's have finally been used to verify that the mesh set $\Delta_{k\rightarrow k+2}$ lies within the asymptotic range of convergence \cite{roache97}: $\frac{\mathrm{GCI}_{k+1,k+2}}{\mathscr{R}^{\mathfrak{p}}\cdot \mathrm{GCI}_{k,k+1}}\cong 1$. It is found that asymptotic convergence of the set $\Delta_{k\rightarrow k+2}$ indeed holds with $\frac{\mathrm{GCI}_{2,3}}{\mathscr{R}^{\mathfrak{p}}\cdot \mathrm{GCI}_{1,2}}=1.0046$ and $\frac{\mathrm{GCI}_{3,4}}{\mathscr{R}^{\mathfrak{p}}\cdot \mathrm{GCI}_{2,3}}=0.9979$ for {\bf{SH}} and case {\bf{C}} waves respectively.    
\end{enumerate}
\begin{table}[!ht]
\caption{Average wave height $\overline{H}$ calculated within: $\mathsf{WM}+4\lambda \leq x \leq \mathbb{L}$ for the mesh configurations $\Delta_{1\rightarrow 4}$ alongwith the medium and fine grid GCI's for {\textbf{SH}} and case {\textbf{C}} waves.}
\begin{tabularx}{\textwidth}{X c c c c c c}
\toprule
{} & {$\bm{\overline{H}_{\Delta_4}}$} & {$\bm{\overline{H}_{\Delta_3}}$} & {$\bm{\overline{H}_{\Delta_2}}$} & {$\bm{\overline{H}_{\Delta_1}}$} & {$\mathrm{GCI}_{k+1,k+2}$} & {$\mathrm{GCI}_{k,k+1}$} \\
\midrule
{{\bf{SH}}} & {$0.02579\,m$} & {$0.02442\,m$} & {$0.02504\,m$} & {$0.02516\,m$} & {$0.6997\%$} & {$0.1280\%$}\\
{{\bf{C}}} & {$0.08732\,m$} & {$0.08592\,m$} & {$0.08567\,m$} & {$0.08997\,m$} & {$0.4430\%$} & {$0.0793\%$}\\ 
\bottomrule
\end{tabularx}
\label{tab:tab4}
\end{table}
This concludes the GCI-based verification of the proposed NWT model. It is demonstrated that $\mathfrak{p}=2.444$ ($+0.732\%$ error in $\overline{H}$) and $\mathfrak{p}=2.485$ ($-4.867\%$ error in $\overline{H}$) for regular {\bf{SH}} and case {\bf{C}} waves respectively. Being atleast second-order accurate, the generation of regular waves using the modified-inflow technique is indeed of high fidelity. The NWT is evaluated for polychromatic wave generation in the next section. 

\section{Polychromatic wave generation over flat bottoms} \label{sec:polywaves}
In the present section, the proposed PLIC-VOF NWT is benchmarked against analytical, experimental and numerical studies reported in the literature for polychromatic wave trains propagating over even (flat) bottoms. Two scenarios have been considered: (a) simulation of free-wave generation during piston-type {\textsf{WM}} motion in near-shallow water \cite{madsen71,dong04} and (b) simulation of a short irregular wave train in deep water $(kh>10)$ \cite{peric15}. 
\subsection{Regular wave train superimposed with free harmonics} \label{ssec:freewaves}
The first case under consideration is free-wave generation occurring when a piston-type wave paddle executes sinusoidal motion in near-shallow water $(kh<1)$. This scenario is challenging to accurately simulate due to the added task of capturing free harmonics that are super-imposed on carrier waves. Piston-type motion of a wave paddle in an experimental wave tank can be mathematically prescribed as,
\begin{equation} \label{eq:paddle-trajec}
\xi(t)=-\xi_{o}\cos(\omega t)
\end{equation}
where, $\omega$ is the angular frequency and $\xi_{o}$ is the amplitude. Then, the free-surface profile $\eta(x,t)$ of waves generated in the flume would be a super-position of (a) the fundamental harmonic, (b) a bound {\texttt{Stokes II}} harmonic and (c) a free harmonic. Using wavemaker theory (WMT), Madsen \cite{madsen71} developed an analytic (inviscid) solution to this problem that yielded the three harmonic amplitudes. Following Madsen's solution, $\eta(x,t)$ would be governed by the expression,  
\begin{flalign} \label{eq:madsen-surfel}
\eta(x,t)=-a\sin\left(k_p x-\omega t\right)-a_{2p}\cos2\left(k_p x-\omega t\right)+a_{2f}\cos\left(k_f x-2\omega t\right)
\end{flalign}
where $a$, $a_{2p}$ and $a_{2f}$ are amplitudes of the primary, second bound and second free harmonics respectively. The harmonic amplitudes are obtained using \cite{madsen71},
\begin{flalign} \label{eq:madsen-amps}
&a= \dfrac{\xi_{o} \tanh\left(k_p h\right)}{n_1}\\
&a_{2p}= \dfrac{k_p a^2}{4}\cdot\dfrac{\left(2+\cosh\left(2k_p h\right)\right)\cosh\left(k_p h\right)}{\sinh^3\left(k_p h\right)}\\
&a_{2f}= \dfrac{a^2\coth\left(k_p h\right)}{2h}\cdot \left\{\dfrac{3}{4\sinh^2\left(k_p h\right)}-\dfrac{n_1}{2}\right\}\cdot\dfrac{\tanh\left(k_f h\right)}{n_2}
\end{flalign}
where, $n_1\equiv \frac{1}{2}\left\{1+\frac{2k_p h}{\sinh\left(2k_p h\right)}\right\}$, $n_2\equiv \frac{1}{2}\left\{1+\frac{2k_f h}{\sinh\left(2k_f h\right)}\right\}$ and $k_p$, $k_f$ are wavenumbers of the primary and second free harmonics respectively governed by the dispersion relation: $(n\omega)^2=gk\tanh(kh)$. Solving the dispersion relation for the fundamental yields $k_p h\cong 0.47$. Then, the task at hand is a NSE-based simulation of the piston-type \textsf{WM} motion and comparison of the resultant $\eta(x,t)$ profile against \autoref{eq:madsen-surfel}. The inflow velocity for piston-type {\textsf{WM}} motion is obtained from,
\begin{equation} \label{eq:madsen-WMvel}
U(t)=\dfrac{\mathrm{d}\xi(t)}{\mathrm{d}t}=\omega \xi_{o} \sin(\omega t)
\end{equation}     
It is evident from \autoref{eq:madsen-WMvel} that $\frac{\partial U}{\partial y}=0$ and $\int\limits_{T} U(t)\,\mathrm{d}t=0$ (for piston-type motion) and, as a consequence, $\mathcal{V}_+=0$ from the inflow boundary; baseline inflow is thus retained for wave generation. The following NWT setup (cf. \autoref{fig:dommesh}) has been considered: $\mathbb{L}=40.0\,m$, $\mathbb{H}=0.6\,m$, $\ell_d\sim 5.0\lambda_p$; $nx_{\lambda_p} \approx 180$, $ny_H \approx 10\,(H\equiv 2a)$; $T/\Delta t=5000$, $R=1$; $\wp\equiv \eta(t)$ (baseline inflow) and $\mathscr{S}=1$. 
\begin {figure}[!ht]
\begin {center}
\begin {tabular}{c c}
\multicolumn{2}{c}{
\centering
{\textbf{(a)}} \includegraphics[trim=5mm 35mm 0mm 35mm, clip, width = 14cm]{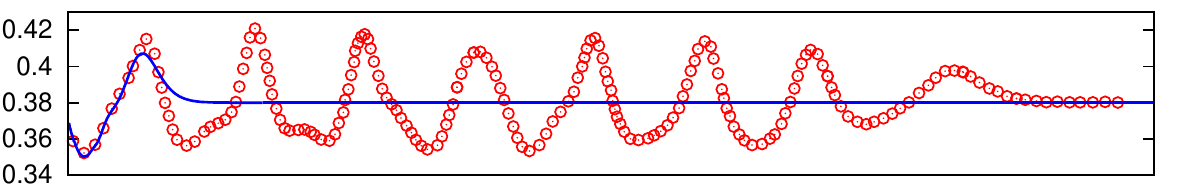}
}\\
\multicolumn{2}{c}{
\centering
{\textbf{(b)}} \includegraphics[trim=5mm 35mm 0mm 36mm, clip, width = 14cm]{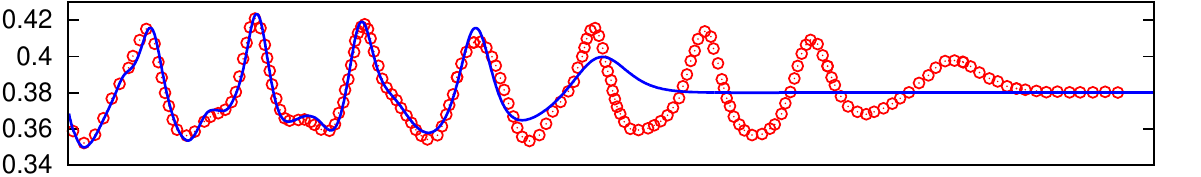}
}\\
\multicolumn{2}{c}{
\centering
{\textbf{(c)}} \includegraphics[trim=5mm 33mm 0mm 35mm, clip, width = 14cm]{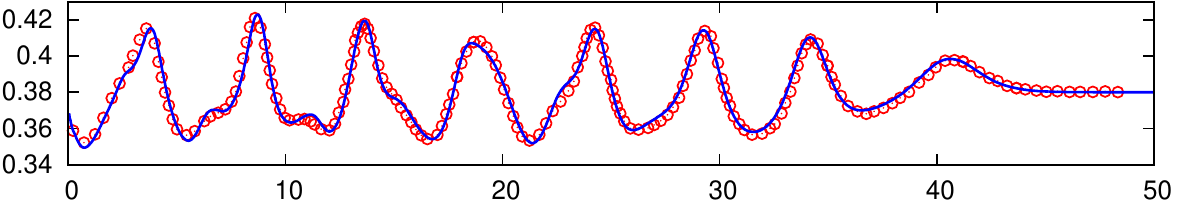}
}\\
\midrule
{\centering
\includegraphics[trim=15mm 0mm 10mm 0mm, clip, height = 6.8cm]{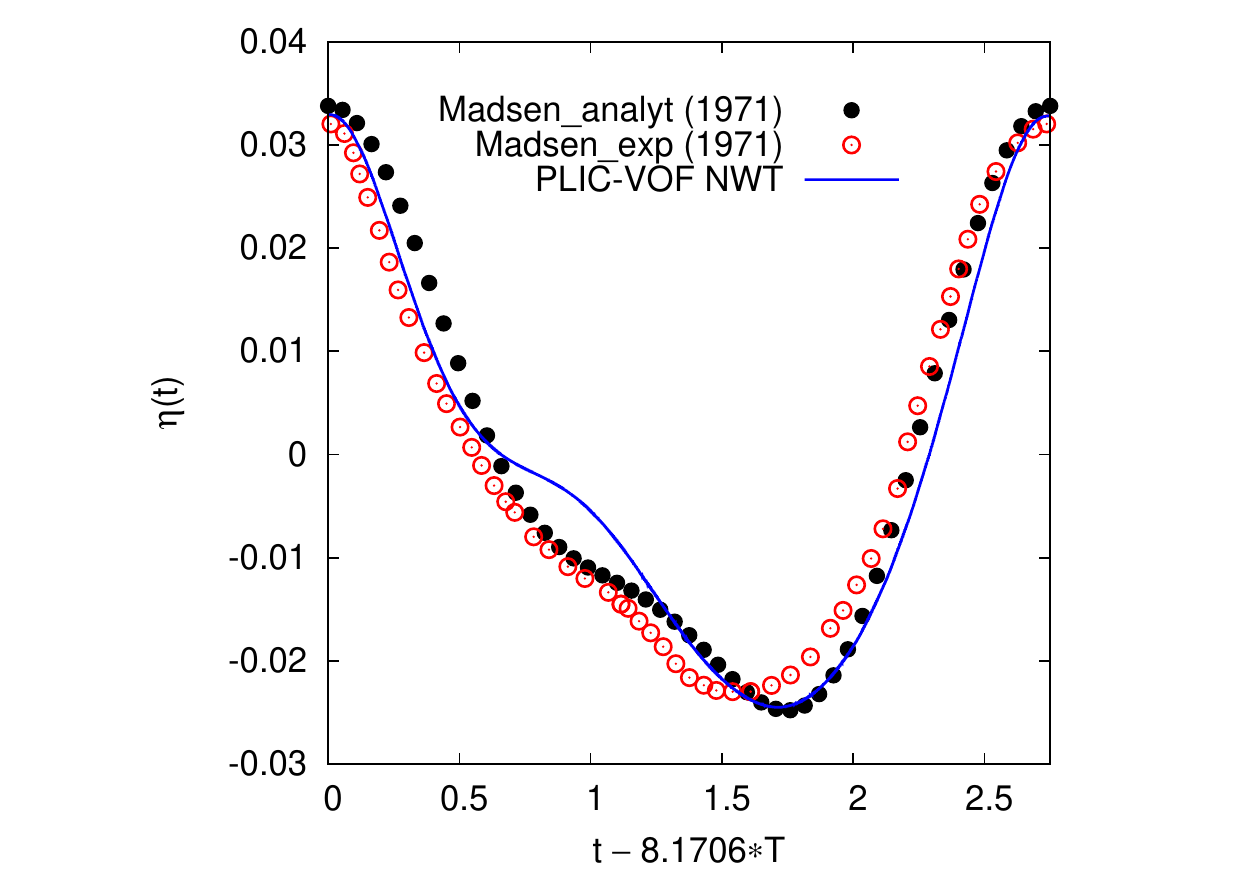}
}&{\centering
\includegraphics[trim=15mm 0mm 10mm 0mm, clip, height = 6.8cm]{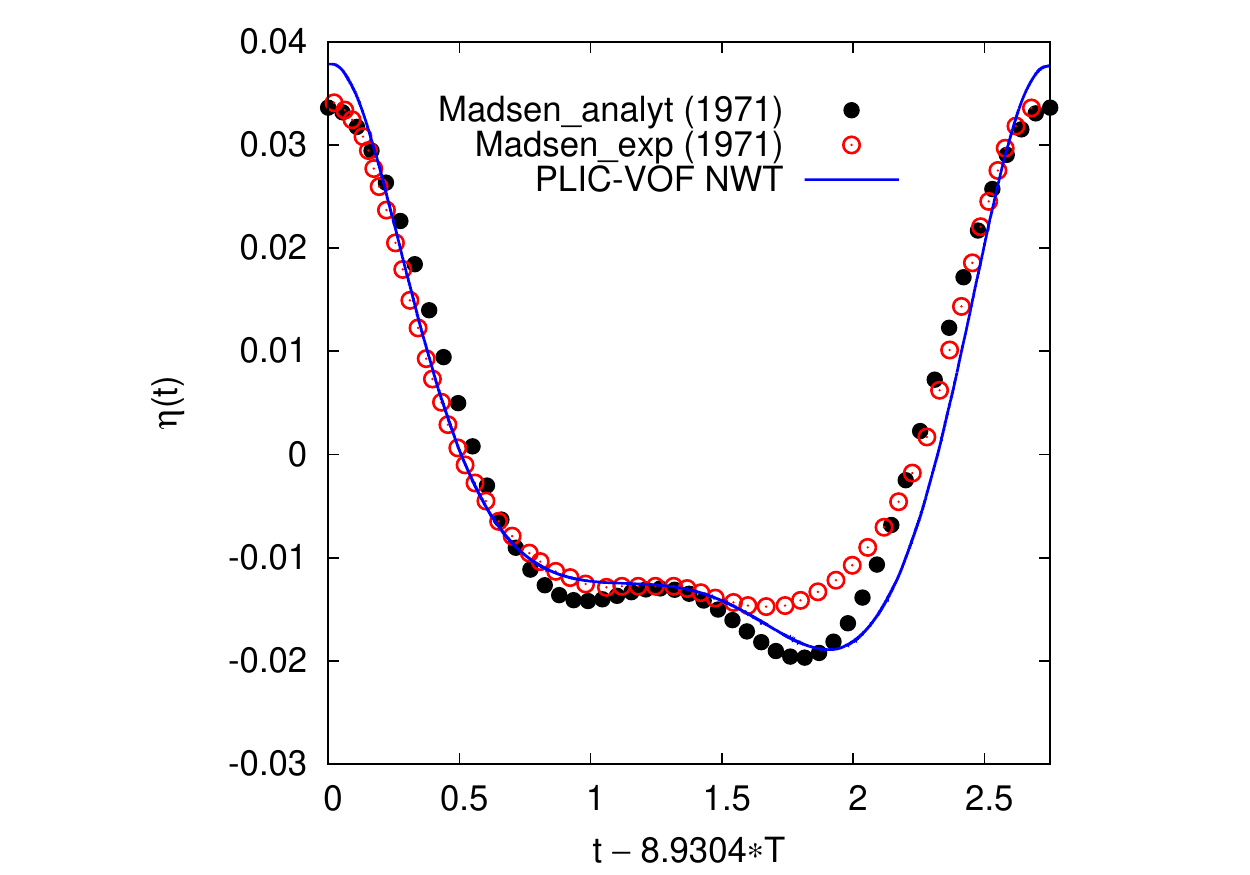}
}\\
{$x=\mathsf{WM}+4.9\,m$}&{$x=\mathsf{WM}+8.7\,m$}
\end{tabular}
\end {center}
\caption{\emph{{\emph{(top)}}Spatial $(\eta(x))$ profiles in the PLIC-VOF NWT (lines) for piston-type $\mathsf{WM}$ motion validated against \cite{dong04} (circles) at $t=2.55,13.55,21.8\,s$; {\emph{(bottom)}}temporal $(\eta(t))$ profiles validated against \cite{madsen71} at two stations, $3.8\,m$ apart.}}
\label{fig:madsen}
\end{figure}\\
The validation study itself is split into two sub-problems: the first part deals with comparison of $\eta(x)$ profiles with the simulations of \cite{dong04} whilst the second part involves validating local $\eta (t)$ signals with those measured experimentally by \cite{madsen71}. It is worth noting at this juncture that waves generated in a experimental flume are slightly smaller than those predicted by wavemaker theory owing to leakage of water past the wave paddle \cite{madsen71}. Thus, one is forced to consider separate validation studies because experimental wave generation entails a ``loss'' in wave-height whilst numerical wave generation is ``lossless'' with regard to {\emph{no}} water leaking past the wave paddle.\\
For the {\textbf{first sub-problem}}, the {\textsf{WM}} parameters are selected as $\xi_{o}=6.1\,cm$ and $T=2.75\,s$ \cite{dong04}; the resulting $U(t)$ (cf. \autoref{eq:madsen-WMvel}) is directly input to the inflow boundary. Dong and Huang \cite{dong04} have reported their simulation results in the form of $\eta(x)$ profiles at $t=21.8\,s$. For the sake of validation, their $\eta(x)$ plot was spatially shifted against the present simulation results such that the first troughs matched in both cases. The resulting (superimposed) $\eta(x)$ profiles are shown in \autoref{fig:madsen}(a-c); an excellent agreement is observed between the present simulations and those reported in \cite{dong04}. However, when the aforementioned simulation framework was directly applied to the {\textbf{second sub-problem}}, it was observed that the numerical waves were $\approx16\%$ larger than experimental measurements reported in \cite{madsen71}. This happens because for a given $\xi_{o}$, wave generation in the NWT would be ``lossless'' whilst that in an actual flume would be ``lossy''. Hence, ``$a$'' appearing in Madsen's theory actually denotes different amplitudes; $a$ in \autoref{eq:madsen-surfel} is the actual amplitude of the fundamental whilst \autoref{eq:madsen-amps} yields a theoretical carrier amplitude $a_{th}$ corresponding to lossless generation. Thus, in order to validate Madsen's experiments, {\emph{slightly smaller}} waves need to be generated in the NWT. Madsen proposed the following modified equation for determining actual amplitude $a$ of the first harmonic \cite{madsen71};
\begin{equation} \label{eq:madsen-losses}
a=a_{th}\cdot \underbrace{\left\{1-C\sqrt{\frac{h}{a_{th}}}\right\}}_{\textsf{leakage}}\cdot \underbrace{\left(1+\epsilon_R \cos\left(2k_o x +\delta\right)+\epsilon_R \cos\delta\right)}_{\textsf{reflections}}
\end{equation}  
where $a_{th}$ is obtained from \autoref{eq:madsen-amps} and $C(=0.04)$, $\epsilon_R(=0.057)$, $\delta(=0.96\,rad)$ are experimentally determined variables \cite{madsen71}. The actual amplitude $a$ is then re-substituted on the LHS of \autoref{eq:madsen-amps} to yield $\xi_{*}$ as a ``diminished amplitude'' of the wave paddle. It is noteworthy that the present calculations yield $\xi_{*}=5.55\,cm$ as opposed to $\xi_{o}=6.1\,cm$ used by Dong and Huang \cite{dong04}. Comparison of presently obtained $\eta(t)$ signals with analytic and experimental profiles of Madsen \cite{madsen71} is shown at the bottom in \autoref{fig:madsen}. A decent match is obtained at both stations with the PLIC-VOF NWT signals following the analytic curve (\autoref{eq:madsen-surfel}) more closely than Madsen's experimental signals. Validation of the NWT against the generation of a deep-water irregular wave-train over a flat bottom is presented in the next section.  
\subsection{Irregular wave train in deep water} \label{ssec:irregwave}
The proposed NWT model is validated against a polychromatic deep-water wave propagation scenario conceived by Peri\'{c} and Abdel-Maksoud \cite{peric15} which was simulated using their novel ``over-designed source'' (OVD) technique. Analytically, the free surface elevation of the irregular wave train at various locations in the NWT is given by,
\begin{equation} \label{eq:irreg-surfel}
\eta(x,t)= \dfrac{1}{2}\sum\limits_{\mathscr{N}=1}^{3} H_\mathscr{N}\cos(\omega_{\mathscr{N}} t-k_{\mathscr{N}} x+\delta_{\mathscr{N}})
\end{equation}
The characteristics of component harmonics (governed by the deep-water dispersion relation: $\omega=\sqrt{gk}$) of the wave train are listed in \autoref{tab:tab5}. The following expressions for velocities from Airy theory \cite{dean91} are used as input to the {\textsf{WM}} $(x=0)$;
\begin{flalign} \label{eq:irreg-WMvel}
U= \dfrac{1}{2}\sum\limits_{\mathscr{N}=1}^{3} H_\mathscr{N} \omega_\mathscr{N}\,e^{k_\mathscr{N}(y-h)} \cos(\omega_{\mathscr{N}} t+\delta_{\mathscr{N}})\,\,;\,\,
V= \dfrac{1}{2}\sum\limits_{\mathscr{N}=1}^{3} H_\mathscr{N} \omega_\mathscr{N}\,e^{k_\mathscr{N}(y-h)} \sin(\omega_{\mathscr{N}} t+\delta_{\mathscr{N}})
\end{flalign}
where a deep-water approximation $(\tanh{k_{\mathscr{N}}h}\rightarrow 1)$ has been invoked. The NWT configuration and a {\emph{local-clustering-based}} vertical meshing strategy adopted for the simulation have been reported in \autoref{fig:dommesh}. The following setup is adopted for simulation: $\mathbb{L}=60.0\,m$, $\mathbb{H}=28.0\,m$, $\ell_d=10.0\lambda_1$; $nx_{\lambda_1}= 88$, $ny_{H_1}=10$; ${T_1}/\Delta t=1000$ (following \cite{peric15}), $R=1$.  
\begin{table}[h]
\caption{\emph{Characteristics of component harmonics of deep-water irregular wave-train \cite{peric15}.}}
\begin{tabularx}{\textwidth}{X c c c c c c c c c}
\toprule
{\textsc{Harmonic}}&{$h\,(m)$}&{$H\,(cm)$}&{$T\,(s)$}&{$\delta\,(rad)$}&{$\lambda\,(m)$}&{$H/\lambda$}&{$\mathsf{Ur}$}\\
\midrule
{$\bm{\mathscr{N}=1}$}&{$26.0$}&{$4.00$}&{$1.6$}&{$0$}&{$4.00$}&{$\color{ForestGreen}\bm{0.0100}$}&{$3.64e-05$}\\
{$\bm{\mathscr{N}=2}$}&{$26.0$}&{$6.00$}&{$2.4$}&{$0$}&{$9.00$}&{$\color{ForestGreen}\bm{0.0067}$}&{$2.77e-04$}\\
{$\bm{\mathscr{N}=3}$}&{$26.0$}&{$8.00$}&{$3.2$}&{$\pi/2$}&{$16.0$}&{$\color{ForestGreen}\bm{0.0050}$}&{$1.17e-03$}\\
\bottomrule
\end{tabularx}
\label{tab:tab5}
\end{table}
Evidently, the length of the domain has been selected based on the second harmonic $(\lambda_2)$ to reduce computation time. Further, considering the fact that the component harmonics are low steepness and (very) low Ursell number waves, $\wp\equiv \eta(t)$ (baseline inflow) and $\mathscr{S}=1$ have been retained. The wave train is simulated upto a physical time of $t=42T_1$. It is also worth mentioning that the {\textsf{WM}} strength is gradually increased over a duration of $t=2T_1$ using a cosine ramping function to maintain numerical stability during the initial stages of the simulation \cite{sas17a}. 
\begin {figure}[!ht]
\begin {center}
\begin {tabular}{c}
{
\centering
\includegraphics[trim=0mm 36mm 0mm 36mm, clip, width = 13cm]{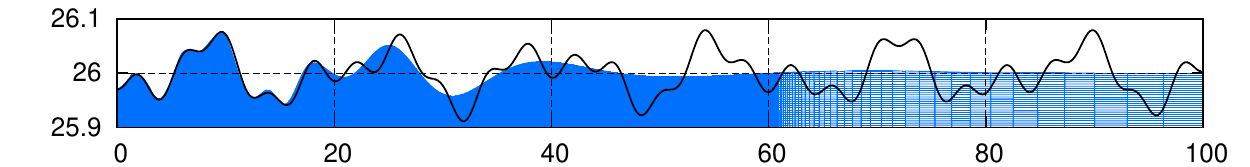}
}\\
{
\centering
\includegraphics[trim=0mm 36mm 0mm 36mm, clip, width = 13cm]{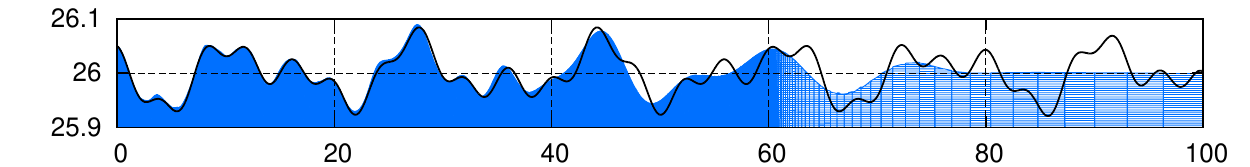}
}\\
{
\centering
\includegraphics[trim=0mm 36mm 0mm 36mm, clip, width = 13cm]{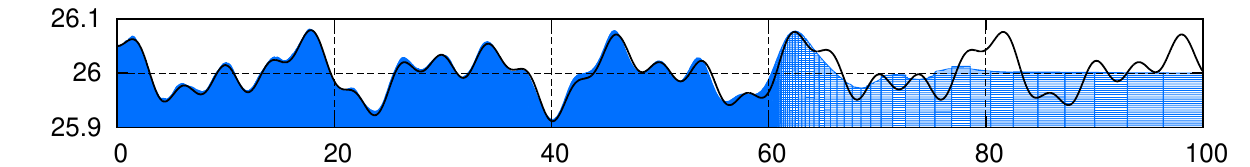}
}\\
{
\centering
\includegraphics[trim=0mm 36mm 0mm 36mm, clip, width = 13cm]{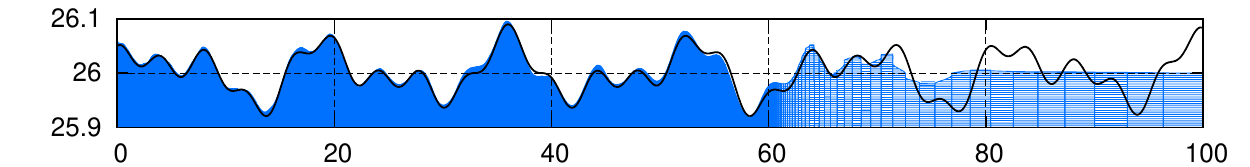}
}\\
\midrule
{
\centering
\includegraphics[trim=0mm 4mm 5mm 5mm, clip, width = 14cm]{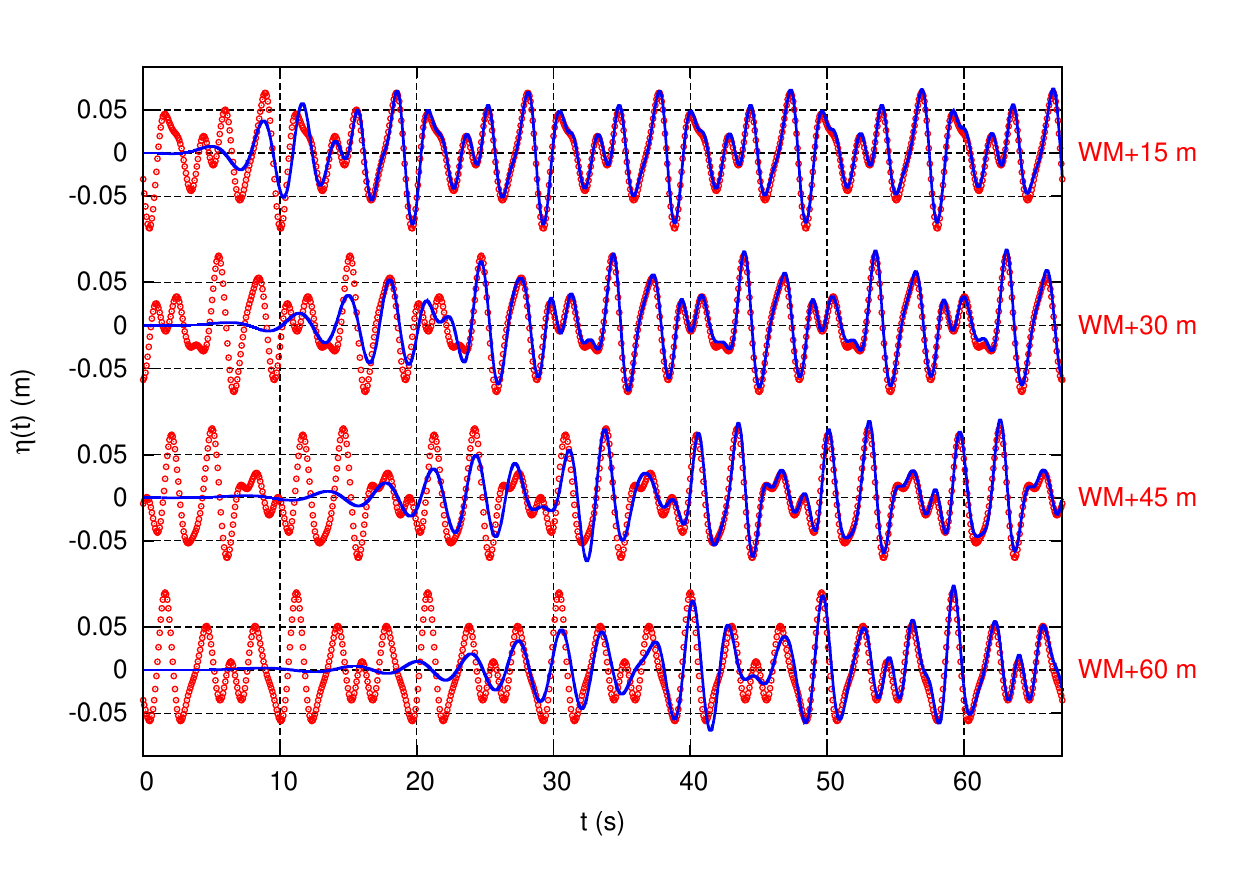}
}
\end{tabular}
\end {center}
\caption{\emph{Irregular wave propagation problem of \cite{peric15} simulated using proposed NSE-based NWT model and validated against Airy theory; {\emph{(top)}} spatial $(\eta(x))$ profiles plotted every $16.8\,s$ and {\emph{(bottom)}} temporal $(\eta(t))$ profiles measured every $15\,m$ from the {\textsf{WM}}.}}
\label{fig:peric}
\end{figure}
Results of irregular wave generation have been validated against Airy theory in terms of both spatial $(\eta(x))$ and temporal $(\eta(t))$ wave profiles. In \autoref{fig:peric}, $\eta(x)$ profiles are depicted at intervals of $t=16.8\,s$ whilst $\eta(t)$ profiles have been recorded at four locations uniformly-spaced every $15.0\,m$ from the {\textsf{WM}}. The spatio-temporal intervals are selected such that neither of the three component periods $(T_1,T_2,T_3)$ nor component wavelengths $(\lambda_1,\lambda_2,\lambda_3)$ be multiples of $16.8$ or $15.0$ respectively. Then, owing to frequency dispersion, the $\eta(x)$ and $\eta(t)$ profiles of interest would themselves vary amongst the chosen space-time intervals; this makes validation exceptionally challenging. Nonetheless, the simulations show excellent agreement with Airy theory in all cases (cf. \autoref{fig:peric}). It is also worth mentioning that, even with baseline inflow, the net percentage change in water volume at $t=67.2\,s$ was minimal: $\mathcal{VE}\cong -1.38e-02\%$ which is jointly attributable to $H/\lambda\leq 0.01$ and $\mathsf{Ur} \lll 1$ for individual harmonics (cf. \autoref{tab:tab5}). \\ 
Thus, the simulations reported in \autoref{sec:polywaves} demonstrate that the proposed NWT algorithm is capable of generating high-fidelity polychromatic waves over flat bottoms. 

\section{Wave transformation over an inclined bottom} \label{sec:wavetrans}
The proposed NWT algorithm is finally benchmarked against scenarios that involve a natural transition of monochromatic waves into a polychromatic wave train (comprised of both free as well as bound harmonics) owing to bathymetric variations. To this effect, the problem of wave transformation over a submerged trapezoidal bar has been considered \cite{beji94}. The problem represents an effective coastal protection strategy in which the structural design is aimed towards transforming incident long waves to transmitted short waves thereby arresting beach erosion. In simulating such scenarios, the challenge lies in that the structure (which generally has a non-Cartesian geometry) needs to be modeled in the NWT and appropriate momentum and pressure boundary conditions need to be prescribed at the structure. Further, the wave topology itself undergoes dramatic changes over the structure. Achieving a simultaneous phase-agreement amongst vastly different wave topologies coexisting in the NWT during wave transformation is challenging. \\
Description of the problem, corresponding NWT setup and strategy adopted for treating non-Cartesian submerged/emergent boundaries are presented in \autoref{ssec:wavetrans_setup} and \autoref{ssec:imm_bound_treat}. This is followed by systematic validation of the NWT simulations against weak as well as strong wave transformation in \autoref{ssec:SHwaves} through \autoref{ssec:vort_dyn}.  
\subsection{Problem description and NWT setup} \label{ssec:wavetrans_setup}
\begin{figure}[!ht]
\begin {center}
\includegraphics[trim=0mm 0mm 0mm 0mm, clip, width = 15cm]{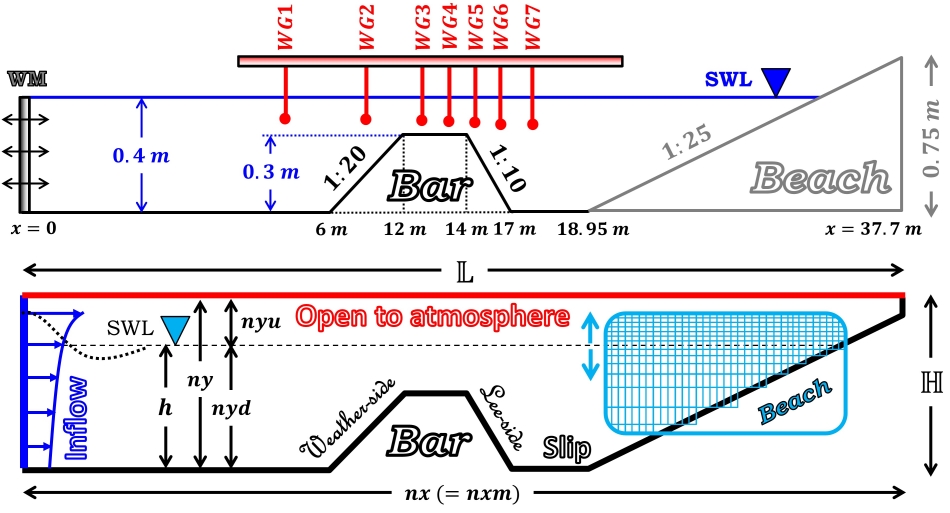}
\end {center}
\caption{\emph{{\emph{(top)}} Original experimental setup \cite{beji94} and {\emph{(bottom)}} resultant NWT model considered for simulation of weak and strong wave transformation.}}
\label{fig:wavetrans_setup}
\end{figure}

A representative sketch of Beji and Battjes' \cite{beji94} experimental setup is shown in \autoref{fig:wavetrans_setup}. A servo-controlled piston-type wave paddle was employed for generating both regular as well as JONSWAP-spectrum based wave trains. The waves pass over a submerged trapezoidal bar with a $1:20$ upslope and a $1:10$ downslope. A $2\,m$ long stretch of water over the bar crest acts as a non-dispersive medium for the waves which eventually get dissipated over a beach with a gentle $1:25$ slope. In the experiments, topological changes occurring in the wave train were recorded by an array of seven wave gauges (see \autoref{fig:wavetrans_setup}): WG1 $(x=5.7\,m)$, WG2 $(x=10.5\,m)$, WG3 $(x=12.5\,m)$, WG4 $(x=13.5\,m)$, WG5 $(x=14.5\,m)$, WG6 $(x=15.7\,m)$ and WG7 $(x=17.3\,m)$. A constant water depth of $h=0.4\,m$ was maintained during the experiments \cite{beji94}. For a rigorous validation of the NWT, three different incident wave designs have been considered which are reported in \autoref{tab:tab6}. These include: (a) short, sinusoidal high frequency {\bf{(SH)}} waves \cite{bejid94} and (b) comparatively longer, sinusoidal low frequency {\bf{(SL)}} waves \cite{bejid94}. We further classify the latter category as: (a) {\bf{SL}} ``low'' $\equiv$ {\bf{SLL}} waves and (b) {\bf{SL}} ``high'' $\equiv$ {\bf{SLH}} waves. The {\bf{SLH}} waves were introduced by Huang and Dong \cite{huang99} as a ``steeper version'' of {\bf{SLL}} waves and were thus not a part of the original experimental paradigm of Beji and Battjes \cite{beji94}. Further, it is clear that $\mathsf{Ur}_{SH}<\mathsf{Ur}_{SLL}<\mathsf{Ur}_{SLH}$; the relative amplitudes $\left(a^{(n)}/H\right)$ of higher order bound harmonics also increase in that order \cite{fenton85}. 
\begin{table}[h]
\caption{\emph{Wave characteristics selected for wave transformation simulations.}}
\begin{tabularx}{\textwidth}{X c c c c c c c}
\toprule
{\textsc{Case}}&{$h\,(m)$}&{$T\,(s)$}&{$H\,(cm)$}&{$\lambda\,(m)$}&{$kh$}&{$H/\lambda$}&{$\mathsf{Ur}$}\\
\midrule
{\textbf{SH} \cite{beji94}}&{$0.40$}&{$1.25$}&{$2.50$}&{$2.055$}&{$1.22$}&{$\color{Cerulean}\bm{0.012}$}&{$\mathsf{1.65}$}\\
{\textbf{SLL} \cite{beji94}}&{$0.40$}&{$2.00$}&{$2.00$}&{$3.699$}&{$0.68$}&{$\color{ForestGreen}\bm{0.005}$}&{$\mathsf{4.28}$}\\
{\textbf{SLH} \cite{huang99}}&{$0.40$}&{$2.00$}&{$4.00$}&{$3.711$}&{$0.68$}&{$\color{Cerulean}\bm{0.011}$}&{$\mathsf{8.61}$}\\
\bottomrule
\end{tabularx}
\label{tab:tab6}
\end{table}
Thus, $\mathsf{Ur}$ of the upstream wave train governs the ``intensity'' of wave transformation over the submerged obstacle \cite{beji94,huang99}. The upslope/downslope of the bar and upslope of the beach represent non-Cartesian immersed boundaries in the NWT. A mesh stair-stepping strategy has been adopted for immersed boundary treatment \cite{sas17b} which is presented in the next subsection. 
\subsection{Numerical treatment of immersed boundaries} \label{ssec:imm_bound_treat}
\begin {figure}[!ht]
\begin {center}
\begin {tabular}{c c}
\multicolumn{2}{c}{
\centering
\includegraphics[trim=0mm 0mm 0mm 0mm, clip, width = 10cm]{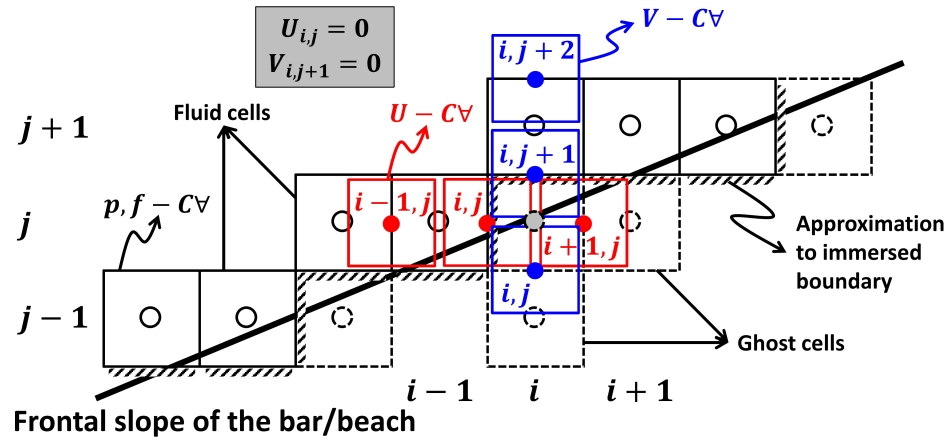}
}\\
{
\centering
\includegraphics[trim=0mm 0mm 0mm 0mm, clip, width = 8cm]{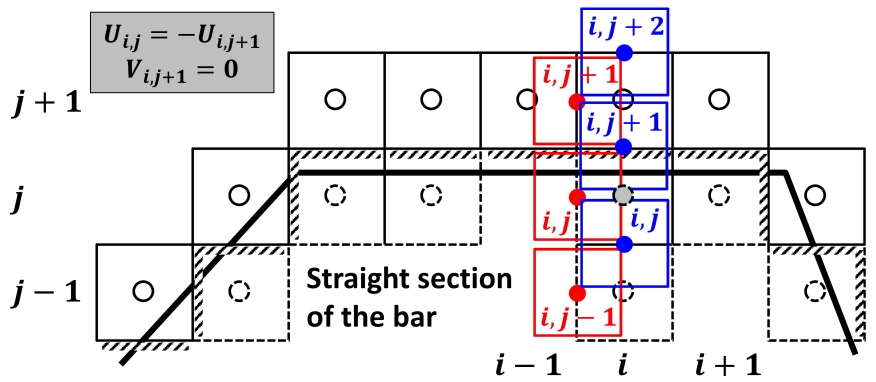}
}&
{
\centering
\includegraphics[trim=0mm 0mm 0mm 0mm, clip, width = 7cm]{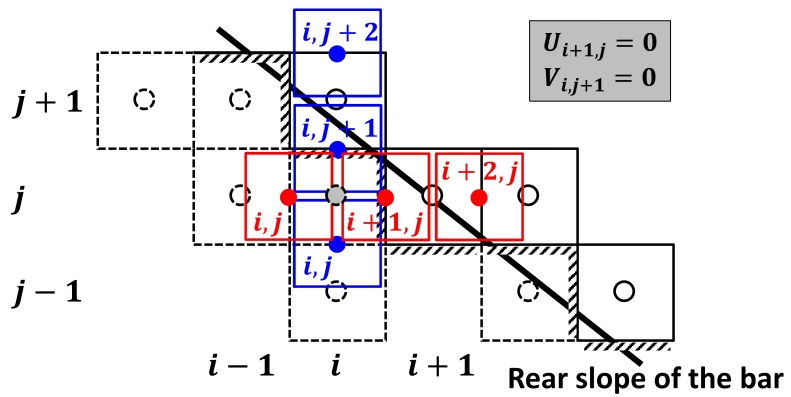}
}
\end {tabular}
\end {center}
\caption{\emph{Mesh stair-stepping approach adopted for treating non-Cartesian immersed boundaries on a staggered $U,V,p$ arrangement for wave transformation simulations.}}
\label{fig:imm_bound_treat}
\end{figure}

Immersed boundaries have been modeled following an ``obstacle approach'' which is illustrated in \autoref{fig:imm_bound_treat}. Immersed boundary treatment involves the following steps:
\begin{itemize}[noitemsep]
\item pressure/volume-fraction $(p,f)$ $C\forall$s falling ``inside'' the bar/beach boundary $(\Omega)$ are flagged; the flagged cells are skipped during $U,V,p,f$ calculations;
\item elimination of flagged cells from the calculation leads to a characteristic ``stair-stepped'' approximation \cite{huang99} $\Omega^{\sqcap}$ to $\Omega$ (cf. \autoref{fig:imm_bound_treat});
\item backward staggering of $U,V$ $C\forall$s with respect to $p,f$ $C\forall$s ensures an exact placement of momentum cell-centers along $\Omega^{\sqcap}$ (cf. \autoref{fig:imm_bound_treat});
\item no-slip boundary conditions are locally imposed at $U,V$ cell-centers lying on $\Omega^{\sqcap}$;
\item the flagged $p,f$ $C\forall$s adjacent to $\Omega^{\sqcap}$ are employed as ``ghost cells'' (see cell $i,j$ in \autoref{fig:imm_bound_treat}) for imposing zero gradient conditions in $p',f$ over $\Omega^{\sqcap}$ ($\mathcal{G}^{\perp}_{p'_I,f_I}=0$).
\end{itemize} 
It should be noted that no local modifications in cell sizes were attempted for improving the approximation $\Omega^{\sqcap}$ and hence $\Omega^{\sqcap}\rightarrow \Omega$ would result only upon mesh refinement. The proposed methodology facilitates a simplified treatment of non-Cartesian geometries in the NWT even when the placement of solution variables is staggered. Results of wave transformation simulations are discussed in the sections that follow. 
\subsection{Weak transformation: SH waves $(T=1.25\,s;H=2.5\,cm)$} \label{ssec:SHwaves}
\begin {figure}[!ht]
\begin {center}
\begin {tabular}{c c c}
{\centering
\includegraphics[trim=20mm 0mm 26mm 0mm, clip, height = 5.25cm]{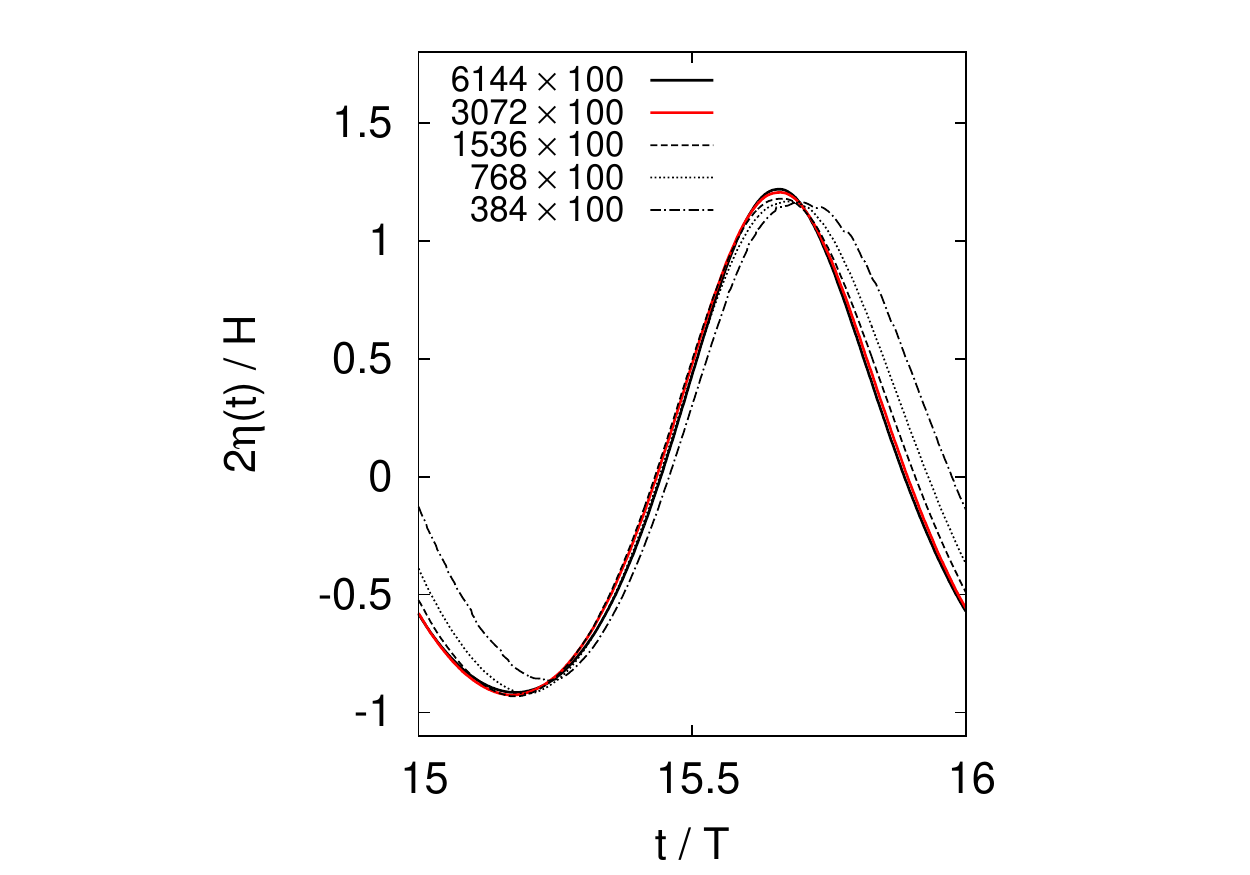}
}&{
\centering
\includegraphics[trim=20mm 0mm 26mm 0mm, clip, height = 5.25cm]{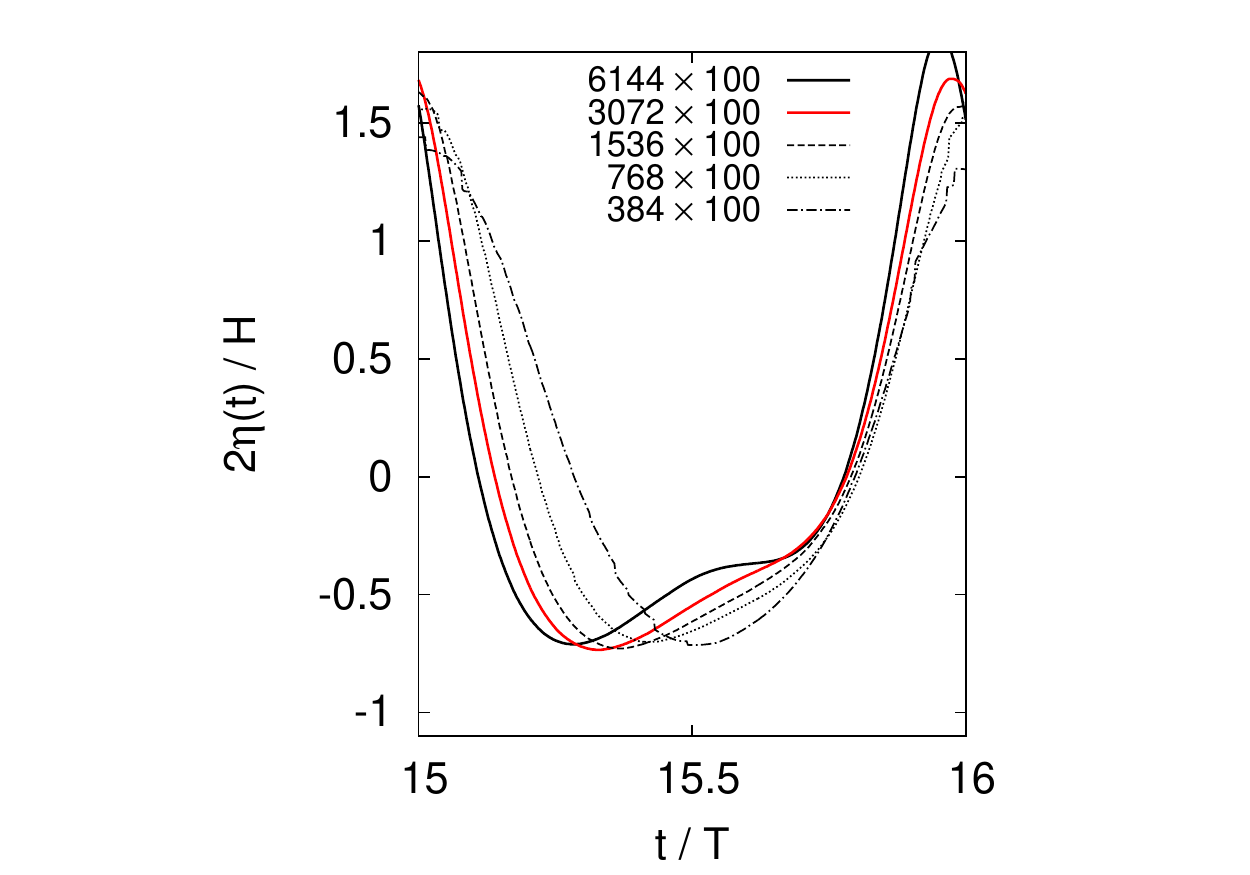}
}&{
\centering
\includegraphics[trim=20mm 0mm 26mm 0mm, clip, height = 5.25cm]{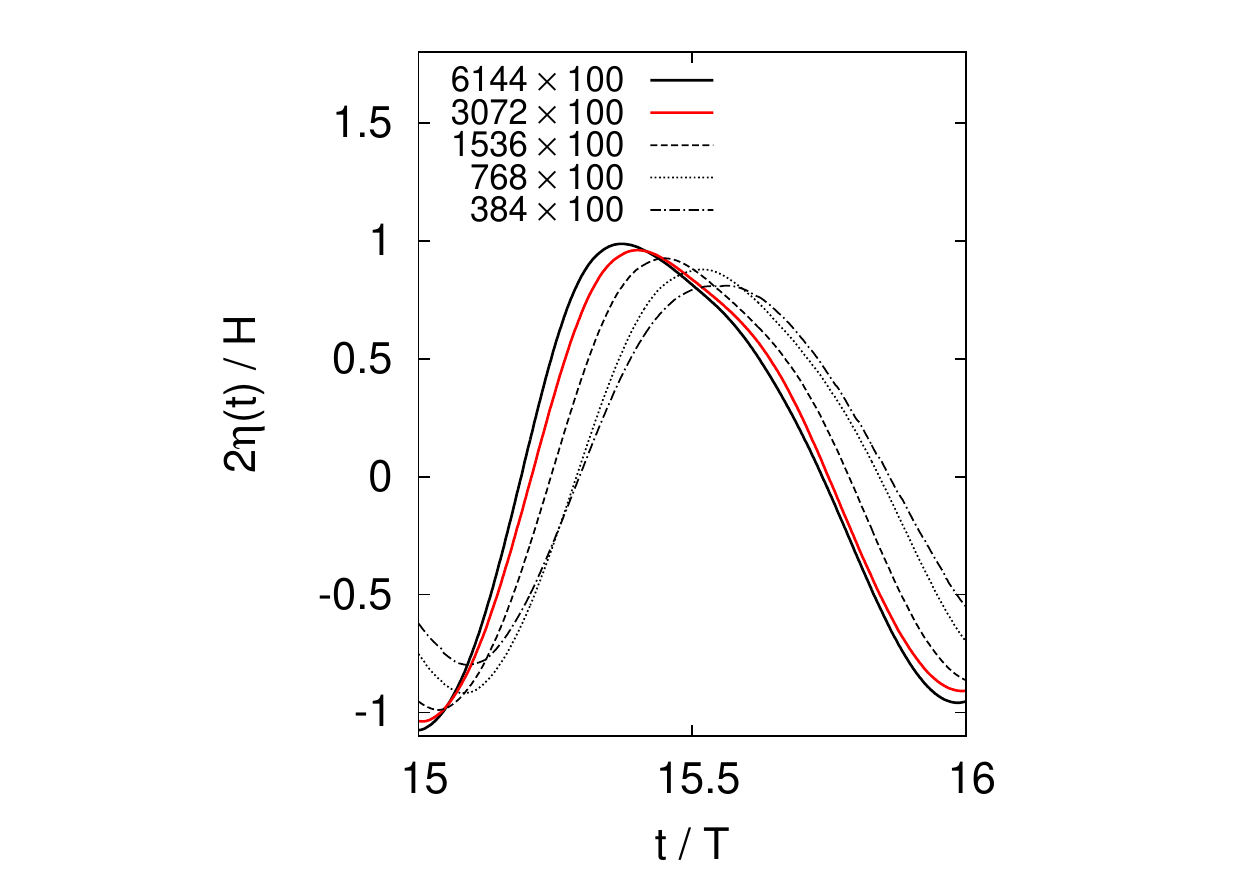}
}\\
{\bf{WG2} $(x=10.5\,m)$} & {\bf{WG4} $(x=13.5\,m)$} & {\bf{WG6} $(x=15.7\,m)$}\\
\end {tabular}
\end {center}
\caption{\emph{Mesh dependence for {\bf{SH}} waves; $\color{red}\bm{3072\times 100}$ is selected for validation.}}
\label{fig:SHgind}
\end{figure}

\begin {figure}[!ht]
\begin {center}
\begin {tabular}{c c}
\multicolumn{2}{c}{
\centering
\includegraphics[trim=0mm 39mm 0mm 37mm, clip, width = 14cm]{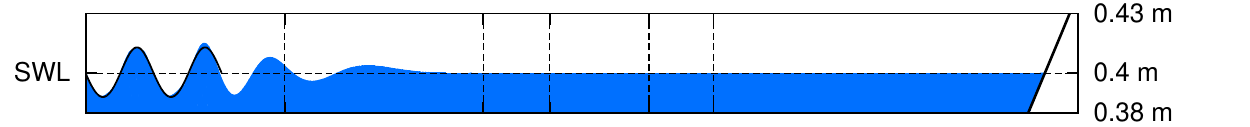}
}\\
\multicolumn{2}{c}{
\centering
\includegraphics[trim=0mm 39mm 0mm 37mm, clip, width = 14cm]{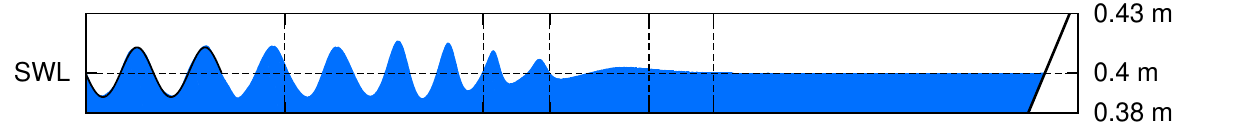}
}\\
\multicolumn{2}{c}{
\centering
\includegraphics[trim=0mm 39mm 0mm 37mm, clip, width = 14cm]{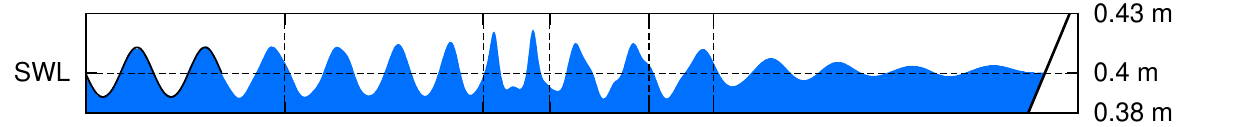}
}\\
\multicolumn{2}{c}{
\centering
\includegraphics[trim=0mm 39mm 0mm 37mm, clip, width = 14cm]{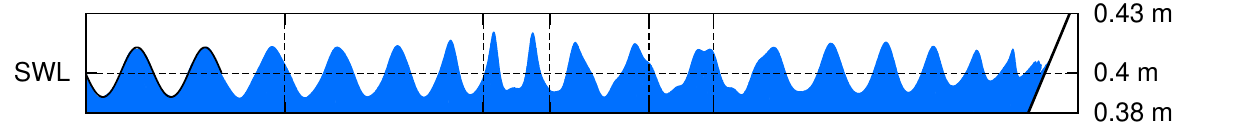}
}\\
\multicolumn{2}{c}{
\centering
\includegraphics[trim=1mm 33mm 0mm 34mm, clip, width = 13.75cm]{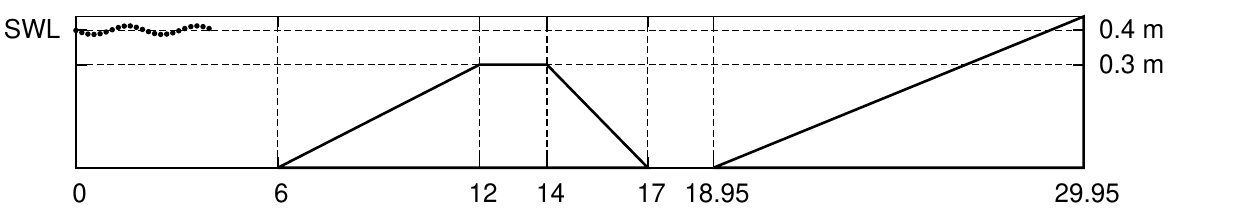}
}\\
\midrule
\multicolumn{2}{c}{
\centering
\includegraphics[trim=10mm 0mm 10mm 0mm, clip, width = 15cm]{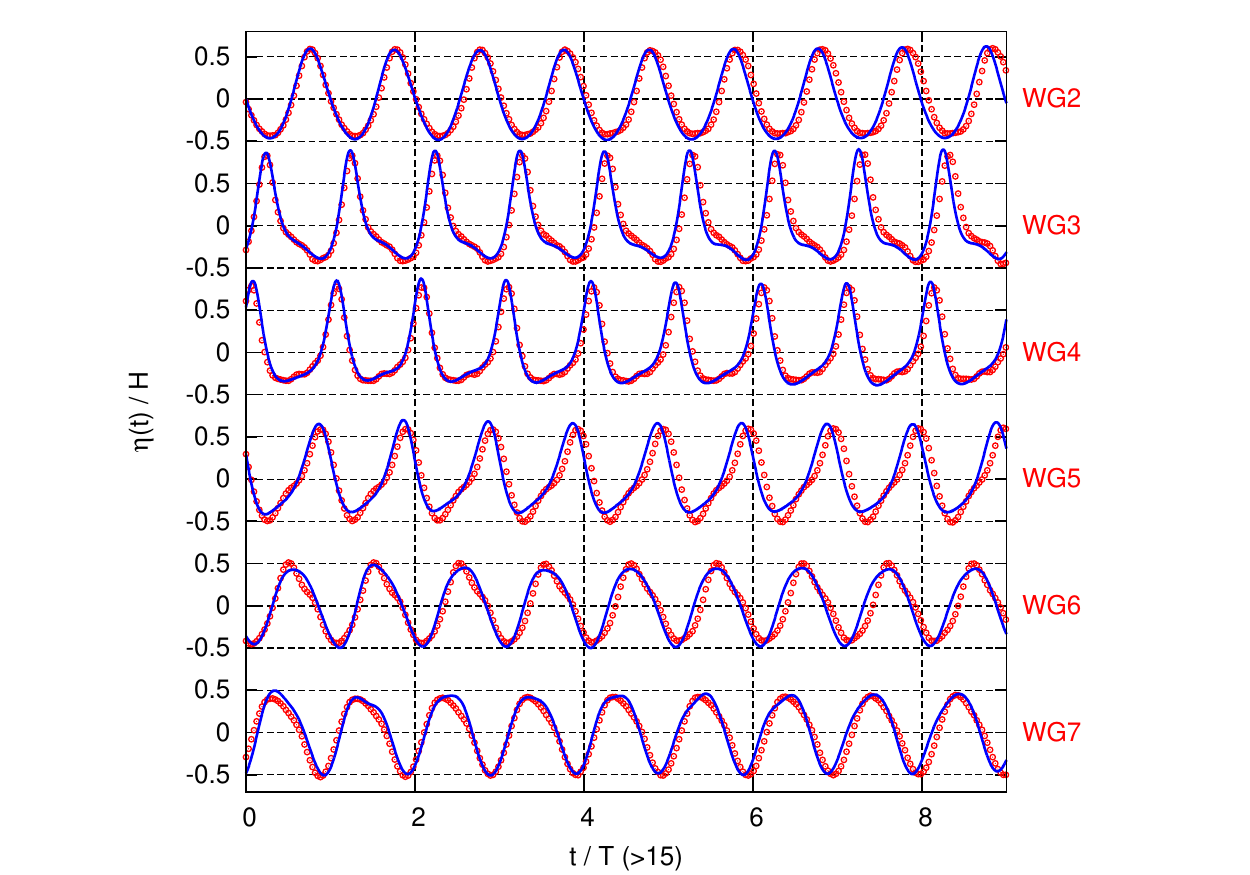}
}
\end {tabular}
\end {center}
\caption{\emph{{\emph{(top)}}Time series of $\eta(x)$ profiles (at $t=4T,9T,15T,20T$) showing ``weak'' transformation of {\bf{SH}} waves; the coordinates are in $\bm{m}$. {\emph{(bottom)}} Validation of normalized $\eta(t)$ signals against the experiments reported in \cite{beji94} for nine wave periods.}}
\label{fig:SHval}
\end{figure}

The first scenario considered is that of {\bf{SH}} wave transformation over the trapezoidal bar. As waves transform over the submerged obstacle, $H/\lambda \propto x$ owing to generation of free and bound harmonics. Small, moderate and large $\mathsf{Ur}$ waves would thus co-exist along the NWT. In such a situation, an optimal spatial resolution $(nx_\lambda)$ cannot be decided solely based on the regular wave generation guidelines provided in \cite{sas17a}; a mesh dependence analysis becomes necessary. In order to assess the mesh dependence of the solution, normalized free surface elevation $(\eta(t)/(0.5H))$ profiles of {\bf{SH}} waves, recorded at WG2, WG4 and WG6, have been compared amongst five spatial resolutions: $nx_\lambda \cong 26,53,105,211,422$ during the interval $t\in [15T:16T]$. The following NWT framework has been considered (cf. \autoref{fig:wavetrans_setup}): $\mathbb{L}=29.95\,m$, $\mathbb{H}=0.44\,m$; $ny=100$, $ny_H \approx 12$; $T/\Delta t=5000$, $\wp=0.5H$; $\mathscr{S}=0.8$ is selected to prevent height damping of high $\mathsf{Ur}$ waves propagating over the bar crest. Results of the parametric analysis are shown in \autoref{fig:SHgind}. As anticipated, $H/\lambda \propto x$ means that ``early'' mesh independence is achieved at WG2 but not at WG4 where the waves are steeper. Interestingly, the $\eta(t)$ profiles in \autoref{fig:SHgind} indicate that $\left.\frac{H}{\lambda}\right|_{\mathrm{WG2}} \cong \left.\frac{H}{\lambda}\right|_{\mathrm{WG6}}$ yet solution dependence on grid size is stronger at WG6. This finding establishes that mesh selection for wave transformation cannot be solely based upon guidelines extrapolated from regular wave generation simulations. From \autoref{fig:SHgind}, the $3072\times 100$ mesh is selected for further simulations and validation. 
\\
Spatial development of the {\bf{SH}} wave train is presented at the top in \autoref{fig:SHval}. It is apparent from the $\eta(x)$ profiles that the spatial evolution of {\bf{SH}} waves is slow. This is because the ``envelope'' of {\bf{SH}} waves evolves with a group velocity $(C_G)$ which is less than the phase velocity $(C)$ of the individual wave packets; $C/C_G=1.4$ (cf. \autoref{tab:tab3}). In the initial stages of development, the train is essentially sinusoidal. For $t \geq 9T$, the {\bf{SH}} wave train begins to shoal over the bar upslope which is characterized by $\lambda \downarrow$ and $H\uparrow$ and there is a visible onset of non-linearity. The wave topology resolved at $t=15T$ over the bar crest reaffirms the observation of \cite{beji94} that {\bf{SH}} waves shoal to more closely resemble higher-order {\texttt{Stokes}} waves. Subsequently, the waves ``de-shoal'' \cite{beji94} over the lee-side of the bar with negligible super-harmonic generation. Nonetheless, the de-shoaling is accompanied by topological distortion $(t=20T)$ which is inturn attributed to a superposition of the transmitted carrier wave with free waves (that have gained amplitude owing to energy transfer from bound harmonics \cite{grue92}). However, in the absence of harmonic generation, the lee-side wave train is essentially in intermediate water $(kh>1)$ and the free waves gradually dissipate. The {\bf{SH}} wave train {\textbf{tends to regain it's incident character}} after the bar downslope; wave transformation is hence termed ``weak'' in this case.
\\
For the sake of validation, normalized $\eta(t)$ profiles recorded in the NWT are compared against experimental measurements of \cite{beji94} during $t\in[15T:25T]$; this is reported at the bottom in \autoref{fig:SHval}. The simulations demonstrate good agreement with experiments. The proposed NWT model is benchmarked against strong wave transformation in the following subsections.
\subsection{Strong transformation: SLL waves $(T=2.0\,s;H=2.0\,cm)$} \label{ssec:SLLwaves}
\begin {figure}[!ht]
\begin {center}
\begin {tabular}{c c c}
{\centering
\includegraphics[trim=20mm 0mm 26mm 0mm, clip, height = 5.25cm]{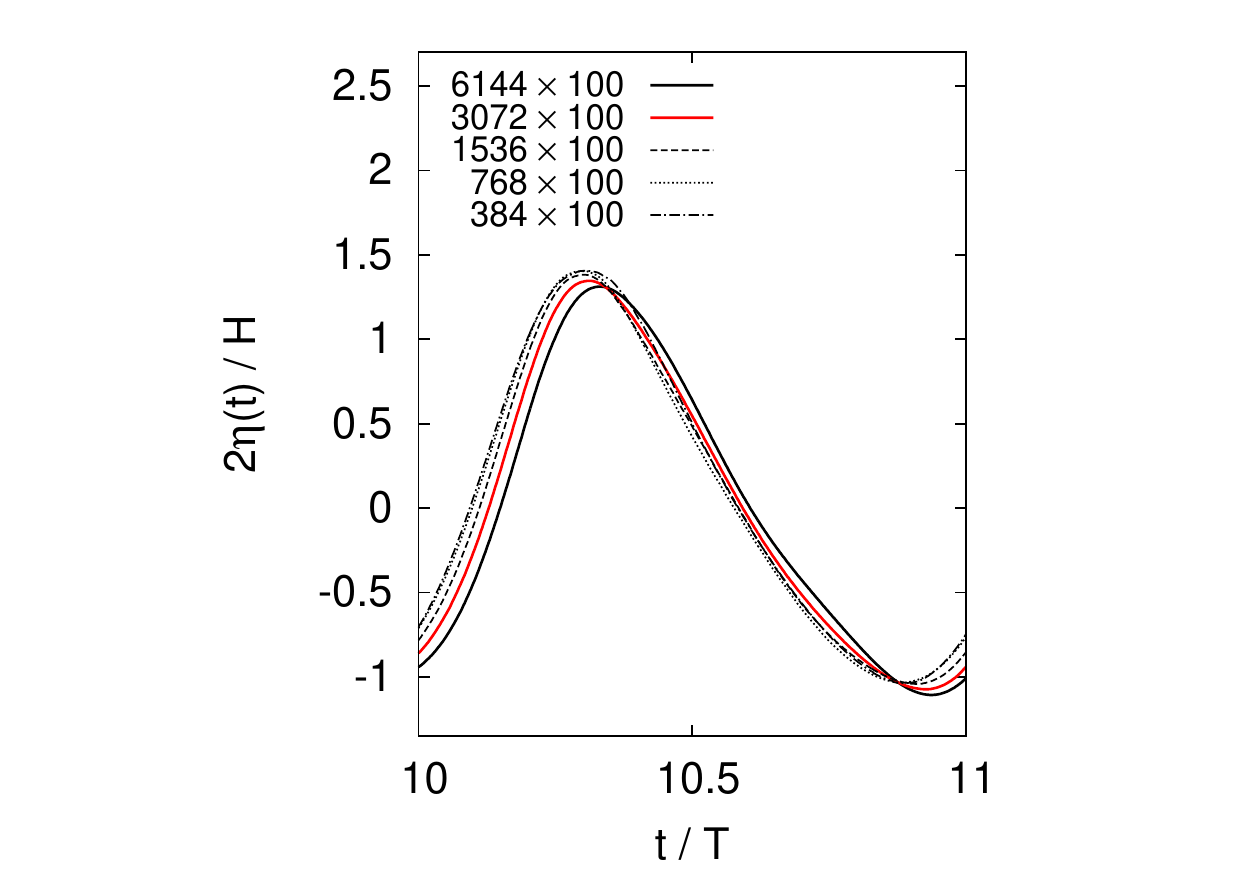}
}&{
\centering
\includegraphics[trim=20mm 0mm 26mm 0mm, clip, height = 5.25cm]{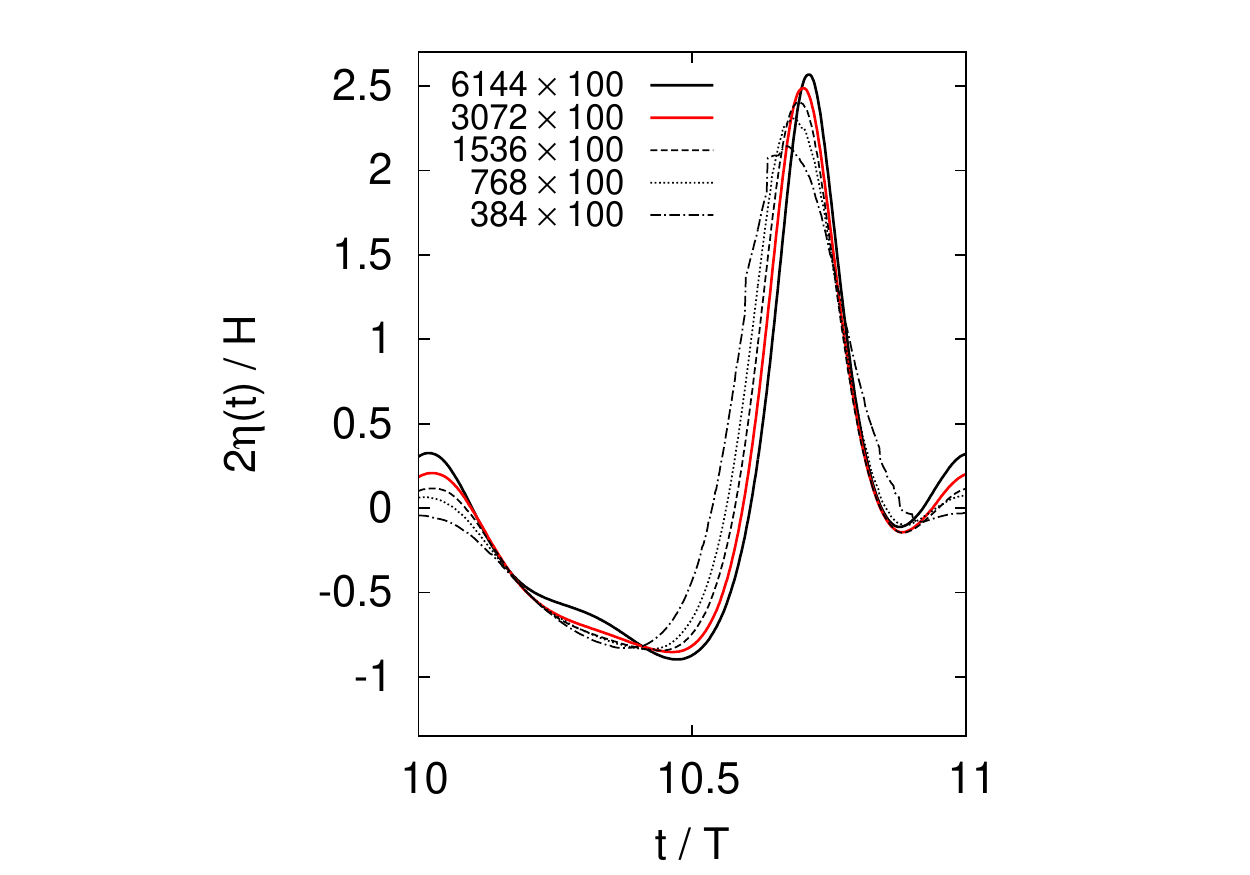}
}&{
\centering
\includegraphics[trim=20mm 0mm 26mm 0mm, clip, height = 5.25cm]{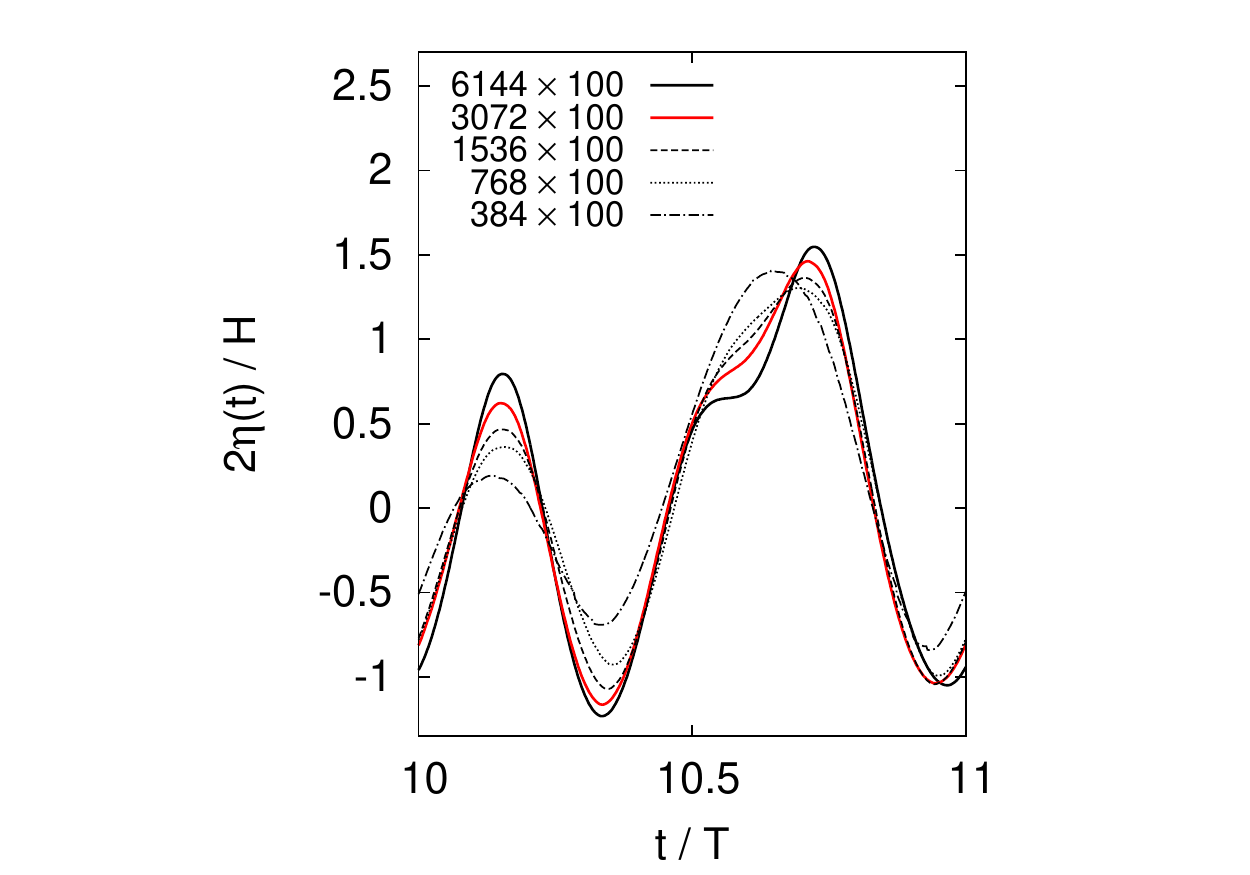}
}\\
{\bf{WG2} $(x=10.5\,m)$} & {\bf{WG4} $(x=13.5\,m)$} & {\bf{WG6} $(x=15.7\,m)$}\\
\end {tabular}
\end {center}
\caption{\emph{Mesh dependence for {\bf{SLL}} waves; $\color{red}\bm{3072\times 100}$ is selected for validation.}}
\label{fig:SLLgind}
\end{figure}

\begin {figure}[!ht]
\begin {center}
\begin {tabular}{c c}
\multicolumn{2}{c}{
\centering
\includegraphics[trim=0mm 39mm 0mm 37mm, clip, width = 14cm]{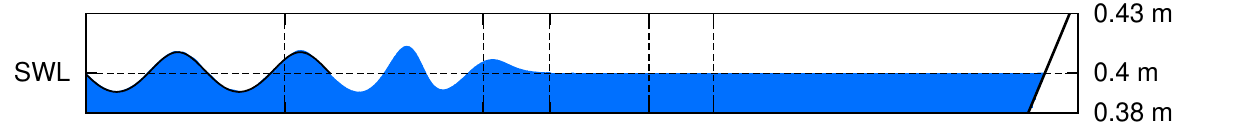}
}\\
\multicolumn{2}{c}{
\centering
\includegraphics[trim=0mm 39mm 0mm 37mm, clip, width = 14cm]{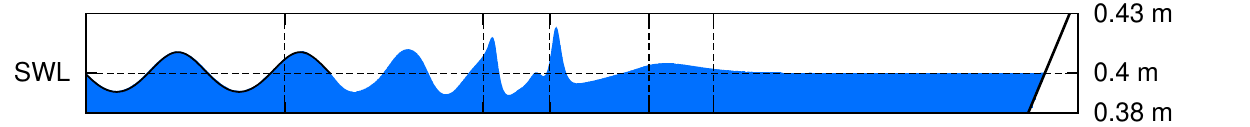}
}\\
\multicolumn{2}{c}{
\centering
\includegraphics[trim=0mm 39mm 0mm 37mm, clip, width = 14cm]{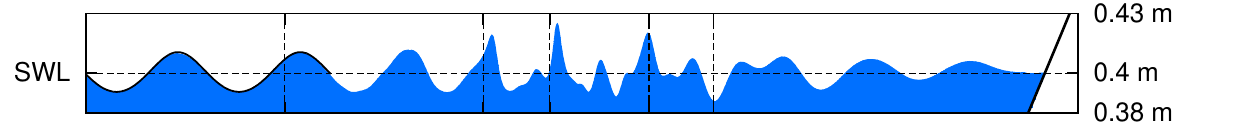}
}\\
\multicolumn{2}{c}{
\centering
\includegraphics[trim=0mm 39mm 0mm 37mm, clip, width = 14cm]{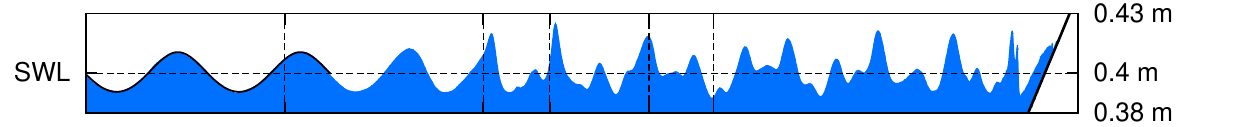}
}\\
\multicolumn{2}{c}{
\centering
\includegraphics[trim=1mm 33mm 0mm 34mm, clip, width = 13.75cm]{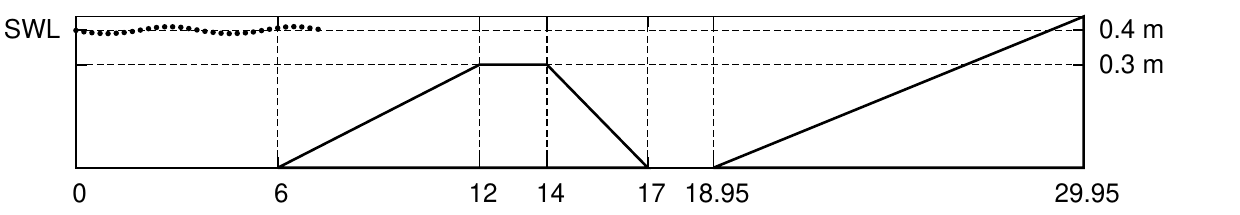}
}\\
\midrule
\multicolumn{2}{c}{
\centering
\includegraphics[trim=10mm 0mm 10mm 0mm, clip, width = 15cm]{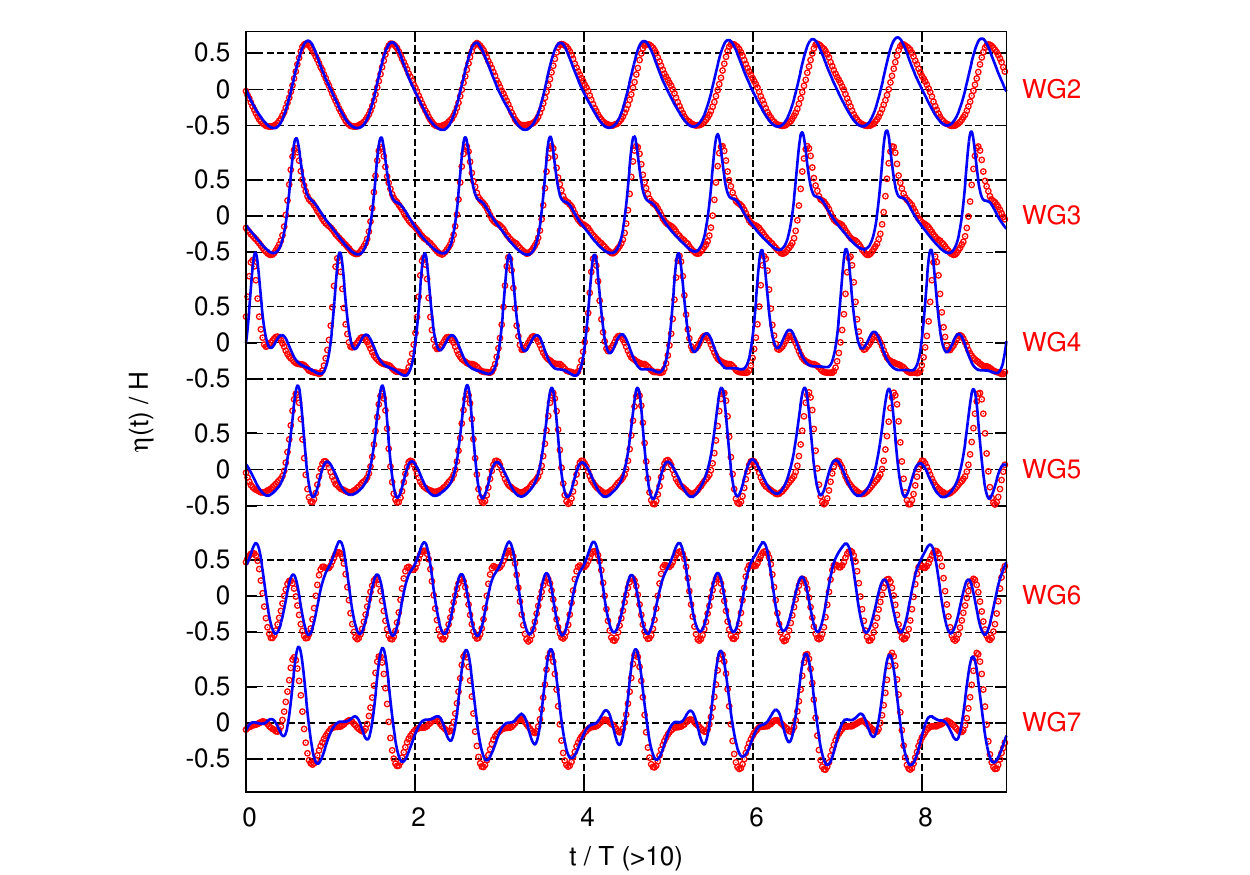}
}
\end {tabular}
\end {center}
\caption{\emph{{\emph{(top)}}Time series of $\eta(x)$ profiles (at $t=4T,6T,9T,15T$) showing ``strong'' transformation of {\bf{SLL}} waves; the coordinates are in $\bm{m}$. {\emph{(bottom)}} Validation of normalized $\eta(t)$ signals against the experiments reported in \cite{beji94} for nine wave periods.}}
\label{fig:SLLval}
\end{figure}

The {\bf{SL}} waves are approximately $1.8\times$ longer than the {\bf{SH}} waves (cf. \autoref{tab:tab6}). Long wave propagation over a submerged obstacle is characterized by extensive short wave generation on the lee side \cite{grue92,beji94}; wave transformation is thus ``strong'' in the case of both {\textbf{SLL}} as well as {\textbf{SLH}} waves. Both $H/\lambda\propto x$ and $kh \propto x$ hold during strong transformation owing to extensive harmonic generation on the lee side of the obstacle. Thus, the {\bf{SLL}} waves have also been subjected to a mesh dependence analysis using the same NWT setup and range of mesh sizes as considered in the {\bf{SH}} case. The normalized free surface elevation $(\eta(t)/(0.5H))$ profiles of {\bf{SLL}} waves, recorded at WG2, WG4 and WG6, have been compared for different mesh sizes during the interval $t\in [10T:11T]$ and reported at the bottom in \autoref{fig:SLLgind}. It should be noted that the {\bf{SLL}} waves propagate in near-shallow water with $\approx 40\%$ larger group celerity $(C_G\approx 1.62\,m/s)$ in comparison to the {\bf{SH}} waves $(C_G\approx 1.17\,m/s)$. Thus, local $\eta (t)$ signals reach a ``fully developed state'' much earlier (in terms of $t/T$) in the {\bf{SLL}} case. It is evident from \autoref{fig:SLLgind} that the wave topology becomes mesh independent beyond $1536\times 100$ at WG2 and WG4 but $\eta(t)$ continues to evolve beyond $3072\times 100$ in the downslope region at WG6. Stronger mesh dependence observed at WG6 is largely attributable to: (a) larger steepness and (b) polychromatic character of waves transmitted to the lee side of the bar \cite{qchen16}. Nonetheless, in an attempt to maintain consistency with the {\bf{SH}} setup, $3072\times 100$ is selected as the independent mesh for {\bf{SLL}} wave simulations. 
\\
The spatial development of the {\bf{SLL}} wave train is depicted at the top in \autoref{fig:SLLval}. The incident waves initially shoal over the windward/weather face of the bar and become progressively asymmetrical $(t>4T)$. This indicates an energy-gain by higher bound harmonics. 
As the waves travel over the bar crest (which is a non-dispersive medium) triplet resonance occurs \cite{beji93} and a prominent free-wave appears behind the steepened crest $(t=6T)$. As the waves propagate over the leeward face of the bar, the train ``de-shoals'' \cite{beji93} and the primary wave breaks up into several smaller amplitude waves. Through Fourier decomposition, Huang and Dong \cite{huang99} discovered that during de-shoaling, energy transfer is largely directed from the higher bound harmonics to free waves owing to a dramatic reduction in topological non-linearity in deep(er) water. This amplitude-gain by free waves on the lee side of the obstacle is clearly observed in \autoref{fig:SLLval} for $t\geq 9T$. Unlike the bound harmonics, the free waves propagate {\emph{relative}} to the carrier waves and are governed by the dispersion relation $(n\omega)^2=gk_n \tanh(k_n h)$ \cite{grue92}. The leeward side is thus characterized by extensive relative motion amongst the waves. \\
The {\bf{SLL}} simulations have also been validated against experiments \cite{beji94,bejid94} in terms of normalized $\eta(t)$ profiles measured at six wave gauge locations (WG2-WG7); the results are reported at the bottom in \autoref{fig:SLLval}. The overall agreement between the proposed NWT model and experiments is observed to be good. As a matter of fact, long-time $(>5T)$ validation against wave transformation experiments is seldom attempted in the literature (\cite{bihs16} being an exception). Thus, the consistency retained between the present NWT simulations and experiments \cite{beji94} over nine wave periods is commendable and is, in fact, superior to some of the recent simulations reported in the literature \cite{zhang16,qchen16,ji17}.  
\subsection{Strong transformation: SLH waves $(T=2.0\,s;H=4.0\,cm)$} \label{ssec:SLHwaves}
\begin {figure}[!ht]
\begin {center}
\begin {tabular}{c}
{\centering
\includegraphics[trim=0mm 0mm 0mm 0mm, clip, width = 15cm]{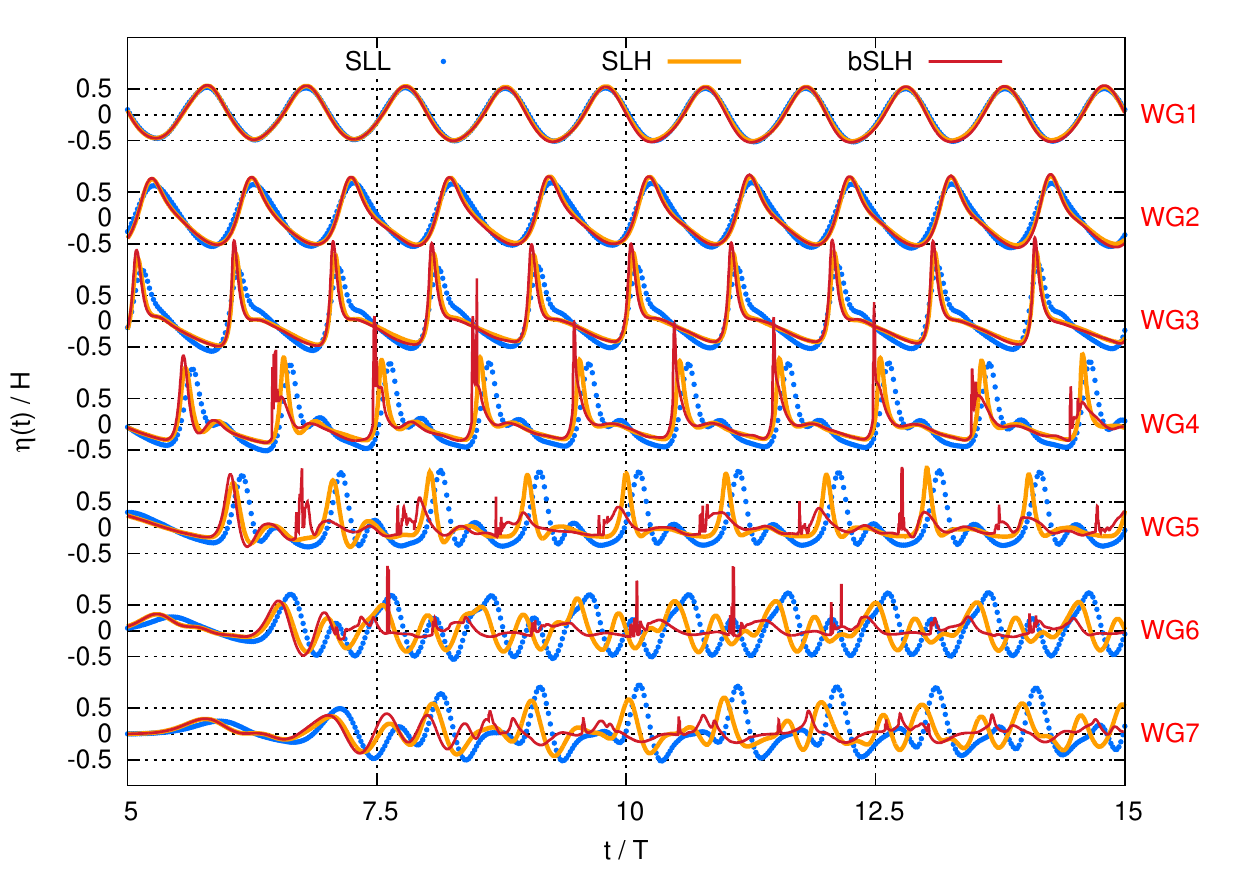}
}
\end {tabular}
\end {center}
\caption{\emph{Normalized $\eta(t)$ profiles of {\bf{SLL}}, {\bf{SLH}} and breaking {\bf{SLH}} (``bSLH'') waves recorded at seven wave gauge locations (WG1-WG7) during $t \in [5T:15T]$; amplitude dispersion between non-breaking {\bf{SLL}} and {\bf{SLH}} waves is evident at WG4-WG7.}}
\label{fig:SLcomp_t}
\end{figure}

Sinusoidal low frequency {\emph{high}} {\bf{(SLH)}} wave propagation over the trapezoidal bar is one of several cases considered by Huang and Dong \cite{huang99} whilst numerically simulating wave deformation and vortex generation over submerged breakwaters. The {\bf{SLH}} waves are twice as high as the {\bf{SLL}} waves (cf. \autoref{tab:tab6}). Hence, stronger wave transformation and greater topological non-linearity (over the bar crest) are expected in the {\bf{SLH}} case. In fact, there is some evidence that the {\bf{SLH}} waves are susceptible to breaking during transformation. This is because the {\bf{SLH}} waves are equivalent in steepness $(H/\lambda={0.011})$ to another wave design $(T=2.5\,s,H=5.2\,cm,h=0.4\,m\,\vdash {\mathsf{Ur}}\approx 18, H/\lambda={0.011})$ considered by Beji and Battjes \cite{beji93} in their experiments, that yielded plunging breakers over the bar. Unfortunately, there is a lack of evidence whether breaking occurred in Huang and Dong's \cite{huang99} simulations or not; this seems highly unlikely given a (relatively) coarse spatial resolution of $nx_{\lambda}=93$ adopted by them in the NWT. In this respect, we demonstrate here that breaking of {\bf{SLH}} waves in the NWT is suppressed by numerical damping of wave height over the bar crest and breaking can (rather) be ``triggered'' by increasing the blending parameter $\mathscr{S}$ (cf. \autoref{eq:schbln}). Existence of such a relationship is expected; accurate simulation of the extent of wave-steepening over the bar crest necessitates higher-order treatment of momentum advection (cf. \autoref{ssec:schbln}). 
\begin {figure}[!ht]
\begin {center}
\begin {tabular}{c c}
\multicolumn{2}{c}{
\centering
\includegraphics[trim=0mm 39mm 0mm 37mm, clip, width = 15cm]{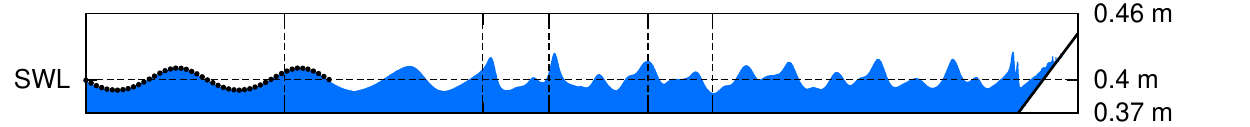}
}\\
\multicolumn{2}{c}{
\centering
\includegraphics[trim=0mm 39mm 0mm 37mm, clip, width = 15cm]{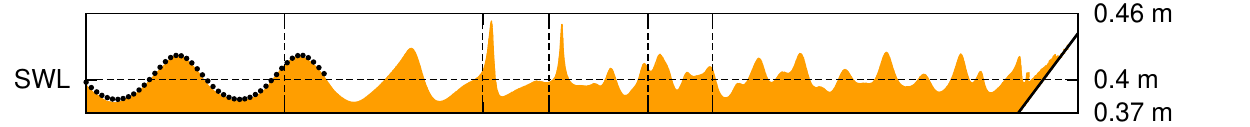}
}\\
\multicolumn{2}{c}{
\centering
\includegraphics[trim=0mm 39mm 0mm 37mm, clip, width = 15cm]{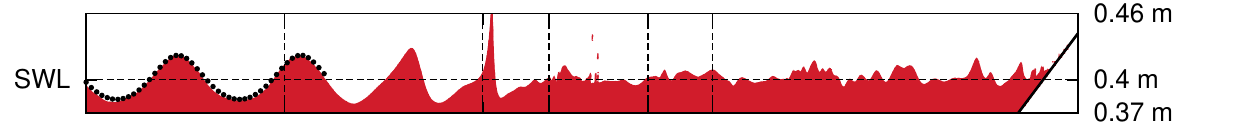}
}\\
\multicolumn{2}{c}{
\centering
\includegraphics[trim=0mm 33mm 0mm 35mm, clip, width = 15cm]{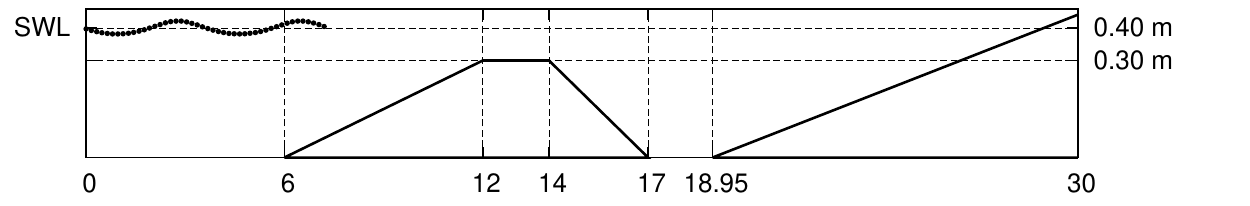}
}\\
\midrule
{
\centering
\includegraphics[trim=10mm 0mm 18mm 0mm, clip, height = 6cm]{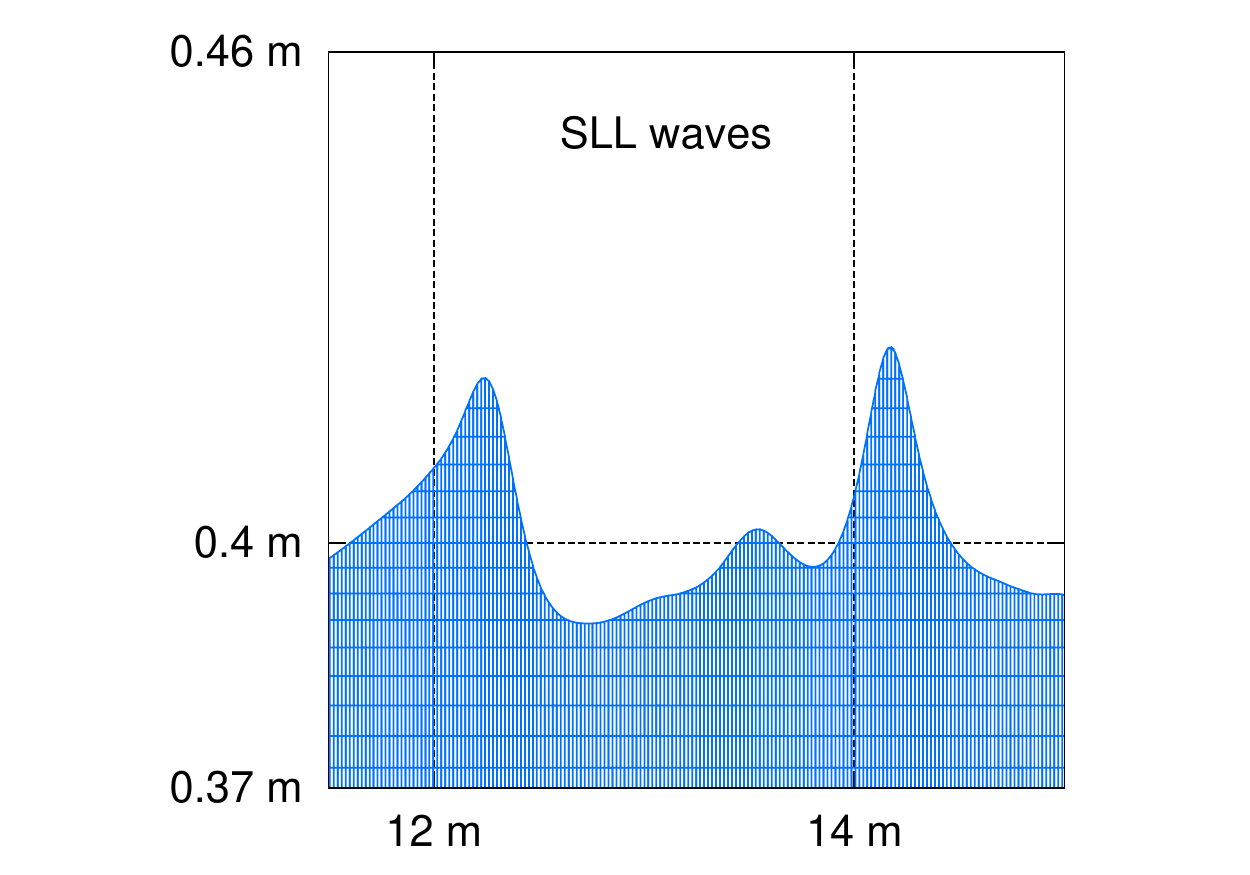}
}&
{
\centering
\includegraphics[trim=15mm 0mm 18mm 0mm, clip, height = 6cm]{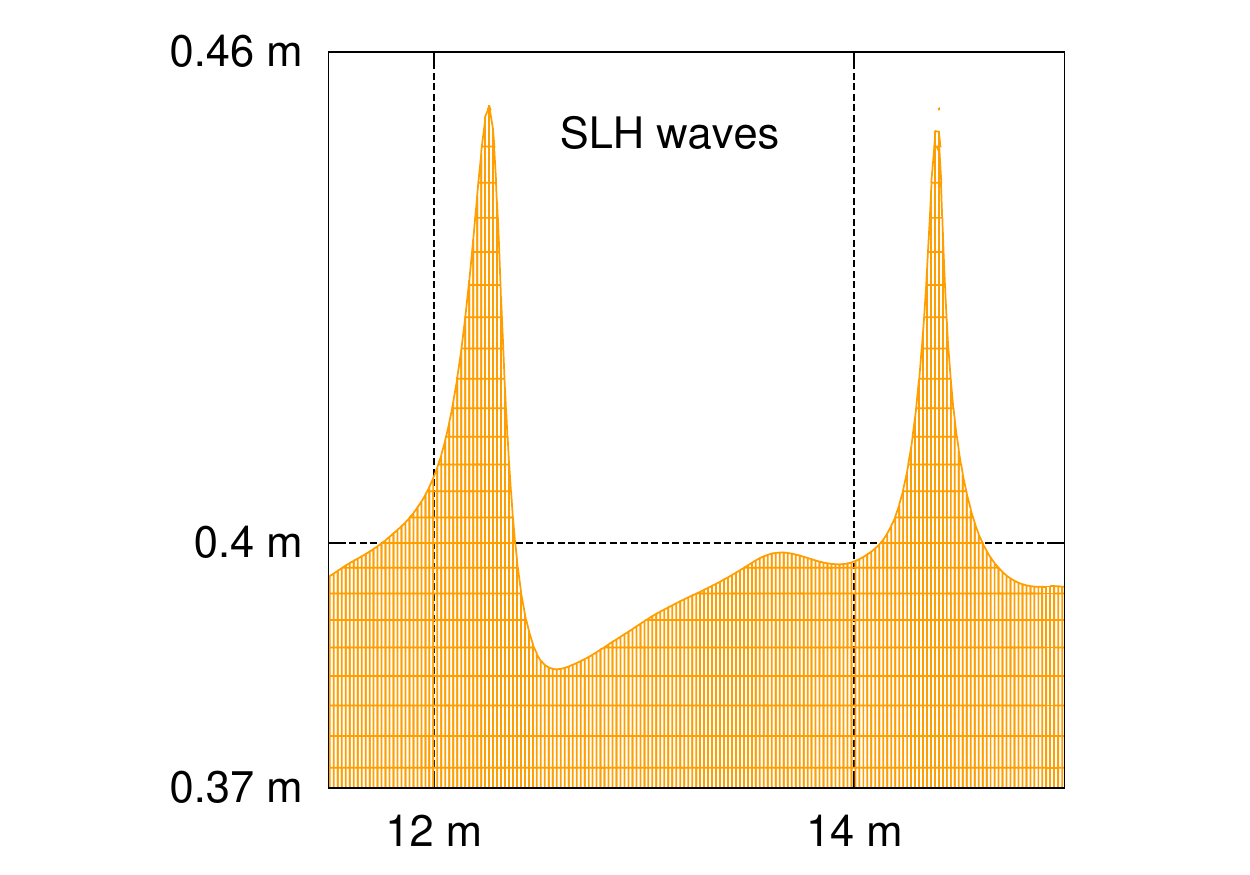}
}
\end {tabular}
\end {center}
\caption{\emph{{\emph{(top)}} Spatial profiles of {\bf{SLL}}, {\bf{SLH}} and breaking {\bf{SLH}} (``bSLH'') waves; {\emph{(bottom)}} Topological detail of the waves steepening over the bar crest.}}
\label{fig:SLcomp_x}
\end{figure}\\
Given the non-availability of $\eta(x)$ or $\eta(t)$ signals for {\bf{SLH}} wave transformation in \cite{huang99}, it would be contributive to first illustrate spatio-temporal development of the {\bf{SLH}} wave train against {\bf{SLL}} waves in a comparative framework. Given much higher incident waves, the domain height is increased to $\mathbb{H}=0.75\,m$ which involves lengthening the propagation region to $\mathbb{L}=37.7\,m$ (which is the actual length of the experimental wave tank used in \cite{beji94}). Rather than resorting to mesh dependence assessment, we directly select: $nx=1992\vdash nx_{\lambda}\approx 196$ and $ny=100\vdash ny_H\approx 13$. Further, $\left(\begin{matrix} \wp \\ \mathscr{S} \end{matrix} \right) \equiv \left(\begin{matrix} 0.18 \\ 1 \end{matrix} \right)$ and $\left(\begin{matrix} 0.18 \\ 0.6 \end{matrix} \right)$ have been considered for non-breaking and breaking {\bf{SLH}} simulations respectively; the value of $\wp$ is adopted from the parametric selection procedure reported in \autoref{fig:smlstptopl}. Time is uniformly advanced using the forward Euler method with $\Delta t=T/5000$ retained from {\bf{SH}} and {\bf{SLL}} simulations. The above simulation setup helped ensure that $C_{max} \leq 0.25$ even when waves broke (as weak plungers \cite{brucker14}) over the bar crest.  \\
Spatial topologies $(\eta(x))$ of both breaking and non-breaking {\bf{SLH}} wave trains at $t=15T$ are shown in \autoref{fig:SLcomp_x} and compared with {\bf{SLL}} waves generated using the same NWT setup (albeit with $\wp=0.5H$). It is demonstrated that, during {\bf{SLH}} wave transformation, much stronger topological non-linearity is induced over the bar crest. Zoomed-in views of wave profiles over the bar crest (see bottom in \autoref{fig:SLcomp_x}) indicate that, unlike {\bf{SH}} waves, shoaled {\bf{SL}} trains resemble shallow-water {\texttt{cnoidal}} waves \cite{beji94}. Interestingly, doubling the incident wave height yields very similar $\eta(x)$ distribution on the lee side of the obstacle; albeit with higher waves. However, an examination of normalized $\eta(t)$ signals in both cases (cf. \autoref{fig:SLcomp_t}) reveals that $\eta(t)/H$ actually decreases on the lee side of the bar in the {\bf{SLH}} case. This finding is substantiated by a general observation made in \cite{grue92,huang99} and recently in \cite{kamath17} that $\eta(t)/H \downarrow$ on the lee side as (incident) $H\uparrow$ on the weather side of a submerged obstacle.  
Following an assessment of vorticity fields (presented later in \autoref{ssec:vort_dyn}) it has been hypothesized that $\eta(t)/H \downarrow$ on the lee side stems from rotational dissipation of packet energy towards inducing pervasive vortex generation in air, especially over the bar crest. Furthermore, \autoref{fig:SLcomp_x} and \autoref{fig:SLcomp_t} jointly support the claim that breaking can be triggered in the {\bf{SLH}} wave train by $\downarrow {\mathscr{S}}$. It is obvious that wave breaking induces considerable energy dissipation and height damping past the breaking point ($x \gtrapprox 12\,m$; cf. \autoref{fig:SLcomp_x}). Moreover, as evidenced from \autoref{fig:SLcomp_x} and \autoref{fig:SLcomp_t}, breaking induces extensive short wave generation over the leeward-side of the bar \cite{beji93,kamath17}. Short wave formation is manifested in packet energy transfer from $n\omega\,\,(n=1- 3)$ to higher harmonic frequencies $n\geq 4$ \cite{kamath17} and since $C_{n \geq 4} \ll C_{1}$, the higher frequency waves almost appear ``frozen in space'' when visualized against the carrier wave train. 
\subsection{Vorticity dynamics and harmonic decomposition} \label{ssec:vort_dyn}
One of the key advantages of the proposed NWT model is the two-phase NSE philosophy adopted for modeling the waves. As evident from our previous work \cite{sas17a}, the NSE paradigm facilitates interrogation of the solution in both air and water phases, thus providing greater insight into momentum dynamics. In context to submerged bars, the design goal is to induce short wave behavior on the lee side, thereby preventing beach erosion \cite{kamath17}. In addition, vorticity dynamics induced over the weather and lee faces as well as over the bar crest is of great interest from a design point of view since persistent vortical activity (in water) would induce hydrodynamic scour in the long-term. Hence, for the strong wave transformation scenarios, it is contributive to investigate the vorticity dynamics in air as well as the water phases. The spatial variation of amplitudes $a^{(n)}\,\,;\,\,n=1- 3$ of the first three harmonics is also presented as a rigorous validation exercise. \\  
As the incident waves encounter the obstacle, energy transfer to higher harmonics occurs from the fundamental; the higher harmonics consist of both bound as well as free waves \cite{huang99,grue92}. However, separation of free and bound components for $n>2$ is not our concern here; a {\emph{single wave gauge}} based Discrete Fourier Transform (DFT) is hence sufficient.  For DFT, the free surface elevation $\eta(x,t)$ has been recorded in each cell falling within $x\in [4\,m:22.5\,m]$ every $\Delta t=0.0004\,s$ during the time interval $t\in[14T:15T]$. We proceed by defining the Fourier transform $\hat{\eta}^{(n)}(x)$ of the free-surface elevation,
\begin{equation} \label{eq:DFT_surfel}
\hat{\eta}^{(n)}(x)=\dfrac{\omega}{2\pi} \int_{0}^{2\pi/\omega} \eta(x,t)\exp(-\mathrm{i} n\omega t)\,\mathrm{d}t
\end{equation} 
where $\mathrm{i}=\sqrt{-1}$ and $n=1,2,3,\dotsc,$. The integral in \autoref{eq:DFT_surfel} is evaluated using the composite Simpson's rule. Then, $a^{(n)}(x)$ could be directly obtained using \cite{dean91},
\begin{equation} \label{eq:harmonic_amps}
a^{(n)}(x)=2\sqrt{\left[\operatorname{\mathbb{R}e}\left(\hat{\eta}^{(n)}(x)\right)\right]^2+\left[\operatorname{\mathbb{I}m}\left(\hat{\eta}^{(n)}(x)\right)\right]^2}
\end{equation}
where the real and imaginary parts of the complex elevation $\hat{\eta}^{(n)}(x)$ are given by, 
\begin{flalign} \label{eq:surfel_Re_Im}
&\operatorname{\mathbb{R}e}\left(\hat{\eta}^{(n)}(x)\right)= \nonumber \\
&\dfrac{\Delta t}{3T}\left\{\eta(x,0)+2\displaystyle\sum\limits_{i=1}^{N/2-1} \eta(x,t_{2i})\cos\left(n\omega t_{2i}\right)+4\displaystyle\sum\limits_{i=1}^{N/2} \eta(x,t_{2i-1})\cos\left(n\omega t_{2i-1}\right)+\eta(x,T)\right\} \nonumber \\
&\operatorname{\mathbb{I}m}\left(\hat{\eta}^{(n)}(x)\right)= \nonumber \\
&\dfrac{\Delta t}{3T}\left\{2\displaystyle\sum\limits_{i=1}^{N/2-1} \eta(x,t_{2i})\sin\left(n\omega t_{2i}\right)+4\displaystyle\sum\limits_{i=1}^{N/2} \eta(x,t_{2i-1})\sin\left(n\omega t_{2i-1}\right)\right\} 
\end{flalign}
where $N\equiv T/\Delta t$ is the number of samples and $t \equiv N\Delta t$ is the physical time.
\begin{landscape}
\thispagestyle{empty}
\begin {figure}[!ht]
\begin {center}
\begin {tabular}{c c c}
{\rotatebox[origin=l]{90}{\textsc{SLL waves}}} & {\centering
\includegraphics[trim=0mm 0mm 0mm 0mm, clip, height = 4.5cm]{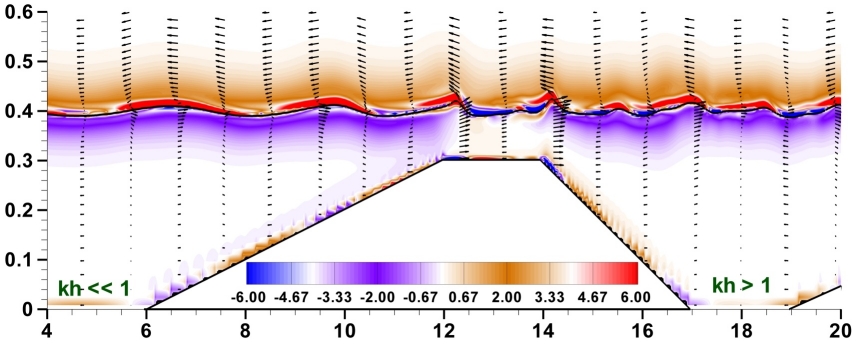}
}&{\centering
\includegraphics[trim=0mm 23mm 0mm 10mm, clip, height = 4.5cm]{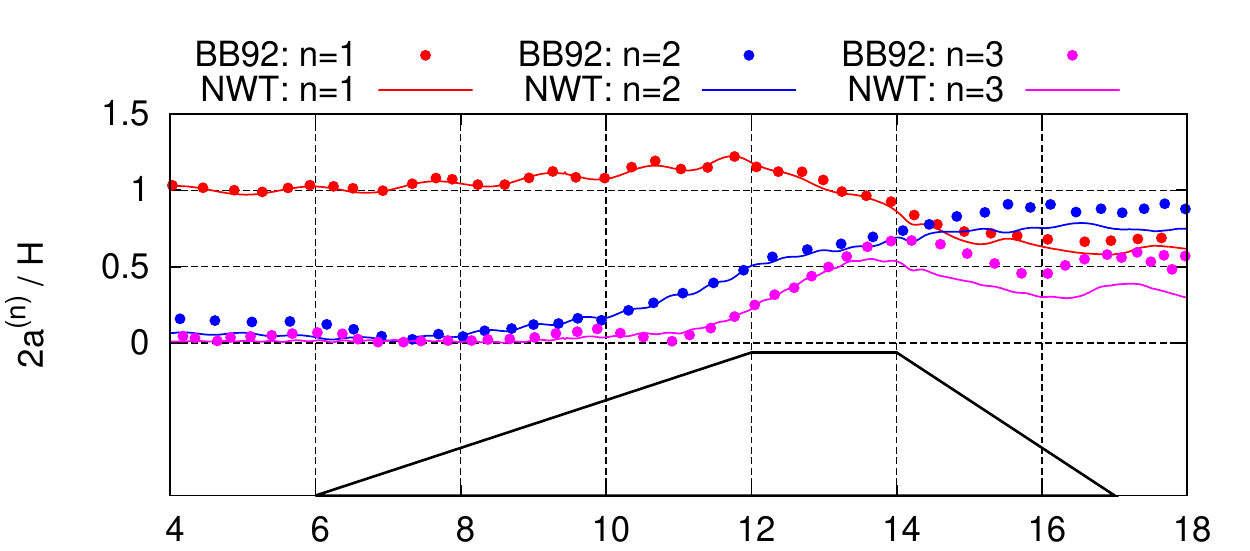}
}\\
\midrule
{\rotatebox[origin=l]{90}{\textsc{SLH waves}}} & {\centering
\includegraphics[trim=0mm 0mm 0mm 0mm, clip, height = 4.5cm]{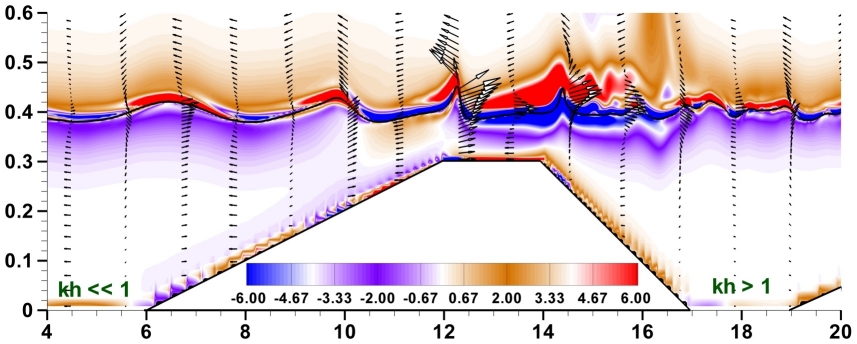}
}&{\centering
\includegraphics[trim=0mm 23mm 0mm 10mm, clip, height = 4.5cm]{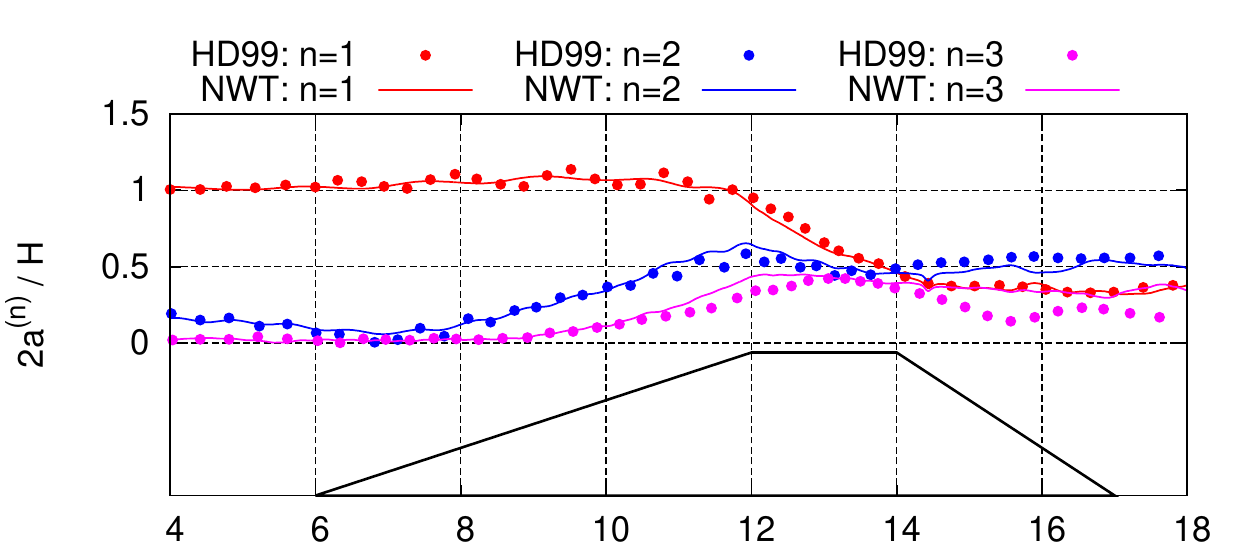}
}\\
\midrule
{\rotatebox[origin=l]{90}{\textsc{bSLH waves}}} & {\centering
\includegraphics[trim=0mm 0mm 0mm 0mm, clip, height = 4.5cm]{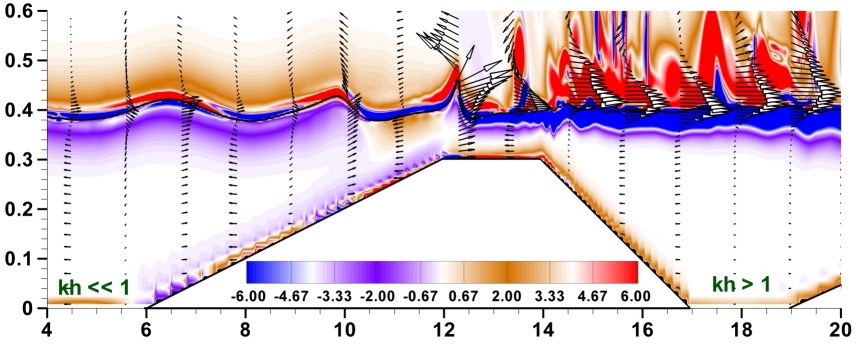}
}&{\centering
\includegraphics[trim=0mm 23mm 0mm 10mm, clip, height = 4.5cm]{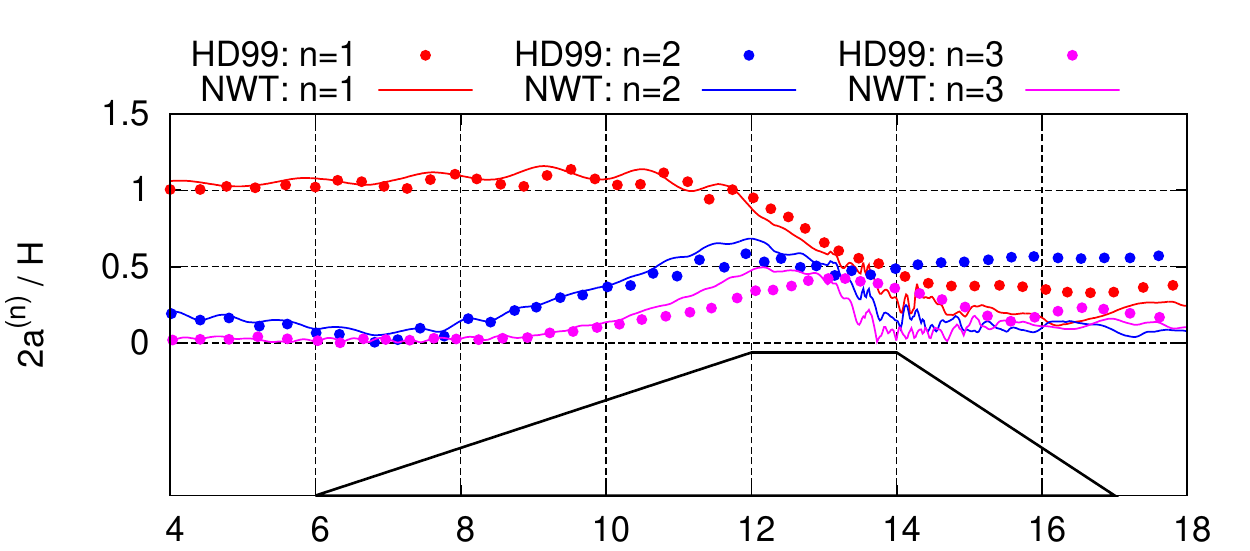}
}
\end {tabular}
\end {center}
\caption{\emph{{\emph{(left)}} Contours of vorticity (in $s^{-1}$) superimposed with velocity vectors and the air-water interface $(f=0.5)$ contour (thick black line) for {\bf{SLL}}, {\bf{SLH}} and breaking {\bf{SLH}} waves at $t=15T$. {\emph{(right)}} Fourier decomposition of harmonic amplitudes $a^{(n)}$ validated against the simulations of BB92--Beji {\emph{et al.}} \cite{beji92} and HD99--Huang and Dong \cite{huang99}.}}
\label{fig:vortdyn}
\end{figure}
\end{landscape}
The spatial variation of harmonic amplitudes $a^{(n)}$ obtained using \autoref{eq:harmonic_amps} is reported in \autoref{fig:vortdyn}. Evidently, an acceptable validation is obtained against the simulations reported in the literature \cite{beji92,huang99}. Energy transfer from the fundamental $(n=1)$ to higher harmonics $(n=2,3)$ over the bar crest is evident from the plots. Interestingly, a reduction of $\left|\eta(t)/H\right|$ on the lee side for {\bf{SLH}} waves is also reflected in the amplitude decomposition. During breaking, the greatest energy loss occurs from $n=2$ followed by the fundamental $(n=1)$. It is reaffirmed from \autoref{fig:vortdyn} that wave-breaking does not induce any upstream influence in the NWT.   \\  
The contours of vorticity $\vec{\Omega}$ (superimposed with velocity vectors and the free-surface contour) alongwith $a^{(n)}(x)$ decomposition are shown in \autoref{fig:vortdyn} for {\bf{SLL}}, {\bf{SLH}} and breaking {\bf{SLH}} waves. The velocity vectors clearly indicate a region of intermediate/deep water $(kh>1)$ on the lee side of the bar in all cases (which is desired). Unlike rectangular dikes, gradual shoaling over the bar upslope precludes flow separation on the lee side; this exhibits reduced susceptibility of the lee side to hydrodynamic scour \cite{huang99}. The $\vec{\Omega}$ contours indicate that the {\bf{SLH}} waves act to expand the region of interaction between air and water by inducing strong regions of rotation, especially over the bar crest. It is hypothesized that the observed vortical activity would drain packet energy to viscous effects thereby causing relatively smaller waves to be transmitted to the lee side of the bar $(\uparrow H \vdash \downarrow \left|\eta(t)/H\right|)$ \cite{grue92}. The region of interaction around the air-water interface gets further expanded as {\bf{SLH}} waves break over the bar crest with consecutive breaking events leading to the formation of a surface current (cf. bottom in \autoref{fig:vortdyn}). \\
To the best of the authors' knowledge, the vorticity dynamics of breaking/non-breaking wave transformation has never been reported in the literature and is therefore one of the novel aspects of the present work. At this juncture, the authors would also like to assert the validity of the vortex dynamics reported in \autoref{fig:vortdyn} from the standpoint of the CFD paradigm used for simulation. Based on the {\emph{instantaneous}} Navier-Stokes equations, the NWT solves for the instantaneous velocity field $(\vec{V})$ induced by the waves and not the mean ($\overline{V}$; unlike RANS) or filtered ($\tilde{V}$; unlike LES) fields. It is obvious that both $\overline{V}$ and $\tilde{V}$ are extracted from $\vec{V}$ with the residual terms being accounted for using a turbulence closure model. Hence, it is argued that the present NWT model automatically accounts for the generation of ``two-phase turbulence'' at the wave surface and within the air and water phases during breaking. Thus, there has been a deliberate effort to ``not model'' the effects of turbulence but rather to allow the flow solver to capture turbulent and vortical scales to the extent that is allowed by the mesh. Whilst this approach may strike one as being overtly simplistic, the same in reality preserves the flow simulation from turbulence-modeling-induced artifacts such as over-production of turbulent viscosity \cite{germano99} and unrealistic dampening of fluctuations caused as a consequence.  

\section{Summary and outlook} \label{sec:summary}
We propose a robust numerical wave tank (NWT) algorithm (and accompanying design paradigm) for simulating ocean wave propagation and wave-structure interaction scenarios in a two-phase, Navier-Stokes framework. Robustness of the proposed algorithm is primarily manifested in: 
\begin{itemize}[noitemsep]
\item a novel kinematic-stretching based modified inflow boundary wavemaker which is volume-preserving and is characterized by a single design variable,
\item a strategy for blending low and higher-order momentum advection schemes that is effective in restricting numerical damping of steep waves without inducing additional wave distortion at the air-water interface and,
\item a simplified methodology for treating non-Cartesian submerged/emergent boundaries on staggered grids which ensures sustenance of a tight coupling between computed pressure and velocity fields. 
\end{itemize}
The proposed NWT model has been successfully benchmarked against several problems including generation of monochromatic waves of varying steepness, simulation of free-wave generation during piston-type wavemaker motion, generation of a deep-water irregular wave train and (weak and strong) wave transformation (including weak plunging of waves) occurring over a submerged trapezoidal bar. Very good agreement with analytical, numerical and experimental studies reported in the literature is obtained. The proposed NWT algorithm has already been parallelized using MPI \cite{sasLNCE19} and is currently being applied toward simulating the hydrodynamics of OWC devices.

\footnotesize
\bibliography{references}
\bibliographystyle{elsarticle-num}

\end{document}